\documentclass{article}
\usepackage[dvips]{graphicx}
\usepackage{latexsym}

\renewcommand{\arraystretch}{1.3}
\makeatletter
\newdimen\normalarrayskip              
\newdimen\minarrayskip                 
\normalarrayskip\baselineskip \minarrayskip\jot
\newif\ifold             \oldtrue            \def\new{\oldfalse}
\def\arraymode{\ifold\relax\else\displaystyle\fi} 
\def\eqnumphantom{\phantom{(\theequation)}}     
\def\@arrayskip{\ifold\baselineskip\z@\lineskip\z@
     \else
     \baselineskip\minarrayskip\lineskip2\minarrayskip\fi}
\def\@arrayclassz{\ifcase \@lastchclass \@acolampacol \or
\@ampacol \or \or \or \@addamp \or
   \@acolampacol \or \@firstampfalse \@acol \fi
\edef\@preamble{\@preamble
  \ifcase \@chnum
     \hfil$\relax\arraymode\@sharp$\hfil
     \or $\relax\arraymode\@sharp$\hfil
     \or \hfil$\relax\arraymode\@sharp$\fi}}
\def\@array[#1]#2{\setbox\@arstrutbox=\hbox{\vrule
     height\arraystretch \ht\strutbox
     depth\arraystretch \dp\strutbox
     width\z@}\@mkpream{#2}\edef\@preamble{\halign
\noexpand\@halignto
\bgroup \tabskip\z@ \@arstrut \@preamble \tabskip\z@ \cr}%
\let\@startpbox\@@startpbox \let\@endpbox\@@endpbox
  \if #1t\vtop \else \if#1b\vbox \else \vcenter \fi\fi
  \bgroup \let\par\relax
  \let\@sharp##\let\protect\relax
  \@arrayskip\@preamble}
\def\eqnarray{\stepcounter{equation}%
              \let\@currentlabel=\theequation
              \global\@eqnswtrue
              \global\@eqcnt\z@
              \tabskip\@centering
              \let\\=\@eqncr
 \halign to \displaywidth\bgroup
    \eqnumphantom\@eqnsel\hskip\@centering
    $\displaystyle \tabskip\z@ {##}$%
    \global\@eqcnt\@ne \hskip 2\arraycolsep
         $\displaystyle\arraymode{##}$\hfil
    \global\@eqcnt\tw@ \hskip 2\arraycolsep
         $\displaystyle\tabskip\z@{##}$\hfil
         \tabskip\@centering
    &{##}\tabskip\z@\cr}
\begingroup\ifx\undefined\newsymbol \else\def\input#1 {\endgroup}\fi

\catcode`\@=11
\def\marginnote#1{}

\newcount\hour
\newcount\minute
\newtoks\amorpm
\hour=\time\divide\hour by60 \minute=\time{\multiply\hour by60
\global\advance\minute by-\hour}
\edef\standardtime{{\ifnum\hour<12 \global\amorpm={am}%
        \else\global\amorpm={pm}\advance\hour by-12 \fi
        \ifnum\hour=0 \hour=12 \fi
        \number\hour:\ifnum\minute<10 0\fi\number\minute\the\amorpm}}
\edef\militarytime{\number\hour:\ifnum\minute<10 0\fi\number\minute}

\def\draftlabel#1{{\@bsphack\if@filesw {\let\thepage\relax
      \xdef\@gtempa{\write\@auxout{\string
          \newlabel{#1}{{\@currentlabel}{\thepage}}}}}\@gtempa \if@nobreak
    \ifvmode\nobreak\fi\fi\fi\@esphack} \gdef\@eqnlabel{#1}}
    \def\@eqnlabel{}
\def\@vacuum{}
\def\draftmarginnote#1{\marginpar{\raggedright\scriptsize\tt#1}}

\def\draft{
  \oddsidemargin -.5truein
  \def\@oddfoot{\footnotesize \sl preliminary draft \hfil
    \rm\thepage\hfil\sl\today\quad\militarytime}
  \let\@evenfoot\@oddfoot \overfullrule 3pt
    \let\label=\draftlabel
    \let\marginnote=\draftmarginnote
  \def\@eqnnum{(\theequation)\rlap{\kern\marginparsep\tt\@eqnlabel}%
    \global\let\@eqnlabel\@vacuum}

  }

\def\nn{\nonumber}
\def\beq{\begin{equation}}
\def\eeq{\end{equation}}

\def\ba{\beq\new\begin{array}{c}}
\def\ea{\end{array}\eeq}
\def\be{\ba}
\def\ee{\ea}

\newfont{\alef}{msbm10 at 12pt}
\newfont {\goth}{eufm10 at 11pt}
\def\mathbb#1{\hbox{{\alef #1}}}

\let\@@savethanks\thanks
\def\thanks#1{\gdef\thefootnote{\alph{footnote}}\@@savethanks{#1}}

\def\theequation{\arabic{section}.\arabic{equation}}

\title{{\bf Boundary Ring: a way to construct approximate
NG solutions with polygon boundary conditions \\I. $Z_n$-symmetric
configurations} \vspace{.5cm}}
\author{{\bf H. Itoyama}\thanks{E-mail: \
itoyama@sci.osaka-cu.ac.jp}\date{ } \\
{\small {\it Osaka City University, Japan}} \\ \\
{\bf A. Mironov}\footnote{E-mail: \ mironov@itep.ru; mironov@lpi.ru}
\date{ } \\
{\small {\it Lebedev Physics Institute}
and {\it ITEP, Moscow, Russia}}\\ \\
{\bf A. Morozov}\thanks{E-mail: \ morozov@itep.ru}
\date{ } \\ {\small {\it ITEP, Moscow, Russia}}
}

\begin{document}

\maketitle

\vspace{-11.5cm}

\begin{center}
\hfill OCU-PHYS 282\\
\hfill FIAN/TD-23/07\\
\hfill ITEP/TH-56/07\\
\end{center}

\vspace{9.0cm}

\begin{abstract}
\noindent We describe an algebro-geometric construction of
polygon-bounded minimal surfaces in $ADS_5$, based on consideration
of what we call the "boundary ring" of polynomials. The first
non-trivial example of the Nambu-Goto (NG) solutions for
$Z_6$-symmetric hexagon is considered in some detail. Solutions are
represented as power series, of which only the first terms are
evaluated. The NG equations leave a number of free parameters (a
free function).
Boundary conditions, which fix the free parameters, are imposed on
truncated series.
It is still unclear if explicit analytic formulas can be found in
this way, but even approximate solutions, obtained by truncation of
power series, can be sufficient to investigate the Alday-Maldacena
-- BDS/BHT version of the string/gauge duality.
\end{abstract}

\bigskip

\def\thefootnote{\arabic{footnote}}

\newpage

\tableofcontents

\section{Introduction}

\subsection{BDS/BHT conjecture}

One of the most important discoveries of the last years in modern
quantum field theory is the BDS conjecture \cite{BDS}, which --
based on extensive investigations of many people during the last
decades -- claims that the (MHV?) amplitude of the $n$-gluon
scattering in the planar limit of $N=4$ SYM theory factorizes and
exponentiates: \be {\cal A}({\bf p_1}, \ldots {\bf p_n}|\lambda) =
{\cal A}_{tree}{\cal A}_{IR}{\cal A}_{finite} \ee where $\lambda$ is
the t'Hooft's coupling constant, ${\cal A}_{tree}$ and ${\cal
A}_{IR}$ are the tree and IR-divergent amplitudes (the latter one is
explicitly expressed through the celebrated anomalous dimension
function $\gamma(\lambda)$ -- a subject of intensive but still
unfinished research of the last years, an eigenvalue of a yet
sophisticated integrable problem and a solution to an integral
Bethe-Anzatz equation \cite{BA}) and \be {\cal A}_{finite} = \exp
\left(\frac{1}{4}\gamma(\lambda) F_n^{(1)}({\bf p_1}, \ldots {\bf
p_n})+ g_n(\lambda)\right) \ee where \cite{BHT} \be F_n^{(1)} =
\oint\oint_\Pi \frac{dy^\mu dy'_\mu} {(y-y')^{2+\epsilon}}
\label{BHT} \ee In this spectacular formula $\Pi$ is a polygon in
the $4d$ Minkovski space with coordinates $y_0,y_1,y_2,y_3$, which
is formed by $n$ null vectors ${\bf p_1},\ldots, {\bf p_n}$. Polygon
is closed because of the energy-momentum conservation, ${\bf
p_1}+\ldots+{\bf p_n}={\bf 0}$. See \cite{mmt1} for a more detailed
presentation of the BDS/BHT conjecture.

If BDS/BHT conjecture is true, it is the first exhaustive solution
of perturbative quantum field theory problem in $4d$. Today it is
constrained by a few restrictions:

-- the theory has maximal supersymmetry ($N=4$),

-- only planar limit is considered,

-- only MHV (maximal helicity violating) amplitudes are carefully
analyzed,

-- the answer is conjectured only for scattering amplitudes, not for
generic correlators of Wilson loops,

-- there is no proof of the conjecture and there are even doubts
that it is fully correct.

\subsection{Alday-Maldacena conjecture}

If BDS conjecture is true, the amplitude should have the same
momentum-dependence in the strong-coupling regime. This means that
the function $F_n^{(1)}\big({\bf p_1}, \ldots {\bf p_n}\big)$ should
be also reproduced at the string side of the string/gauge (AdS/CFT)
duality in all orders of the strong-coupling expansion. In
particular, since in the leading order it is given by a regularized
minimal area of world-sheet embedding into the $AdS_5$ space, one
expects that \be F_n^{(1)}\big({\bf p_1}, \ldots {\bf p_n}\big) \
\stackrel{(\ref{BHT})}{=}\ \oint\oint_\Pi \frac{dy^\mu dy'_\mu}
{(y-y')^{2+\epsilon}} = {\rm Minimal\ Area}_\epsilon \label{FMS} \ee
where the set of momenta at the l.h.s. specifies the boundary
conditions at the r.h.s. In a recent breakthrough made in the paper
\cite{am1}, see also \cite{amfirst}-\cite{amlast} and
\cite{BHT,mmt1}, the first steps are done towards accurate
formulation and proof of (\ref{FMS}). The most important step of
\cite{am1} is a Kallosh-Tseytlin (KT) \cite{KT} $T$-duality
transformation (involving transition from NG to $\sigma$-model
actions on the world sheet and back, since KT transformation can be
performed only in the $\sigma$-model with no Virasoro-like
constraints imposed), which allows to formulate boundary conditions
at the r.h.s. of (\ref{FMS}) in elegant way: the boundary of the
$2d$ surface in $AdS_5$ is the same polygon $\Pi$ which appeared in
(\ref{BHT}). In \cite{am1} explicit solution for the minimal surface
is found in the particular case of $n=4$ (using the previous results
of \cite{pream} and especially \cite{ka}), see \cite{mmt1,mmt2,popo}
for more -- sometime intriguing -- details about these solutions.
The next steps of \cite{am1} involve regularization of the minimal
action and KLOV-style \cite{KLOV} interpolation of the functions
$\gamma(\lambda)$ and $g_n(\lambda)$, but these steps are beyond our
discussion in the present paper.

\subsection{The goal of this paper}

We are going to concentrate here on the minimal surface problem: on
search of solutions to the NG and $\sigma$-model equations in AdS
background with boundary conditions requiring that the corresponding
$2d$ surface ends on a polygon $\Pi$ located at the boundary of AdS
space. Note, that it makes sense to speak about a polygon (with the
boundary made of straight segments), provided it is located at the
AdS boundary since AdS space is asymptotically flat (our problem
could not be equally well formulated , say, in the spherical
geometry). Construction of minimal surfaces with given boundary
conditions is a classical and difficult problem (known as the
Plateau problem in mathematical literature). Still, if both the
BDS/BHT and AdS/CFT conjectures are true, this problem should
possess a more or less explicit solutions for the particular case of
polygons at the boundary of $AdS_5$. A kind of explicit solution
seems needed because what we need is {\it regularized} area, which
is somewhat difficult to evaluate (and even define) without knowing
the solution. We shall not solve this problem to the end in this
paper, only the first step will be done, but this seems to be a
decisive step, opening the way to analyze many other examples.

In what follows we use the notation of papers \cite{mmt1,mmt2} and
also refer to those papers for detailed description of our
understanding of Alday-Maldacena program. The $AdS_5$ space of
interest (it is actually a $T$-dual of the "physical" one) has the
metric \be \frac{-dy_0^2+dy_1^2+dy_2^2+dy_3^2+dr^2}{r^2} \ee which
is induced from the flat one in $R_6^{--++++}$ on the hypersurface
\be -Y_{-1}^2-Y_0^2+Y_1^2+Y_2^2+Y_3^2+Y_4^2 = -1 \ee where $Y_i =
v_i = zy_i$, $i=0,1,2,3$, $Y_{-1}+Y_4=z=1/r$ and $Y_{-1}-Y_4=w=qz$,
see \cite{am1} and s.3.3 of \cite{mmt1} for more details.

\subsection{Suggested approach}

Our first suggestion is to begin with solving the NG equations for
the functions $r(y_1,y_2)$, $y_0(y_1,y_2)$, and $y_3(y_1,y_2)$, and
only after that proceed to solution of the $\sigma$-model ones for
two more functions $y_1(\vec u)$ and $y_2(\vec u)$. This allows one
to minimize the number of unknown functional dependencies at the
first stage of calculations.

The second suggestion is to assume, at least temporarily, that
minimal surfaces in question are algebraic surfaces, described by
polynomial equations. Then the question is reduced to the search of
appropriate ansatze for these equations. We did not justify this
assumption in this paper: it ends with description of a power series
solution and it is yet unclear whether the series is ever reduced to
a ratio of polynomials. However, the algebraic assumption, even if
not {\it a posteriori} true, plays an important role in arriving to
this power-series ansatz.

The third suggestion is to begin with the simplified boundary
conditions. We actually {\it over}simplify them in the present
paper, since our main task is to show the way to solve the NG
equations beyond the "classical" examples. In order to compare with
BDS/BHT formulas one needs rather general boundary conditions, but
this is a rather straightforward generalization which would,
however, obscure the main message of this paper and these
generalizations will be discussed elsewhere.

We use three levels of simplification.

$\bullet$ First, we put $y_3=0$ (this was generic b.c. for $n\leq
4$, but is no longer the case for $n>4$). Of course, if boundary
polygon $\Pi$ lies at $y_3=0$, we are allowed to look for a solution
which entirely lies in this hyperplane (even if the $Z_2$ symmetry
$y_3 \rightarrow -y_3$ is spontaneously broken, there should still
be a symmetric solution: an extremum if not the minimum). This
allows to eliminate one of the three unknown functions and
considerably simplifies the problem. Now it will be also convenient
to consider projection of $\Pi$ on the $y_1,y_2$ plane, which will
be again a polygon, which we denote by $\bar \Pi$.

\begin{figure}
\unitlength 1mm 
\linethickness{0.4pt}
\ifx\plotpoint\undefined\newsavebox{\plotpoint}\fi 
\begin{picture}(136.389,74.578)(0,0)
\put(78.202,42.649){\line(0,1){1.0837}}
\put(78.178,43.733){\line(0,1){1.0817}}
\put(78.109,44.814){\line(0,1){1.0777}}
\multiput(77.993,45.892)(-.032403,.214346){5}{\line(0,1){.214346}}
\multiput(77.831,46.964)(-.029685,.151972){7}{\line(0,1){.151972}}
\multiput(77.623,48.028)(-.031649,.13174){8}{\line(0,1){.13174}}
\multiput(77.37,49.082)(-.033124,.115789){9}{\line(0,1){.115789}}
\multiput(77.072,50.124)(-.031137,.093488){11}{\line(0,1){.093488}}
\multiput(76.729,51.152)(-.032188,.084396){12}{\line(0,1){.084396}}
\multiput(76.343,52.165)(-.0330234,.076559){13}{\line(0,1){.076559}}
\multiput(75.914,53.16)(-.0336829,.0697111){14}{\line(0,1){.0697111}}
\multiput(75.442,54.136)(-.0320594,.0596781){16}{\line(0,1){.0596781}}
\multiput(74.929,55.091)(-.0325528,.054823){17}{\line(0,1){.054823}}
\multiput(74.376,56.023)(-.0329349,.0504122){18}{\line(0,1){.0504122}}
\multiput(73.783,56.93)(-.0332195,.046378){19}{\line(0,1){.046378}}
\multiput(73.152,57.811)(-.0334176,.0426662){20}{\line(0,1){.0426662}}
\multiput(72.483,58.665)(-.0335384,.0392333){21}{\line(0,1){.0392333}}
\multiput(71.779,59.489)(-.0335894,.0360436){22}{\line(0,1){.0360436}}
\multiput(71.04,60.282)(-.0335769,.033068){23}{\line(-1,0){.0335769}}
\multiput(70.268,61.042)(-.0365524,.0330351){22}{\line(-1,0){.0365524}}
\multiput(69.464,61.769)(-.0397409,.0329354){21}{\line(-1,0){.0397409}}
\multiput(68.629,62.461)(-.0431716,.0327622){20}{\line(-1,0){.0431716}}
\multiput(67.766,63.116)(-.0468799,.0325074){19}{\line(-1,0){.0468799}}
\multiput(66.875,63.733)(-.0509093,.0321613){18}{\line(-1,0){.0509093}}
\multiput(65.959,64.312)(-.0587708,.0336938){16}{\line(-1,0){.0587708}}
\multiput(65.018,64.851)(-.0641715,.0332206){15}{\line(-1,0){.0641715}}
\multiput(64.056,65.35)(-.0702173,.0326144){14}{\line(-1,0){.0702173}}
\multiput(63.073,65.806)(-.0770543,.0318505){13}{\line(-1,0){.0770543}}
\multiput(62.071,66.22)(-.092594,.033704){11}{\line(-1,0){.092594}}
\multiput(61.052,66.591)(-.103348,.032676){10}{\line(-1,0){.103348}}
\multiput(60.019,66.918)(-.116282,.031352){9}{\line(-1,0){.116282}}
\multiput(58.972,67.2)(-.132208,.029633){8}{\line(-1,0){.132208}}
\multiput(57.915,67.437)(-.177809,.03192){6}{\line(-1,0){.177809}}
\multiput(56.848,67.629)(-.214816,.029126){5}{\line(-1,0){.214816}}
\put(55.774,67.774){\line(-1,0){1.0793}}
\put(54.694,67.874){\line(-1,0){1.0826}}
\put(53.612,67.927){\line(-1,0){1.0839}}
\put(52.528,67.934){\line(-1,0){1.0832}}
\put(51.445,67.894){\line(-1,0){1.0805}}
\multiput(50.364,67.808)(-.26895,-.0331){4}{\line(-1,0){.26895}}
\multiput(49.289,67.675)(-.178189,-.029727){6}{\line(-1,0){.178189}}
\multiput(48.219,67.497)(-.151501,-.032002){7}{\line(-1,0){.151501}}
\multiput(47.159,67.273)(-.131242,-.033657){8}{\line(-1,0){.131242}}
\multiput(46.109,67.004)(-.103743,-.0314){10}{\line(-1,0){.103743}}
\multiput(45.072,66.69)(-.093002,-.032561){11}{\line(-1,0){.093002}}
\multiput(44.049,66.332)(-.083894,-.033473){12}{\line(-1,0){.083894}}
\multiput(43.042,65.93)(-.0706139,-.0317466){14}{\line(-1,0){.0706139}}
\multiput(42.053,65.485)(-.064576,-.0324273){15}{\line(-1,0){.064576}}
\multiput(41.085,64.999)(-.0591816,-.032967){16}{\line(-1,0){.0591816}}
\multiput(40.138,64.472)(-.0543195,-.0333862){17}{\line(-1,0){.0543195}}
\multiput(39.214,63.904)(-.0499034,-.0337009){18}{\line(-1,0){.0499034}}
\multiput(38.316,63.297)(-.043572,-.0322277){20}{\line(-1,0){.043572}}
\multiput(37.445,62.653)(-.0401437,-.0324431){21}{\line(-1,0){.0401437}}
\multiput(36.601,61.971)(-.0369567,-.0325821){22}{\line(-1,0){.0369567}}
\multiput(35.788,61.255)(-.0339819,-.0326517){23}{\line(-1,0){.0339819}}
\multiput(35.007,60.504)(-.0325514,-.034078){23}{\line(0,-1){.034078}}
\multiput(34.258,59.72)(-.032473,-.0370526){22}{\line(0,-1){.0370526}}
\multiput(33.544,58.905)(-.0323247,-.0402392){21}{\line(0,-1){.0402392}}
\multiput(32.865,58.06)(-.0320991,-.0436668){20}{\line(0,-1){.0436668}}
\multiput(32.223,57.186)(-.0335537,-.0500025){18}{\line(0,-1){.0500025}}
\multiput(31.619,56.286)(-.033226,-.0544177){17}{\line(0,-1){.0544177}}
\multiput(31.054,55.361)(-.0327924,-.0592785){16}{\line(0,-1){.0592785}}
\multiput(30.529,54.413)(-.0322368,-.0646713){15}{\line(0,-1){.0646713}}
\multiput(30.046,53.443)(-.0315384,-.0707072){14}{\line(0,-1){.0707072}}
\multiput(29.604,52.453)(-.033226,-.083993){12}{\line(0,-1){.083993}}
\multiput(29.206,51.445)(-.032287,-.093098){11}{\line(0,-1){.093098}}
\multiput(28.851,50.421)(-.031094,-.103835){10}{\line(0,-1){.103835}}
\multiput(28.54,49.382)(-.03327,-.13134){8}{\line(0,-1){.13134}}
\multiput(28.273,48.332)(-.031555,-.151595){7}{\line(0,-1){.151595}}
\multiput(28.053,47.271)(-.029201,-.178276){6}{\line(0,-1){.178276}}
\multiput(27.877,46.201)(-.0323,-.26904){4}{\line(0,-1){.26904}}
\put(27.748,45.125){\line(0,-1){1.0807}}
\put(27.665,44.044){\line(0,-1){2.1672}}
\put(27.638,41.877){\line(0,-1){1.0824}}
\put(27.695,40.794){\line(0,-1){1.079}}
\multiput(27.797,39.715)(.029759,-.214729){5}{\line(0,-1){.214729}}
\multiput(27.946,38.642)(.032444,-.177714){6}{\line(0,-1){.177714}}
\multiput(28.141,37.575)(.030023,-.13212){8}{\line(0,-1){.13212}}
\multiput(28.381,36.518)(.031695,-.116189){9}{\line(0,-1){.116189}}
\multiput(28.666,35.473)(.03298,-.103251){10}{\line(0,-1){.103251}}
\multiput(28.996,34.44)(.031146,-.084786){12}{\line(0,-1){.084786}}
\multiput(29.37,33.423)(.0320774,-.0769601){13}{\line(0,-1){.0769601}}
\multiput(29.787,32.422)(.0328212,-.0701209){14}{\line(0,-1){.0701209}}
\multiput(30.246,31.441)(.0334096,-.0640733){15}{\line(0,-1){.0640733}}
\multiput(30.747,30.48)(.0318747,-.05522){17}{\line(0,-1){.05522}}
\multiput(31.289,29.541)(.0323112,-.0508143){18}{\line(0,-1){.0508143}}
\multiput(31.871,28.626)(.0326454,-.0467839){19}{\line(0,-1){.0467839}}
\multiput(32.491,27.737)(.0328893,-.0430748){20}{\line(0,-1){.0430748}}
\multiput(33.149,26.876)(.0330524,-.0396436){21}{\line(0,-1){.0396436}}
\multiput(33.843,26.043)(.0331426,-.0364549){22}{\line(0,-1){.0364549}}
\multiput(34.572,25.241)(.0331668,-.0334793){23}{\line(0,-1){.0334793}}
\multiput(35.335,24.471)(.0361425,-.033483){22}{\line(1,0){.0361425}}
\multiput(36.13,23.735)(.039332,-.0334226){21}{\line(1,0){.039332}}
\multiput(36.956,23.033)(.0427645,-.0332917){20}{\line(1,0){.0427645}}
\multiput(37.811,22.367)(.0464757,-.0330827){19}{\line(1,0){.0464757}}
\multiput(38.695,21.738)(.0505091,-.0327862){18}{\line(1,0){.0505091}}
\multiput(39.604,21.148)(.0549187,-.0323911){17}{\line(1,0){.0549187}}
\multiput(40.537,20.598)(.0597724,-.0318834){16}{\line(1,0){.0597724}}
\multiput(41.494,20.087)(.06981,-.0334773){14}{\line(1,0){.06981}}
\multiput(42.471,19.619)(.0766559,-.0327976){13}{\line(1,0){.0766559}}
\multiput(43.468,19.192)(.08449,-.031939){12}{\line(1,0){.08449}}
\multiput(44.481,18.809)(.09358,-.030861){11}{\line(1,0){.09358}}
\multiput(45.511,18.47)(.115887,-.032783){9}{\line(1,0){.115887}}
\multiput(46.554,18.175)(.131833,-.03126){8}{\line(1,0){.131833}}
\multiput(47.608,17.924)(.152059,-.029237){7}{\line(1,0){.152059}}
\multiput(48.673,17.72)(.214441,-.031771){5}{\line(1,0){.214441}}
\put(49.745,17.561){\line(1,0){1.078}}
\put(50.823,17.448){\line(1,0){1.0819}}
\put(51.905,17.382){\line(1,0){2.1673}}
\put(54.072,17.388){\line(1,0){1.0815}}
\put(55.154,17.461){\line(1,0){1.0773}}
\multiput(56.231,17.58)(.21425,.033034){5}{\line(1,0){.21425}}
\multiput(57.302,17.745)(.151884,.030132){7}{\line(1,0){.151884}}
\multiput(58.365,17.956)(.131646,.032037){8}{\line(1,0){.131646}}
\multiput(59.419,18.212)(.115691,.033466){9}{\line(1,0){.115691}}
\multiput(60.46,18.514)(.093396,.031412){11}{\line(1,0){.093396}}
\multiput(61.487,18.859)(.084301,.032437){12}{\line(1,0){.084301}}
\multiput(62.499,19.248)(.0764613,.0332489){13}{\line(1,0){.0764613}}
\multiput(63.493,19.681)(.0649707,.031629){15}{\line(1,0){.0649707}}
\multiput(64.467,20.155)(.0595834,.0322352){16}{\line(1,0){.0595834}}
\multiput(65.421,20.671)(.0547269,.0327143){17}{\line(1,0){.0547269}}
\multiput(66.351,21.227)(.050315,.0330834){18}{\line(1,0){.050315}}
\multiput(67.257,21.822)(.0462799,.033356){19}{\line(1,0){.0462799}}
\multiput(68.136,22.456)(.0425675,.0335432){20}{\line(1,0){.0425675}}
\multiput(68.987,23.127)(.0391343,.0336539){21}{\line(1,0){.0391343}}
\multiput(69.809,23.834)(.0359445,.0336955){22}{\line(1,0){.0359445}}
\multiput(70.6,24.575)(.0329689,.0336742){23}{\line(0,1){.0336742}}
\multiput(71.358,25.35)(.0329272,.0366496){22}{\line(0,1){.0366496}}
\multiput(72.083,26.156)(.0328181,.0398378){21}{\line(0,1){.0398378}}
\multiput(72.772,26.993)(.0326348,.0432679){20}{\line(0,1){.0432679}}
\multiput(73.425,27.858)(.0323691,.0469755){19}{\line(0,1){.0469755}}
\multiput(74.04,28.75)(.0320111,.0510039){18}{\line(0,1){.0510039}}
\multiput(74.616,29.668)(.0335205,.0588699){16}{\line(0,1){.0588699}}
\multiput(75.152,30.61)(.0330314,.0642691){15}{\line(0,1){.0642691}}
\multiput(75.648,31.574)(.0324074,.0703131){14}{\line(0,1){.0703131}}
\multiput(76.101,32.559)(.0316232,.0771479){13}{\line(0,1){.0771479}}
\multiput(76.512,33.562)(.033431,.092693){11}{\line(0,1){.092693}}
\multiput(76.88,34.581)(.032371,.103444){10}{\line(0,1){.103444}}
\multiput(77.204,35.616)(.03101,.116374){9}{\line(0,1){.116374}}
\multiput(77.483,36.663)(.033421,.151194){7}{\line(0,1){.151194}}
\multiput(77.717,37.722)(.031396,.177902){6}{\line(0,1){.177902}}
\multiput(77.905,38.789)(.028492,.214901){5}{\line(0,1){.214901}}
\put(78.048,39.863){\line(0,1){1.0796}}
\put(78.144,40.943){\line(0,1){1.706}}
\multiput(63.939,67.849)(.0337124879,-.03557599226){1033}{\line(0,-1){.03557599226}}
\multiput(63.755,67.808)(-1.3349,.0315){14}{\line(-1,0){1.3349}}
\multiput(45.066,68.249)(-.0481664659,-.033685743){498}{\line(-1,0){.0481664659}}
\multiput(21.08,51.474)(.0337150943,-.0737773585){371}{\line(0,-1){.0737773585}}
\multiput(33.588,24.103)(.0686359684,-.0337355731){253}{\line(1,0){.0686359684}}
\multiput(98.632,31.166)(-.1032961123,-.0336907127){463}{\line(-1,0){.1032961123}}
\put(39.924,67.635){\makebox(0,0)[cc]{a}}
\put(48.451,69.648){\makebox(0,0)[cc]{a}}
\put(59.341,69.795){\makebox(0,0)[cc]{b}}
\put(72.006,61.697){\makebox(0,0)[cc]{b}}
\put(90.106,44.332){\makebox(0,0)[cc]{c}}
\put(82.876,23.829){\makebox(0,0)[cc]{c}}
\put(55.946,14.558){\makebox(0,0)[cc]{d}}
\put(45.817,15.14){\makebox(0,0)[cc]{d}}
\put(35.516,20.299){\makebox(0,0)[cc]{e}}
\put(30.232,27.056){\makebox(0,0)[cc]{e}}
\put(22.433,42.361){\makebox(0,0)[cc]{f}}
\put(25.355,57.922){\makebox(0,0)[cc]{f}}
\put(136.389,46.826){\makebox(0,0)[cc]{(a+b)+(c+d)+(e+f)=(b+c)+(e+d)+(f+a)}}
\put(54.779,74.578){\makebox(0,0)[cc]{$l_1$}}
\put(85.795,57.126){\makebox(0,0)[cc]{$l_2$}}
\put(78.267,17.104){\makebox(0,0)[cc]{$l_3$}}
\put(39.841,15.231){\makebox(0,0)[cc]{$l_4$}}
\put(21.556,32.921){\makebox(0,0)[cc]{$l_5$}}
\put(30.401,66.427){\makebox(0,0)[cc]{$l_6$}}
\put(135.501,57.6){\makebox(0,0)[cc]{$l_1+l_3+l_5=l_2+l_4+l_6$}}
\thicklines
\multiput(21.125,51.25)(.0473713152,.0337303605){504}{\line(1,0){.0473713152}}
\multiput(45,68.25)(1.583343,-.03125){12}{\line(1,0){1.583343}}
\multiput(64,67.875)(.03372454195,-.03580177823){1023}{\line(0,-1){.03580177823}}
\multiput(98.501,31.25)(-.1017136706,-.0337261118){467}{\line(-1,0){.1017136706}}
\multiput(51,15.5)(-.0684527905,.0337303605){252}{\line(-1,0){.0684527905}}
\multiput(33.75,24)(-.0336668681,.0726671014){375}{\line(0,1){.0726671014}}
\end{picture}
\begin{center}
\unitlength 1mm 
\linethickness{0.4pt}
\ifx\plotpoint\undefined\newsavebox{\plotpoint}\fi 
\begin{picture}(98.818,49.144)(0,0)
\put(41.626,29.543){\line(0,1){.7835}}
\put(41.607,30.326){\line(0,1){.7816}}
\put(41.549,31.108){\line(0,1){.7778}}
\multiput(41.453,31.886)(-.03352,.19304){4}{\line(0,1){.19304}}
\multiput(41.319,32.658)(-.028638,.127444){6}{\line(0,1){.127444}}
\multiput(41.148,33.422)(-.029881,.1079){7}{\line(0,1){.1079}}
\multiput(40.938,34.178)(-.03075,.093015){8}{\line(0,1){.093015}}
\multiput(40.692,34.922)(-.03136,.081238){9}{\line(0,1){.081238}}
\multiput(40.41,35.653)(-.03178,.07164){10}{\line(0,1){.07164}}
\multiput(40.092,36.369)(-.032054,.063631){11}{\line(0,1){.063631}}
\multiput(39.74,37.069)(-.032211,.056815){12}{\line(0,1){.056815}}
\multiput(39.353,37.751)(-.0322727,.0509213){13}{\line(0,1){.0509213}}
\multiput(38.934,38.413)(-.0322531,.0457556){14}{\line(0,1){.0457556}}
\multiput(38.482,39.054)(-.0321634,.0411757){15}{\line(0,1){.0411757}}
\multiput(38,39.671)(-.0320122,.0370751){16}{\line(0,1){.0370751}}
\multiput(37.487,40.264)(-.0318062,.0333727){17}{\line(0,1){.0333727}}
\multiput(36.947,40.832)(-.0334064,.0317708){17}{\line(-1,0){.0334064}}
\multiput(36.379,41.372)(-.037109,.0319729){16}{\line(-1,0){.037109}}
\multiput(35.785,41.884)(-.0412097,.0321198){15}{\line(-1,0){.0412097}}
\multiput(35.167,42.365)(-.0457898,.0322046){14}{\line(-1,0){.0457898}}
\multiput(34.526,42.816)(-.0509555,.0322188){13}{\line(-1,0){.0509555}}
\multiput(33.864,43.235)(-.056849,.032151){12}{\line(-1,0){.056849}}
\multiput(33.181,43.621)(-.063664,.031987){11}{\line(-1,0){.063664}}
\multiput(32.481,43.973)(-.071674,.031704){10}{\line(-1,0){.071674}}
\multiput(31.764,44.29)(-.081271,.031274){9}{\line(-1,0){.081271}}
\multiput(31.033,44.571)(-.093048,.030652){8}{\line(-1,0){.093048}}
\multiput(30.288,44.816)(-.107932,.029767){7}{\line(-1,0){.107932}}
\multiput(29.533,45.025)(-.127474,.028503){6}{\line(-1,0){.127474}}
\multiput(28.768,45.196)(-.19308,.03331){4}{\line(-1,0){.19308}}
\put(27.996,45.329){\line(-1,0){.7779}}
\put(27.218,45.424){\line(-1,0){.7817}}
\put(26.436,45.481){\line(-1,0){1.567}}
\put(24.869,45.479){\line(-1,0){.7815}}
\put(24.088,45.421){\line(-1,0){.7777}}
\multiput(23.31,45.324)(-.19301,-.03372){4}{\line(-1,0){.19301}}
\multiput(22.538,45.189)(-.127413,-.028773){6}{\line(-1,0){.127413}}
\multiput(21.773,45.017)(-.107869,-.029996){7}{\line(-1,0){.107869}}
\multiput(21.018,44.807)(-.092982,-.030849){8}{\line(-1,0){.092982}}
\multiput(20.274,44.56)(-.081205,-.031446){9}{\line(-1,0){.081205}}
\multiput(19.544,44.277)(-.071607,-.031856){10}{\line(-1,0){.071607}}
\multiput(18.828,43.958)(-.063596,-.032121){11}{\line(-1,0){.063596}}
\multiput(18.128,43.605)(-.056781,-.032272){12}{\line(-1,0){.056781}}
\multiput(17.447,43.218)(-.0508871,-.0323267){13}{\line(-1,0){.0508871}}
\multiput(16.785,42.797)(-.0457214,-.0323016){14}{\line(-1,0){.0457214}}
\multiput(16.145,42.345)(-.0411415,-.032207){15}{\line(-1,0){.0411415}}
\multiput(15.528,41.862)(-.0370411,-.0320515){16}{\line(-1,0){.0370411}}
\multiput(14.935,41.349)(-.033339,-.0318415){17}{\line(-1,0){.033339}}
\multiput(14.368,40.808)(-.0337188,-.0355301){16}{\line(0,-1){.0355301}}
\multiput(13.829,40.239)(-.0319336,-.0371428){16}{\line(0,-1){.0371428}}
\multiput(13.318,39.645)(-.0320761,-.0412437){15}{\line(0,-1){.0412437}}
\multiput(12.837,39.026)(-.032156,-.0458239){14}{\line(0,-1){.0458239}}
\multiput(12.387,38.385)(-.0321647,-.0509896){13}{\line(0,-1){.0509896}}
\multiput(11.969,37.722)(-.032091,-.056883){12}{\line(0,-1){.056883}}
\multiput(11.583,37.039)(-.031919,-.063698){11}{\line(0,-1){.063698}}
\multiput(11.232,36.339)(-.031628,-.071708){10}{\line(0,-1){.071708}}
\multiput(10.916,35.622)(-.031188,-.081305){9}{\line(0,-1){.081305}}
\multiput(10.635,34.89)(-.030553,-.09308){8}{\line(0,-1){.09308}}
\multiput(10.391,34.145)(-.029652,-.107963){7}{\line(0,-1){.107963}}
\multiput(10.183,33.39)(-.028368,-.127504){6}{\line(0,-1){.127504}}
\multiput(10.013,32.625)(-.03311,-.19311){4}{\line(0,-1){.19311}}
\put(9.881,31.852){\line(0,-1){.778}}
\put(9.786,31.074){\line(0,-1){.7817}}
\put(9.73,30.292){\line(0,-1){2.3485}}
\put(9.793,27.944){\line(0,-1){.7776}}
\multiput(9.891,27.166)(.027143,-.154378){5}{\line(0,-1){.154378}}
\multiput(10.026,26.394)(.028908,-.127383){6}{\line(0,-1){.127383}}
\multiput(10.2,25.63)(.03011,-.107837){7}{\line(0,-1){.107837}}
\multiput(10.411,24.875)(.030947,-.09295){8}{\line(0,-1){.09295}}
\multiput(10.658,24.132)(.031532,-.081172){9}{\line(0,-1){.081172}}
\multiput(10.942,23.401)(.031932,-.071573){10}{\line(0,-1){.071573}}
\multiput(11.261,22.685)(.032189,-.063562){11}{\line(0,-1){.063562}}
\multiput(11.615,21.986)(.032332,-.056746){12}{\line(0,-1){.056746}}
\multiput(12.003,21.305)(.0323806,-.0508528){13}{\line(0,-1){.0508528}}
\multiput(12.424,20.644)(.03235,-.0456872){14}{\line(0,-1){.0456872}}
\multiput(12.877,20.005)(.0322506,-.0411074){15}{\line(0,-1){.0411074}}
\multiput(13.361,19.388)(.0320908,-.0370071){16}{\line(0,-1){.0370071}}
\multiput(13.874,18.796)(.0318768,-.0333052){17}{\line(0,-1){.0333052}}
\multiput(14.416,18.23)(.0355658,-.0336811){16}{\line(1,0){.0355658}}
\multiput(14.985,17.691)(.0371767,-.0318942){16}{\line(1,0){.0371767}}
\multiput(15.58,17.18)(.0412777,-.0320324){15}{\line(1,0){.0412777}}
\multiput(16.199,16.7)(.0458579,-.0321075){14}{\line(1,0){.0458579}}
\multiput(16.841,16.25)(.0510237,-.0321107){13}{\line(1,0){.0510237}}
\multiput(17.505,15.833)(.056917,-.032031){12}{\line(1,0){.056917}}
\multiput(18.188,15.449)(.063732,-.031852){11}{\line(1,0){.063732}}
\multiput(18.889,15.098)(.071741,-.031552){10}{\line(1,0){.071741}}
\multiput(19.606,14.783)(.081338,-.031102){9}{\line(1,0){.081338}}
\multiput(20.338,14.503)(.093112,-.030454){8}{\line(1,0){.093112}}
\multiput(21.083,14.259)(.107995,-.029538){7}{\line(1,0){.107995}}
\multiput(21.839,14.052)(.127534,-.028233){6}{\line(1,0){.127534}}
\multiput(22.604,13.883)(.19315,-.03291){4}{\line(1,0){.19315}}
\put(23.377,13.751){\line(1,0){.7781}}
\put(24.155,13.658){\line(1,0){.7818}}
\put(24.937,13.603){\line(1,0){1.567}}
\put(26.504,13.608){\line(1,0){.7814}}
\put(27.285,13.668){\line(1,0){.7775}}
\multiput(28.063,13.766)(.154349,.027306){5}{\line(1,0){.154349}}
\multiput(28.834,13.903)(.127352,.029043){6}{\line(1,0){.127352}}
\multiput(29.599,14.077)(.107805,.030224){7}{\line(1,0){.107805}}
\multiput(30.353,14.289)(.092917,.031046){8}{\line(1,0){.092917}}
\multiput(31.096,14.537)(.081138,.031618){9}{\line(1,0){.081138}}
\multiput(31.827,14.822)(.071539,.032008){10}{\line(1,0){.071539}}
\multiput(32.542,15.142)(.063528,.032256){11}{\line(1,0){.063528}}
\multiput(33.241,15.496)(.056712,.032392){12}{\line(1,0){.056712}}
\multiput(33.921,15.885)(.0508185,.0324345){13}{\line(1,0){.0508185}}
\multiput(34.582,16.307)(.0456528,.0323984){14}{\line(1,0){.0456528}}
\multiput(35.221,16.76)(.0410732,.0322942){15}{\line(1,0){.0410732}}
\multiput(35.837,17.245)(.0369731,.03213){16}{\line(1,0){.0369731}}
\multiput(36.429,17.759)(.0332714,.0319121){17}{\line(1,0){.0332714}}
\multiput(36.995,18.301)(.0336434,.0356015){16}{\line(0,1){.0356015}}
\multiput(37.533,18.871)(.0318548,.0372104){16}{\line(0,1){.0372104}}
\multiput(38.043,19.466)(.0319886,.0413116){15}{\line(0,1){.0413116}}
\multiput(38.522,20.086)(.0320588,.0458919){14}{\line(0,1){.0458919}}
\multiput(38.971,20.729)(.0320566,.0510577){13}{\line(0,1){.0510577}}
\multiput(39.388,21.392)(.03197,.056951){12}{\line(0,1){.056951}}
\multiput(39.772,22.076)(.031784,.063766){11}{\line(0,1){.063766}}
\multiput(40.121,22.777)(.031476,.071774){10}{\line(0,1){.071774}}
\multiput(40.436,23.495)(.031016,.08137){9}{\line(0,1){.08137}}
\multiput(40.715,24.227)(.030356,.093145){8}{\line(0,1){.093145}}
\multiput(40.958,24.972)(.029424,.108026){7}{\line(0,1){.108026}}
\multiput(41.164,25.729)(.033717,.153077){5}{\line(0,1){.153077}}
\multiput(41.332,26.494)(.0327,.19318){4}{\line(0,1){.19318}}
\put(41.463,27.267){\line(0,1){.7782}}
\put(41.556,28.045){\line(0,1){1.4978}}
\multiput(59.91,44.636)(-1.20399263,.03344424){37}{\line(-1,0){1.20399263}}
\multiput(15.362,45.874)(-.033700408,-.076352488){235}{\line(0,-1){.076352488}}
\multiput(7.442,27.931)(.0336838116,-.0478664691){349}{\line(0,-1){.0478664691}}
\multiput(19.198,11.225)(.1144102213,.0337209073){411}{\line(1,0){.1144102213}}
\multiput(66.221,25.085)(.03331561,.14278118){78}{\line(0,1){.14278118}}
\multiput(68.819,36.222)(-.0356381818,.0336582828){250}{\line(-1,0){.0356381818}}
\put(36.239,49.144){\makebox(0,0)[cc]{$l_1$}}
\put(66.468,42.957){\makebox(0,0)[cc]{$l_2$}}
\put(70.887,29.168){\makebox(0,0)[cc]{$l_3$}}
\put(44.017,14.496){\makebox(0,0)[cc]{$l_4$}}
\put(9.192,17.324){\makebox(0,0)[cc]{$l_5$}}
\put(7.778,39.775){\makebox(0,0)[cc]{$l_6$}}
\put(98.818,35.179){\makebox(0,0)[cc]{$l_1+l_3+l_5=l_2+l_4+l_6$}}
\end{picture}
\end{center}
\caption{{\footnotesize It rarely happens for $n>3$ that a polygon
has an inscribing circle, which touches all of its $n$ sides.
However, if such circle exists, then it is obvious that each side
length $l_i = l_{i,1}+l_{i,2}$ and $l_{i,2} = l_{i+1,1}$, so that
for even $n$ we have $\sum_{i=1}^n (-)^i l_i = 0$. For $n=4$ inverse
is also true: $l_1+l_3=l_2+l_4$ implies the existence of inscribing
circle, however, this is of course incorrect for $n>4$. }}
\label{quadrilatcircle}
\end{figure}
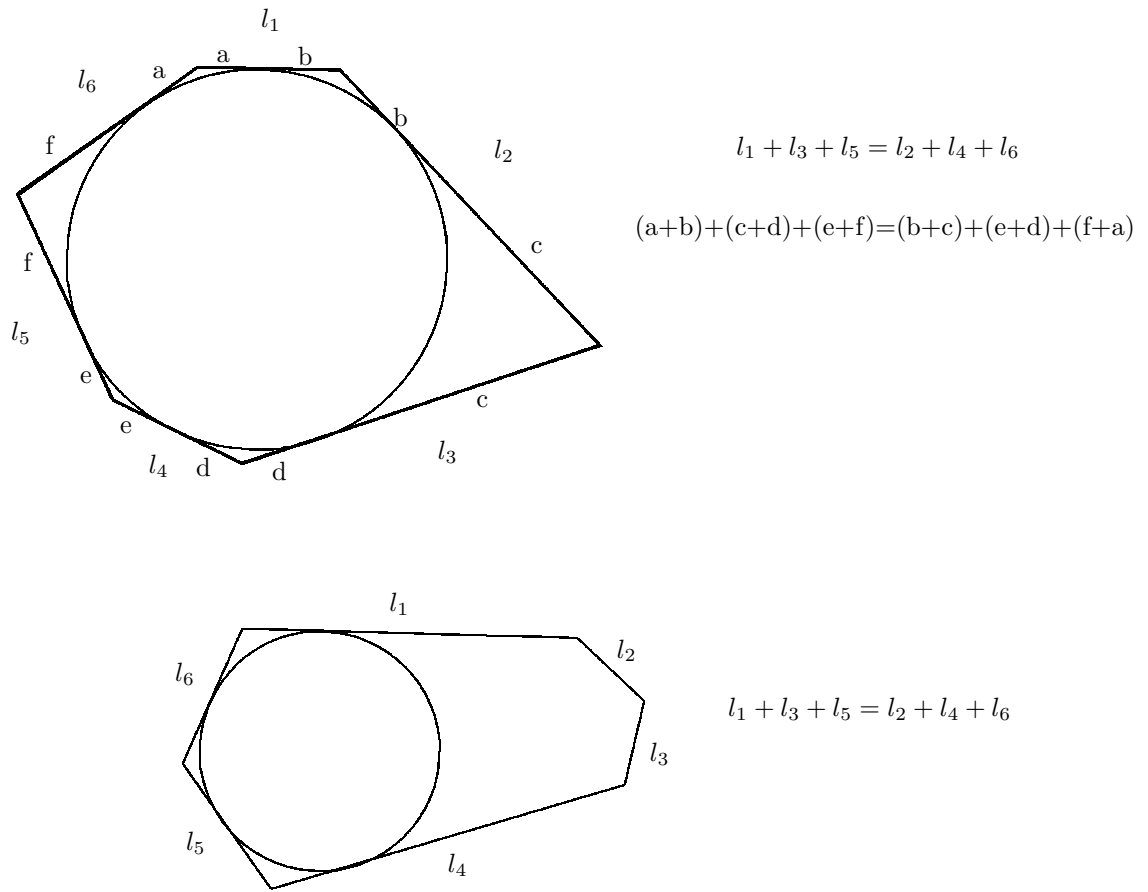

$\bullet$ Second, we assume that $\bar\Pi$ is special: there is an
inscribing circle which touches all of its sides. Such a circle
always exists for a triangle ($n=3$). For quadrilaterals ($n=4$) it
exists when the lengths of the four sides satisfy \be
l_1+l_3=l_2+l_4, \ee see Fig.\ref{quadrilatcircle}, which is exactly
the condition that $\Pi$ (with all sides formed by {\it
null}-vectors) was a closed polygon in $y_0$ direction. Again, for
$n>4$ there is no reason for such a circle to exist: we just
restrict consideration to particular b.c. with this property. The
reason for this is that then all the points of $\Pi$ satisfy \be
-y_0^2+y_1^2+y_2^2=1 \ee (common rescaling is performed to make the
circle radius unit), and in embedding (Poincare) coordinates
$Y_\mu$, $\mu = 0,\ldots,5$ this means that $Y_4=0$ at the boundary.
Like in the case of $Y_3\sim y_3=0$ this implies that we can look
for a solution, which  entirely lies at $Y_4=0$, i.e. has $z=w$ or
$q=z/w=1$ (again we ignore the possibility of spontaneous breakdown
of $Z_2$ symmetry $Y_4 \rightarrow -Y_4$, though in this case such
solutions could be very interesting: described by non-trivial
Riemann surfaces). In still other words, with such boundary
conditions we can impose the ansatz \be y_0^2 = y_1^2+y_2^2 + y_3^2
+ r^2 -1 \label{anzaNG} \ee Of course, it is immediately consistent
with the $\sigma$-model equations: \be
\partial_i\frac{1}{r^2}\partial_i y = 0
\ee together with (\ref{anzaNG}) implies that \be
r\partial_i\frac{1}{r^2}\partial_i r = -L_\sigma = \frac{
(\partial_i y_0)^2 - (\partial_i y_1)^2 - (\partial_i y_2)^2 -
(\partial_i y_3)^2 - (\partial_i r)^2}{r^2} \ee Eq.(\ref{anzaNG}) is
our first ingredient of the algebro-geometric ansatz. Upon putting
$y_3=0$, it leaves us with a single unknown function, which we can
take to be either $r(y_1,y_2)$ or $y_0(y_1,y_2)$.

This remaining function should satisfy boundary conditions given on
a polygon $\Pi$ with light-like edges. We shall look for ansatze
for this remaining function among the elements of the
{\it boundary ring}, to be introduced in s.\ref{bori} below in order
to implement the boundary conditions. The boundary ring can be
constructed for any polygon $\bar\Pi$, still it is greatly
simplified by existence of an inscribing circle. Its analysis is
even further simplified by presence of extra symmetries.

\begin{figure}
\begin{center}
\unitlength 1mm 
\linethickness{0.4pt}
\ifx\plotpoint\undefined\newsavebox{\plotpoint}\fi 
\begin{picture}(122.711,63.287)(0,0)
\multiput(46.343,27.578)(.0487302053,.0336950147){341}{\line(1,0){.0487302053}}
\multiput(62.96,39.068)(.48364151,.03335849){53}{\line(1,0){.48364151}}
\multiput(88.593,40.836)(.0570535117,-.0336989967){598}{\line(1,0){.0570535117}}
\multiput(122.711,20.684)(-.056844523,-.0337314488){283}{\line(-1,0){.056844523}}
\multiput(106.624,11.138)(-.55979167,-.03314583){48}{\line(-1,0){.55979167}}
\multiput(79.754,9.547)(-.0627813084,.0337028037){535}{\line(-1,0){.0627813084}}
\put(122.464,21.32){\line(0,-1){.9775}}
\put(122.464,19.365){\line(0,-1){.9775}}
\put(122.464,17.41){\line(0,-1){.9775}}
\put(122.464,15.456){\line(0,-1){.9775}}
\put(122.464,13.501){\line(0,-1){.9775}}
\put(122.464,11.546){\line(0,-1){.9775}}
\put(122.464,9.591){\line(0,-1){.9775}}
\put(122.464,7.636){\line(0,-1){.9775}}
\put(122.464,5.681){\line(0,-1){.9775}}
\put(88.346,40.942){\line(0,1){.9684}}
\put(88.346,42.879){\line(0,1){.9684}}
\put(88.346,44.816){\line(0,1){.9684}}
\put(88.346,46.753){\line(0,1){.9684}}
\put(88.346,48.69){\line(0,1){.9684}}
\put(88.346,50.627){\line(0,1){.9684}}
\put(88.346,52.564){\line(0,1){.9684}}
\put(88.346,54.5){\line(0,1){.9684}}
\put(88.346,56.437){\line(0,1){.9684}}
\put(88.346,58.374){\line(0,1){.9684}}
\put(88.346,60.311){\line(0,1){.9684}}
\put(88.346,62.248){\line(0,1){.9684}}
\put(46.273,57.382){\line(0,-1){.9751}}
\put(46.273,55.432){\line(0,-1){.9751}}
\put(46.273,53.482){\line(0,-1){.9751}}
\put(46.273,51.532){\line(0,-1){.9751}}
\put(46.273,49.581){\line(0,-1){.9751}}
\put(46.273,47.631){\line(0,-1){.9751}}
\put(46.273,45.681){\line(0,-1){.9751}}
\put(46.273,43.731){\line(0,-1){.9751}}
\put(46.273,41.78){\line(0,-1){.9751}}
\put(46.273,39.83){\line(0,-1){.9751}}
\put(46.273,37.88){\line(0,-1){.9751}}
\put(46.273,35.93){\line(0,-1){.9751}}
\put(46.273,33.979){\line(0,-1){.9751}}
\put(46.273,32.029){\line(0,-1){.9751}}
\put(46.273,30.079){\line(0,-1){.9751}}
\put(46.273,28.129){\line(0,-1){.9751}}
\put(63.067,43.063){\line(0,-1){.8132}}
\put(63.067,41.437){\line(0,-1){.8132}}
\put(63.067,39.811){\line(0,-1){.8132}}
\thicklines
\multiput(79.577,9.723)(.03710150892,.03370644719){729}{\line(1,0){.03710150892}}
\multiput(106.624,34.295)(.0337076271,-.0352055085){472}{\line(0,-1){.0352055085}}
\multiput(122.534,17.678)(-.03372173058,.04189183874){1017}{\line(0,1){.04189183874}}
\multiput(88.239,60.282)(-.0435043328,-.0337019064){577}{\line(-1,0){.0435043328}}
\multiput(63.137,40.836)(-.0485375723,.0337196532){346}{\line(-1,0){.0485375723}}
\multiput(46.343,52.503)(.03373163265,-.04347244898){980}{\line(0,-1){.04347244898}}
\thinlines
\put(106.731,11.067){\line(0,1){.9999}}
\put(106.731,13.067){\line(0,1){.9999}}
\put(106.731,15.067){\line(0,1){.9999}}
\put(106.731,17.067){\line(0,1){.9999}}
\put(106.731,19.066){\line(0,1){.9999}}
\put(106.731,21.066){\line(0,1){.9999}}
\put(106.731,23.066){\line(0,1){.9999}}
\put(106.731,25.066){\line(0,1){.9999}}
\put(106.731,27.065){\line(0,1){.9999}}
\put(106.731,29.065){\line(0,1){.9999}}
\put(106.731,31.065){\line(0,1){.9999}}
\put(106.731,33.065){\line(0,1){.9999}}
\put(106.731,35.064){\line(0,1){.9999}}
\put(106.731,37.064){\line(0,1){.9999}}
\put(106.731,39.064){\line(0,1){.9999}}
\put(106.731,41.064){\line(0,1){.9999}}
\put(14,27.5){\vector(0,1){20}}
\put(14.125,27.5){\vector(1,0){17.5}}
\put(6.375,19.5){\vector(-1,-1){.07}}\multiput(14.125,27.625)(-.03369538,-.035325801){230}{\line(0,-1){.035325801}}
\put(15.75,44.625){\makebox(0,0)[cc]{$y_o$}}
\put(27.625,24.875){\makebox(0,0)[cc]{$y_1$}}
\put(10.5,20.5){\makebox(0,0)[cc]{$y_2$}}
\put(74.391,55.125){\makebox(0,0)[cc]{$\Pi$}}
\put(56.392,17.25){\makebox(0,0)[cc]{$\bar\Pi$}}
\end{picture}
\caption{{\footnotesize Polygon $\Pi$ in the $3d$ space
$(y_0,y_1,y_2)$, located at the boundary $r=0$ of $AdS_5$.
Coordinate $y_3=0$. The sides of polygons are null (light-like) and
$y_0$ switches direction at every corner. Polygon $\bar\Pi$ is the
projection of $\Pi$ onto the plane $(y_1,y_2)$. In most places in
this paper $\bar\Pi$ is assumed to be a $Z_n$ symmetric polygon with
$n$ even. Of course, such $\bar\Pi$ has an inscribing circle, we
assume that it has radius one. }} \label{regpoly0}
\end{center}
\end{figure}
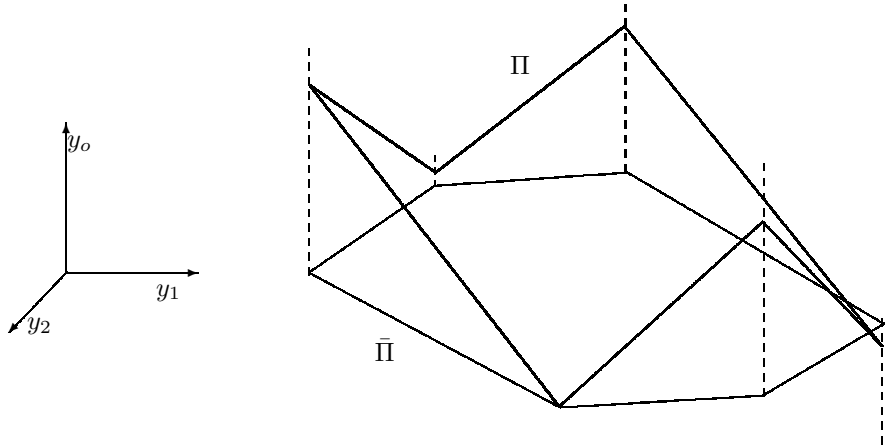

$\bullet$ Therefore, in our examples in s.\ref{exa} we additionally
assume that $\bar\Pi$ is the $Z_n$-symmetric polygon. This leaves no
free parameters (and completely eliminates the possibility of
comparison with BDS/BHT formula, which described the dependence on
the {\it shape} of $\bar\Pi$), but will be enough to illustrate our
approach. Generalizations are relatively straightforward. In this
$Z_n$-symmetric situation we further assume that $y_0$ changes
direction at every vertex, see Fig.\ref{regpoly0}. In this way we
further restrict consideration to {\it even} $n$, instead our entire
problem acquires $Z_{n}$ symmetry (lifting of $Z_n$-symmetric
$\bar\Pi$ to $\Pi$ preserves $Z_{n/2}$ rotational symmetry, while
rotation by an elementary angle $2\pi/n$ should be accompanied by a
$Z_2$ flip $y_0\rightarrow -y_0$), and this considerably simplifies
the boundary ring.

\subsection{Plan of the paper and the main equations}

The remaining part of the paper can be considered as a set of
examples: we begin with the well known ones in s.\ref{oldexa}, use
them to illustrate the concept of the boundary ring, to be
introduced in s.\ref{bori}, and end with truncated power series for
$Z_n$ symmetric polygons $\bar\Pi$ in s.\ref{exa}. As will be
demonstrated, the most crude truncations already provide the
surprisingly good approximations to the true solutions for lower
values of $n$ (like $n=6,8,10$), and further corrections are very
small. This, however, may not be the case in general asymmetric
situation. Also, the deviations from the would-be exact solutions
are concentrated near the {\it angles} of the polygons, which
provide the dominant $1/\epsilon^2$ divergencies in the regularized
action. This is important to keep in mind in further use of our
approximate solutions in studies of the string/gauge duality. A
drastic improvement of behavior near the boundaries can be achieved
by a fuller use of the boundary ring structure, which is suggested
in s.\ref{puzz}. However, it looks like the accuracy of equations of
motion get less controlled in such an approach and it is yet unclear
if accurate estimate of regularized area can be found in this way
(though we do not see any more potential obstacles). A short summary
is presented in Conclusion, s.\ref{conc}.

According to above suggestions, for each example we consider first
the NG action and the NG equations of motion. Since these are
invariant under generic coordinate transformation on the world
sheet, one has a freedom to choose these coordinates in any way that
seems convenient. We take $y_1$ and $y_2$ for the world sheet
coordinates and consider the NG equations for functions
$y_0(y_1,y_2)$ and $r(y_1,y_2)$ (we look for the special solutions
with $y_3\equiv 0$): \be \frac{\partial}{\partial y_1}
\left(\frac{\partial y_0}{\partial y_1}
\frac{H_{22}}{r^2L_{NG}}\right) + \frac{\partial}{\partial y_2}
\left(\frac{\partial y_0}{\partial y_2}
\frac{H_{11}}{r^2L_{NG}}\right) - \frac{\partial}{\partial y_1}
\left(\frac{\partial y_0}{\partial y_2}
\frac{H_{12}}{r^2L_{NG}}\right) - \frac{\partial}{\partial y_2}
\left(\frac{\partial y_0}{\partial y_1}
\frac{H_{12}}{r^2L_{NG}}\right) = 0, \nn\\
\frac{\partial}{\partial y_1} \left(\frac{\partial r}{\partial y_1}
\frac{H_{22}}{r^2L_{NG}}\right) + \frac{\partial}{\partial y_2}
\left(\frac{\partial r}{\partial y_2}
\frac{H_{11}}{r^2L_{NG}}\right) - \frac{\partial}{\partial y_1}
\left(\frac{\partial r}{\partial y_2}
\frac{H_{12}}{r^2L_{NG}}\right) - \frac{\partial}{\partial y_2}
\left(\frac{\partial r}{\partial y_1}
\frac{H_{12}}{r^2L_{NG}}\right) + \frac{2L_{NG}}{r^3} = 0
\label{NGeq} \ee where \be H_{ij} = \frac{ -\frac{\partial
y_0}{\partial y_i} \frac{\partial y_0}{\partial y_j} +
\frac{\partial r}{\partial y_i}\frac{\partial r}{\partial y_j} +
\delta_{ij} }{r^2} \label{NGH} \ee and \be L_{NG} = \sqrt{ \det_{ij}
H_{ij} } = \sqrt{H_{11}H_{22}-H_{12}^2} \ee After substitution of
(\ref{anzaNG}) the NG Lagrangian density turns into \be
L_{NG}dy_1dy_2 = \frac{dy_1dy_2}{r^2} \sqrt{\frac{(y_i\partial_i r -
r)^2 - (\partial_i r)^2 - 1} {y_1^2+y_2^2+r^2-1}} \ee and provides
an equation of motion for the single function $r(y_1,y_2)$.
Similarly one can write the Lagrangian density and the NG equations
for $y_0$ instead of $r$ in the role of a single unknown function:
\be L_{NG}dy_1dy_2 =
\sqrt{\frac{(y_i\partial_i y_0 - y_0)^2 - (\partial_i y_0)^2 +
1}{(1+y_0^2-y_1^2-y_2^2)^3}}\,dy_1dy_2 \ee and for the other pairs
$(y_0,y_1)$ and $(y_0,y_2)$ chosen to play the role of world sheet
coordinates. They look the same as (\ref{NGeq}) with obvious change
of indices and signs in (\ref{NGH}).

After solutions to the NG equations is found, we proceed to the
$\sigma$-model equations, which are no longer invariant under
coordinate transformations on the world sheet. Given an NG solution
these equations can be considered as defining the two additional
functions $y_i(u_1,u_2)$, $i=1,2$. Actually we do not reach this
step in non-trivial examples in s.\ref{exa},
it remains an open problem for future consideration.

\section{The known solutions \label{oldexa}}

\subsection{$n=2$: Two parallel lines}

\subsubsection{NG equations}

If the two parallel lines are directed along the $y_2$ axis and
located at $y_1=\pm 1$, then the NG solution with such boundary
conditions is \be
r^2 = 1-y_1^2, \nn \\
y_0=y_2
\label{n2NGsol}
\ee
Near the boundaries, where $y_1=\pm (1-y_\bot)$ \be r
\sim \sqrt{y_\bot} \ee The NG Lagrangian density is \be L_{NG} = 0
\ee

\subsubsection{$\sigma$-model equations}

The corresponding solution to the $\sigma$-model equations
\be
\partial_i \frac{1}{r^2}\partial_i y_j = 0, \ \ j=1,2,\nn\\
r\partial_i \frac{1}{r^2}\partial_i r = -L_\sigma =
\frac{(\partial_i y_0)^2 - (\partial_i y_1)^2 -
(\partial_i y_2)^2 - (\partial_i r)^2}{r^2}
\label{smeq}
\ee
is given by
\be
y_1 = \tanh u_1, \nn \\
y_0=y_2=\tanh u_2, \nn \\
r = \sqrt{1-y_1^2} = \frac{1}{\cosh u_1}
\label{n2smsol}
\ee
The $\sigma$-model
Lagrangian density is \be L_\sigma = 1 \ee

\subsection{$n=2$: Two intersecting lines ("cusp").
The simplest configuration
\label{cuspsolsi}}

\subsubsection{Boundary conditions: description of
$\bar\Pi_2$ and $\Pi_2$}

In this case the domain of interest -- the would-be polygon -- lies
inside an angle between two straight lines. To begin with, let us
assume that one of them is projected to the horizontal axis, $\tilde
y_2=0$ and another -- to $\tilde y_2=\kappa \tilde y_1$ with $\kappa
= \tan(2\alpha)$. Angle is set to be $2\alpha$ in order to simplify
formulas below, and this is the value of angle $\bar\Pi_2$, obtained
by projection onto the $(y_1,y_2)$ plane. Original angle $\Pi_2$ is
formed by two null rays \be
\left\{\begin{array}{c} \tilde y_2=0, \\
\tilde y_0 = -\tilde y_1 \end{array}\right.
\ee
and
\be
\left\{\begin{array}{l}
\tilde y_2 = \kappa \tilde y_1 = \tilde y_1\tan(2\alpha), \\
\tilde y_0 = - \tilde y_1\sqrt{1+\kappa^2} = -\frac{\tilde
y_1}{\cos(2\alpha)} = -\frac{\tilde y_2}{\sin(2\alpha)}
\end{array}\right.
\label{cuspbc2} \ee We assume here that the two lines intersect in
the origin not only on the plane $(y_1,y_2)$, but in $3d$ Minkovski
space $(y_0,y_1,y_2)$ and that $y_0$ takes maximal value $\tilde
y_{00}=0$ at the vertex and decreases along the rays. In what
follows we also assume that the angle is acute, $2\alpha
<\frac{\pi}{2}$ and $\kappa >0$, otherwise some signs should be
changed.

\subsubsection{NG equations
\label{n2NGsolsi}}

Solution which satisfies our boundary conditions is
\be
r^2 = 2\mu \tilde y_2(\kappa \tilde y_1-\tilde y_2), \nn \\
\tilde y_0 = - \tilde y_1-\mu\kappa \tilde y_2
\label{cuspNGba}
\ee
while (\ref{cuspbc2}) requires that $(1+\mu\kappa) =
\sqrt{1+\kappa^2}$, so that \be \mu =
\frac{\sqrt{1+\kappa^2}-1}{\kappa} =
\frac{\cos(2\alpha)}{2\cos^2\!\!\alpha} \ee Therefore the second
equation in (\ref{cuspNGba}) can be also rewritten as \be \tilde
y_0\cos\alpha + \tilde y_1\cos\alpha + \tilde y_2\sin\alpha = 0
\label{linre} \ee

Solution (\ref{cuspNGba}) satisfies \be \tilde y_0^2 = \tilde
y_1^2+\tilde y_2^2 + r^2, \label{quadre} \ee what is somewhat
different from (\ref{anzaNG}). This is not a surprise because
(\ref{anzaNG}) implies that the origin of coordinate system is
located at the center of a {\it unit} circle, inscribed into
$\bar\Pi$ and $y_0$ vanishes at the tangent points, while in our
example this circle is shrinked to a point at the angle vertex. In
order to recover (\ref{anzaNG}) we need to make an appropriate
change of variables, see s.\ref{cuspsol} below. In anticipation of
this we put tildes over $y$-variables in this section.

\subsubsection{$\sigma$-model equations}

In this case it is convenient to remember that the first equation in
(\ref{smeq}) is true also for $y_0$ and thus for $y_\pm = y_0 \pm
y_1$. Thus $\sigma$-model equations are automatically consistent
with (\ref{linre}), stating that $\tilde y_+ = - \tilde
y_2\tan\alpha$. After that (\ref{quadre}) turns into a product: \be
r^2 = \tilde y_+(\tilde y_- -\tilde y_+\cot^2\!\!\alpha) \ee This
factorization implies separation of variables in the first equations
(\ref{smeq}): \be
\tilde y_+ = 2e^{-2u_1}, \nn \\
\tilde y_- -\tilde y_+\cot^2\!\!\alpha = 2e^{-2u_2} \label{tyvsu}
\ee while the last equation in (\ref{smeq}) is automatically
satisfied with \be L_\sigma = 2 \ee Coefficient $2$ in exponents in
(\ref{tyvsu}) can be changed by rescaling of $u$-variables, it is
chosen so that behavior of $y(u)$ and $r(u)$ in the vicinity of the
boundary $\Pi$ (which lies at infinity in $u$-plane) is the same as
in (\ref{n2smsol}). Pre-exponential constants in (\ref{tyvsu}) are
regulated by shifts of $u$-variables, they are put to $2$ in order
to simplify (\ref{cuspsmssi}) below.

Finally we obtain $\sigma$-model solution in the form: \be \tilde
y_0 = \frac{\tilde y_+ + \tilde y_-}{2} =
e^{-2u_1}(1+\cot^2\!\!\alpha) + e^{-2u_2} =
\frac{1}{\sin^2\!\!\alpha}\,e^{-2u_1} + e^{-2u_2}, \nn \\
\tilde y_1 = \frac{\tilde y_+ - \tilde y_-}{2} =
e^{-2u_1}(1-\cot^2\!\!\alpha) - e^{-2u_2}=
-\frac{\cos(2\alpha)}{\sin^2\!\!\alpha}\,
e^{-2u_1}- e^{-2u_2},\nn \\
\tilde y_2 = -\tilde y_+ \cot\alpha =
-2 e^{-2u_1}\cot\alpha, \nn \\
r = \sqrt{\tilde y_+(\tilde y_- -\tilde y_+\cot^2\!\!\alpha)} =
2e^{-u_1-u_2} \label{cuspsmssi} \ee

\subsection{$n=2$: "Cusp" in
generic configuration which satisfies (\ref{anzaNG})
\label{cuspsol}}

\subsubsection{Coordinate transformation}

\begin{figure}
\begin{center}
\unitlength 1mm 
\linethickness{0.4pt}
\ifx\plotpoint\undefined\newsavebox{\plotpoint}\fi 
\begin{picture}(137.17,96.105)(0,0)
\put(53.049,3.855){\vector(0,1){92.25}}
\put(19.92,13){\vector(1,0){117.25}}
\multiput(53.083,12.993)(.03370128811,.0655417689){855}{\line(0,1){.0655417689}}
\multiput(81.777,68.743)(-.356780336,-.033635857){175}{\line(-1,0){.356780336}}
\put(52.906,12.753){\line(0,1){.9916}}
\put(52.724,14.737){\line(0,1){.9916}}
\put(52.541,16.72){\line(0,1){.9916}}
\put(52.359,18.703){\line(0,1){.9916}}
\put(52.176,20.686){\line(0,1){.9916}}
\put(51.994,22.67){\line(0,1){.9916}}
\put(51.811,24.653){\line(0,1){.9916}}
\put(51.629,26.636){\line(0,1){.9916}}
\put(51.447,28.619){\line(0,1){.9916}}
\put(51.264,30.603){\line(0,1){.9916}}
\put(51.082,32.586){\line(0,1){.9916}}
\put(50.899,34.569){\line(0,1){.9916}}
\put(50.717,36.552){\line(0,1){.9916}}
\put(50.534,38.536){\line(0,1){.9916}}
\put(50.352,40.519){\line(0,1){.9916}}
\put(50.169,42.502){\line(0,1){.9916}}
\put(49.987,44.485){\line(0,1){.9916}}
\put(49.804,46.469){\line(0,1){.9916}}
\put(49.622,48.452){\line(0,1){.9916}}
\put(49.439,50.435){\line(0,1){.9916}}
\put(49.257,52.418){\line(0,1){.9916}}
\put(49.075,54.402){\line(0,1){.9916}}
\put(48.892,56.385){\line(0,1){.9916}}
\put(48.71,58.368){\line(0,1){.9916}}
\put(48.527,60.351){\line(0,1){.9916}}
\put(48.345,62.335){\line(0,1){.9916}}
\put(48.162,64.318){\line(0,1){.9916}}
\put(52.696,12.753){\line(1,0){.1051}}
\put(52.906,12.753){\line(0,1){0}}
\qbezier(56.375,19.125)(60,17.438)(60.125,13)
\qbezier(51.875,25.125)(64.875,24.875)(64.875,13.125)
\qbezier(68.375,21.25)(70.063,18.125)(70,13)
\multiput(81.75,68.625)(.03373405299,-.06108930324){1019}{\line(0,-1){.06108930324}}
\multiput(53.055,12.93)(.0586538,.0317308){15}{\line(1,0){.0586538}}
\multiput(54.814,13.882)(.0586538,.0317308){15}{\line(1,0){.0586538}}
\multiput(56.574,14.834)(.0586538,.0317308){15}{\line(1,0){.0586538}}
\multiput(58.334,15.785)(.0586538,.0317308){15}{\line(1,0){.0586538}}
\multiput(60.093,16.737)(.0586538,.0317308){15}{\line(1,0){.0586538}}
\multiput(61.853,17.689)(.0586538,.0317308){15}{\line(1,0){.0586538}}
\multiput(63.612,18.641)(.0586538,.0317308){15}{\line(1,0){.0586538}}
\multiput(65.372,19.593)(.0586538,.0317308){15}{\line(1,0){.0586538}}
\multiput(67.132,20.545)(.0586538,.0317308){15}{\line(1,0){.0586538}}
\multiput(68.891,21.497)(.0586538,.0317308){15}{\line(1,0){.0586538}}
\multiput(70.651,22.449)(.0586538,.0317308){15}{\line(1,0){.0586538}}
\multiput(72.41,23.401)(.0586538,.0317308){15}{\line(1,0){.0586538}}
\multiput(74.17,24.353)(.0586538,.0317308){15}{\line(1,0){.0586538}}
\multiput(75.93,25.305)(.0586538,.0317308){15}{\line(1,0){.0586538}}
\multiput(77.689,26.257)(.0586538,.0317308){15}{\line(1,0){.0586538}}
\multiput(79.449,27.209)(.0586538,.0317308){15}{\line(1,0){.0586538}}
\multiput(81.209,28.16)(.0586538,.0317308){15}{\line(1,0){.0586538}}
\multiput(82.968,29.112)(.0586538,.0317308){15}{\line(1,0){.0586538}}
\multiput(84.728,30.064)(.0586538,.0317308){15}{\line(1,0){.0586538}}
\multiput(86.487,31.016)(.0586538,.0317308){15}{\line(1,0){.0586538}}
\multiput(88.247,31.968)(.0586538,.0317308){15}{\line(1,0){.0586538}}
\multiput(90.007,32.92)(.0586538,.0317308){15}{\line(1,0){.0586538}}
\multiput(91.766,33.872)(.0586538,.0317308){15}{\line(1,0){.0586538}}
\multiput(93.526,34.824)(.0586538,.0317308){15}{\line(1,0){.0586538}}
\multiput(95.285,35.776)(.0586538,.0317308){15}{\line(1,0){.0586538}}
\multiput(97.045,36.728)(.0586538,.0317308){15}{\line(1,0){.0586538}}
\qbezier(75.875,68.125)(78.125,61)(84.375,63.875)
\qbezier(75,68)(76.625,60.25)(84.75,63)
\put(81.875,59.625){\makebox(0,0)[cc]{$\alpha$}}
\put(73.75,63.125){\makebox(0,0)[cc]{$\alpha$}}
\multiput(42.125,64.75)(.0328947,-.2960526){19}{\line(0,-1){.2960526}}
\multiput(42.75,59.125)(.3157895,.0328947){19}{\line(1,0){.3157895}}
\multiput(96,42.875)(-.06554878,-.03353659){82}{\line(-1,0){.06554878}}
\multiput(90.625,40.125)(.03343023,-.06104651){86}{\line(0,-1){.06104651}}
\put(59.875,7.75){\makebox(0,0)[cc]{$\theta$}}
\put(47,24.625){\makebox(0,0)[cc]{$\phi_2$}}
\put(74.125,18.25){\makebox(0,0)[cc]{$\phi_1$}}
\end{picture}
\caption{{\footnotesize Rotated angle of the size $2\alpha$. Shown
are the direction $\theta$ to the angle vertex and the two
directions $\phi_1$ and $\phi_2$ of normals to two sides of the
angle. Both normals have the same unit length and are the two radii
of inscribed circle with center in the origin (not shown). }}
\label{anglexa}
\end{center}
\end{figure}
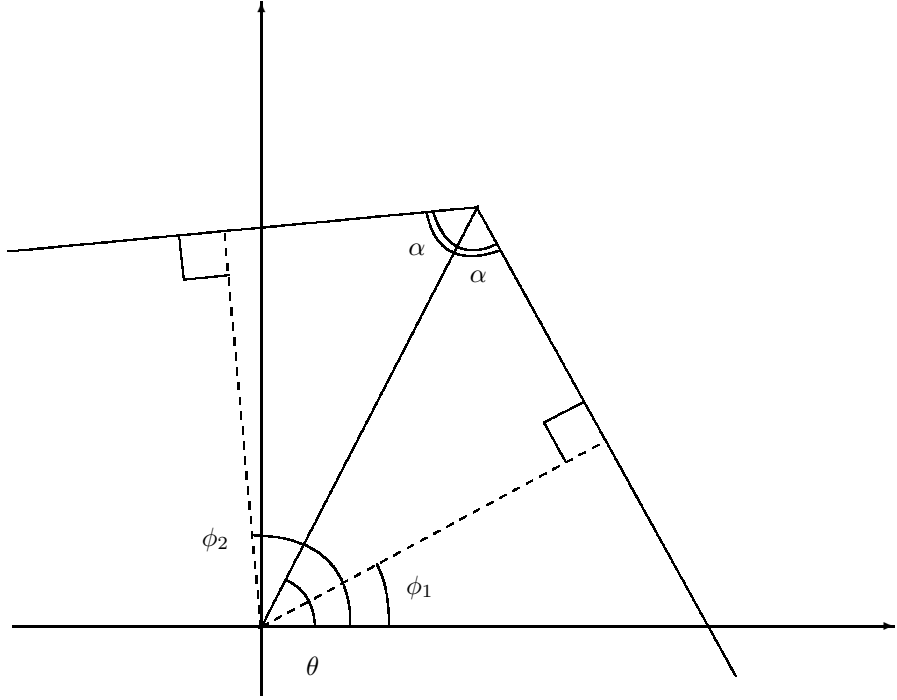

As explained at the end of s.\ref{n2NGsolsi}, in order to represent
the cusp formulas in the same form as all other examples in this
paper, in particular to restore (\ref{anzaNG}), we need to make a
change of $y$-variables. Namely, let the origin of coordinate system
on the $(y_1,y_2)$ plane be a center of inscribed {\it unit} circle,
then the vertex of our angle is at the point $z= y_1+iy_2 =
\frac{e^{i\theta}}{\sin\alpha}$ with some angle $\theta$. These new
coordinates are related to $\tilde y_i$ in s.\ref{cuspsolsi} by a
combination of shift and rotation, see Fig.\ref{anglexa}: $z -
\frac{e^{i\theta}}{\sin\alpha} = \tilde z e^{i(\pi+\theta-\alpha)}$
or \be \left(ze^{-i\theta} + \tilde ze^{-i\alpha}\right)\sin\alpha =
1 \label{ztz} \ee

If eq.(\ref{linre}), a corollary of NG equations, is combined with
(\ref{ztz}), then we obtain: \be \tilde y_0 ^2 - \tilde y_1^2 -
\tilde y_2^2 \ \stackrel{(\ref{ztz}) \& (\ref{linre})}{=}\ (\tilde
y_0+\cot\alpha)^2+1-y_1^2+y_2^2 = y_0^2+1-y_1^2-y_2^2,
\label{quadre2} \ee provided we put \be y_{0} = \tilde y_0 +
\cot\alpha \label{y0vty0} \ee This shift implies that $y_0=0$ at two
points where the unit circle touches the sides of our angle, while
at the vertex of the angle $y_0 = \cot\alpha\neq 0$. We see, that
such choice of $y_0$ is exactly what is needed to reproduce
(\ref{anzaNG}).

\subsubsection{NG equations}

It is now straightforward to convert NG solution (\ref{cuspNGba})
into \be y_0 \ \stackrel{(\ref{y0vty0})}{=}\ \tilde y_0 + \cot\alpha
\ \stackrel{(\ref{linre})}{=}\ - \tilde y_1 - \tilde y_2\tan\alpha +
\cot\alpha \ \stackrel{(\ref{ztz})}{=}\
\frac{1}{\cos\alpha}\left\{{\rm Re}\left(ze^{-i\theta}\right)
-\sin\alpha\right\}, \nn \\
r^2 = \tilde y_0 ^2 - \tilde y_1^2 - \tilde y_2^2  =
y_0^2+1-y_1^2-y_2^2 = 1-y_1^2-y_2^2 + \left( \frac{y_1\cos\theta +
y_2\sin\theta - \sin\alpha}{\cos\alpha} \right)^2 \label{NGcuspneq}
\ee The first formula can be also rewritten as \be y_0\cos\alpha +
\sin\alpha = y_1\cos\theta + y_2\sin\theta \ee

In the particular case of $\theta=0$ solution looks simpler:
\be
y_0\cos\alpha + \sin\alpha = y_1, \nn \\
r^2 = \left(\frac{1-y_1\sin\alpha}{\cos\alpha}\right)^2 - y_2^2 \ee
If instead one of the sides of the angle is the vertical line
$y_1=1$ (such side will exist in most of our examples in this
paper), then $\theta = \frac{\pi}{2} - \alpha$ and we obtain the NG
solution in the form: \be
y_0 = y_2 + (y_1-1)\tan\alpha, \nn \\
r^2 = \frac{1}{\cos^2\!\!\alpha} - 2y_1\tan^2\!\!\alpha
-2y_2\tan\alpha - \frac{\cos(2\alpha)}{\cos^2\!\!\alpha}y_1^2 +
2y_1y_2\tan\alpha \label{cuspNGsol} \ee In particular, in
rectangular case, $2\alpha = \frac{\pi}{2}$, when the angle is
formed by the two lines $y_1=1$ and $y_2=1$, \be
y_0 = y_1+y_2-1, \nn \\
r^2 = 2(1-y_1)(1-y_2) \ee This solution coincides with the limit
$y_1\to +1$, $y_2\to +1$ of (\ref{squareNG}) up to a factor of 2
(which is due to the fact that an arbitrary scaled $r$ is still a
solution in the cusp case). Similarly, choosing $2\alpha =
\frac{\pi}{2}+n\pi$ for various $n$ one can reproduce (\ref{squareNG})
in various limits of $y_1\to \pm 1$, $y_2\to \pm 1$.

If instead $\alpha=0$, then we reproduce (\ref{n2NGsol}).

\subsubsection{$\sigma$-model equations}

After coordinate transformation to (\ref{ztz}) and (\ref{y0vty0})
solution (\ref{cuspsmssi}) turns into \be
y_1={\cos\theta\over\sin\alpha}-2\sin(\theta-\alpha)\cot\alpha
e^{-2u_1}+\cos (\theta-\alpha)\left[e^{-2u_2}+{\cos
2\alpha\over\sin^2\alpha}e^{-2u_1}\right]\\
y_2={\sin\theta\over\sin\alpha}+2\cos(\theta-\alpha)\cot\alpha
e^{-2u_1}+\sin (\theta-\alpha)\left[e^{-2u_2}+{\cos 2\alpha\over\sin^2\alpha}e^{-2u_1}\right] \\
y_0={e^{-2u_1}\over\sin^2\alpha}+e^{-2u_2}+\cot\alpha\\
r=2e^{-u_1-u_2}
 \ee
For $\theta = \frac{\pi}{2} - \alpha$ this turns into \be
y_1=1+\sin 2\alpha e^{-2u_2}\\
y_2=\cot\alpha +\cos 2\alpha e^{-2u_2}+{e^{-2u_1}\over\sin^2\alpha}
\ee and if further $2\alpha=\frac{\pi}{2}$, then \be
y_1=1+e^{-2u_2},\ \ \ \ y_2=1+2e^{-2u_1},\ \ \ \
y_0=1+2e^{-2u_1}+e^{-2u_2},\ \ \ \ \ r=2e^{-u_1-u_2} \ee while at
$\alpha=0$ we reproduce (\ref{n2smsol}).


\subsection{$n=3$: Impossible triangle}

For three null-vectors the conservation condition $p_1+p_2+p_3=0$
implies that they are collinear. Indeed, this condition implies that
$p_1p_2 = |p_1||p_2|(1-\cos\phi) = 0$, i.e. the angle between the
two vectors is zero: $\phi=0$.

\subsection{$n=4$: A square
\label{square}}

"Square" in this section and "rhombus" in the next one refer to the
shapes of $\bar\Pi$. Associated $\Pi$ are not planar and look
slightly more involved.

\subsubsection{NG equations}

In this case the NG solution is \be\label{squareNG}
r^2 = (1-y_1^2)(1-y_2^2), \nn \\
y_0 = \sqrt{y_1^2+y_2^1+r^2-1} = y_1y_2
\ee
The corresponding
\be
L_{NG}dy_1dy_2=\frac{dy_1dy_2}{(1-y_1^2)(1-y_2^2)}
= d\xi_1d\xi_2
\ee
Near the boundaries
\be
r \sim \sqrt{y_\bot}
\ee

\subsubsection{$\sigma$-model equations}

Solution to the $\sigma$-model equations of motion is provided by
identification \be y_i = \tanh u_i, \ \ \ i=1,2 \ee The
corresponding \be L_\sigma = 2 \ee

\subsection{$n=4$: A rhombus}

Deformations of the square into rhombus and other skew
quadrilaterals are described in \cite{am1,mmt1}. Deformed solutions
look simpler in $\sigma$-model terms and this is how they are
usually represented.

\subsubsection{$\sigma$-model solution}

In the case of rhombus this solution is \cite{am1}:
\be
\tilde y_0 =
\frac{B\xi_1\xi_2}{1+b\xi_1\xi_2}, \ \ \
\tilde y_1 =
\frac{\xi_1}{1+b\xi_1\xi_2}, \ \ \
\tilde y_2 =
\frac{\xi_2}{1+b\xi_1\xi_2}, \ \ \
\tilde r =\frac{\sqrt{(1-\xi_1^2)(1-\xi_2^2)}}{1+b\xi_1\xi_2},
\label{rhombsigma}
\ee
where $B^2=1+b^2$, and satisfies
\be
\tilde
y_0^2 - \frac{2b}{B}\tilde y_0 =
\tilde y_1^2+\tilde y_2^2+\tilde
r^2-1
\label{yyr1}
\ee
instead of (\ref{anzaNG}), so that
\be
\tilde r^2 =
\frac{\Big[(\tilde y_1+b\tilde y_2)^2-1\Big] \Big[(b\tilde
y_1+\tilde y_2)^2-1\Big]} {Q(\tilde y_1,\tilde y_2)}, \ \ \ \
Q(\tilde y_1,\tilde y_2) = \frac{(1-b^2\xi_1^2)(1-b^2\xi_2^2)}
{(1+b\xi_1\xi_2)^2} \ee Note that the rhombus exists only for
$|b|<1$, while $|\xi_1|\leq 1$ and $|\xi_2|\leq 1$, so that there
are no poles at $\xi_i = \pm 1/b$ in this formula.

After rescaling $\tilde y = y/B$, $\tilde r= r/B$, which converts
the boundary equation to the form $c y_1+s y_2=1$ with $c^2+s^2=1$,
and additional shift $y_0-b\rightarrow y_0$, which converts
(\ref{rhombsigma})  into \be y_0 =
\frac{\xi_1\xi_2-b}{1+b\xi_1\xi_2}, \ \ \ y_1 =
\frac{B\xi_1}{1+b\xi_1\xi_2}, \ \ \ y_2 =
\frac{B\xi_2}{1+b\xi_1\xi_2}, \ \ \ r =
B\frac{\sqrt{(1-\xi_1^2)(1-\xi_2^2)}}{1+b\xi_1\xi_2},
\label{rhombussigma} \ee eq.(\ref{yyr1}) turns into (\ref{anzaNG})
\be y_0^2 = y_1^2+ y_2^2 +r^2-1 \label{rhombusanzaNG} \ee

\subsubsection{NG solution}

From eqs.(\ref{rhombussigma}) one can express, say, $y_0$ through
$y_1$ and $y_2$, and, together with (\ref{rhombusanzaNG}), this
provides a solution to the NG equations. This formula, however, is
not as nice as the previous ones: \be y_0 = \frac{1-b^2 -
B^2\sqrt{1-\frac{4by_1y_2}{B^2}}}{2b} =
-b+y_1y_2+\frac{b}{B^2}(y_1y_2)^2+
\frac{2b^2}{B^4}(y_1y_2)^3+\frac{5b^3}{B^6}(y_1y_2)^4+
\frac{14b^4}{B^8}(y_1y_2)^5+\ldots \label{rhombusNG} \ee and can be
already considered as an example of a power series solution. Moreover,
already here can construct a plot as a prototype of non-trivial examples in
s.\ref{exa}: it has to show an approximate shape of exact solution
and of its truncated approximations, provided by keeping the first
terms in the power series (\ref{rhombusNG}). The essential
difference with s.\ref{exa} is that there exact solutions are not
yet available, instead the truncations match boundary conditions
much better than in this rhombus case.

\subsubsection{Another description of NG solution:
first appearance of boundary ring}

If the boundary $\Pi$, where
\be
r^2=y_0^2+1-y_1^2-y_2^2=0,
\label{y0r}
\ee
is parameterized as \be \Pi = \Big\{ c_ay_1+s_ay_2=1, \
a=1,\ldots,4\Big\} \ee with $c_a=\cos \alpha_a$, $s_a = \sin
\alpha_a$ and $y_0 = \pm(-s_a y_1+c_ay_2)$, the NG solution is
actually described by \be y_1^{2}r^2 = -\prod_{a=1}^4
\Big(y_1+(-)^{a+1}s_a y_0-c_a\Big), \nn \\
y_2^{2}r^2 = -\prod_{a=1}^4 \Big(y_2+(-)^{a}c_a y_0-s_a\Big)
\label{barin4}
\ee
The sign $(-)^a$ takes into account that $y_0$
switches sign $\pm$ at {\it every} corner of $\Pi$.

For example, at $b=0$ we have: \be y_1^2r^2 =
-(y_1-1)(y_1+y_0)(y_1+1)(y_1-y_0) = (1-y_1^2)(y_1^2-y_0^2) =\nn\\
\stackrel{(\ref{y0r})}{=} \ (1-y_1^2)(1-r^2-y_2^2) = y_1^2r^2 -r^2 +
(1-y_1^2)(1-y_2^2) \ee and \be y_2^2r^2 =
-(y_2-y_0)(y_2-1)(y_2+y_0)(y_2+1) = (1-y_2^2)(y_2^2-y_0^2) =\nn\\ \
\stackrel{(\ref{y0r})}{=} \ (1-y_2^2)(1-r^2-y_1^2) = y_2^2r^2 -r^2 +
(1-y_1^2)(1-y_2^2) \ee

Mixed representations are also possible: \be r^2 = \frac{
(y_1+s_1y_0-c_1)(y_2+c_2y_0-s_2) (y_1+s_3y_0-c_3)(y_2+c_4y_0-s_4) }
{(s_1y_0-c_1)(c_2y_0-s_2)} \ee or (note the change of sign in front
of $y_0$) \be r^2 = \frac{ (y_2-c_1y_0-s_1)(y_1-s_2y_0-c_2)
(y_2-c_3y_0-s_3)(y_1-s_4y_0-c_4) } {(-c_1y_0-s_1)(-s_2y_0-c_2)} \ee

The values of $c_a$ and $s_a$, along with some more details about
geometry of the rhombus are given in the tables:
$$
\begin{array}{|c|c|c|c|c|}
\hline
&&&&\\
a &1&2&3&4\\
&&&&\\
\hline
&&&&\\
c_a &\frac{1}{B} &\frac{b}{B}&-\frac{1}{B}&-\frac{b}{B}\\
&&&&\\
\hline
&&&&\\
s_a&\frac{b}{B}&\frac{1}{B}&-\frac{b}{B}&-\frac{1}{B}\\
&&&&\\\hline
\end{array}
$$

{\bf Vertices:}
$$
\begin{array}{|c|c|c||c|c|c|}
\hline
&&&&&\\
y_1&y_2&y_0 & \frac{y_2-by_1}{B} & \frac{y_1-by_2}{B} &
\frac{(y_2-by_1)(y_1-by_2)}{B^2} \\
&&&&&\\
\hline
&&&&&\\
\frac{B}{1+b}&\frac{B}{1+b}&\frac{1-b}{1+b} &
\frac{1-b}{1+b} & \frac{1-b}{1+b} & \frac{(1-b)^2}{(1+b)^2}\\
-\frac{B}{1-b}&\frac{B}{1-b}&-\frac{1+b}{1-b} &
\frac{1+b}{1-b} & -\frac{1+b}{1-b} & -\frac{(1+b)^2}{(1-b)^2}\\
-\frac{B}{1+b}& -\frac{B}{1+b}&\frac{1-b}{1+b} &
-\frac{1-b}{1+b} & -\frac{1-b}{1+b} & \frac{(1-b)^2}{(1+b)^2}\\
\frac{B}{1-b}&-\frac{B}{1-b}&-\frac{1+b}{1-b} &
-\frac{1+b}{1-b} & \frac{1+b}{1-b} & -\frac{(1+b)^2}{(1-b)^2}\\
&&&&&\\
\hline
\end{array}
$$

{\bf Edges:}

$$
\begin{array}{|c|c|c|c|c|c|c||c}
\hline
&&&&&&\\
a&c_a&s_a&c_ay_1+s_ay_2=1&y_1=(-)^as_ay_0 + c_a &
y_2=(-)^{a-1}c_ay_0 + s_a&{\rm comment}\\
&&&&&&\\
\hline
&&&&&&\\
1 & \frac{1}{B} & \frac{b}{B} & y_1+by_2=B & By_1+by_0 = 1 &
By_2-y_0 = b &
\xi_1=1\\
&&&&&&\\
2 & \frac{b}{B} & \frac{1}{B} &
by_1+y_2=B & By_1-y_0 = b & By_2+by_0 = 1 & \xi_2=1\\
&&&&&&\\
3 & -\frac{1}{B} & -\frac{b}{B} &
y_1+by_2=-B & By_1-by_0 = -1 & By_2+y_0 = -b & \xi_1=-1\\
&&&&&&\\
4 & -\frac{b}{B} & -\frac{1}{B} &
by_1+y_2=-B & By_1+y_0 = -b & By_2-by_0 = -1& \xi_2=-1\\
&&&&&&\\
\hline
\end{array}
$$

For generic $b$ we have also, as a generalization of $y_1y_2=y_0$
at $b=0$,
\be
\frac{(y_1-by_2)(y_2-by_1)}{B^2} = \frac{(1-b^2)y_0 -
2by_0^2+br^2}{B^2}
\label{y0y1y2b}
\ee

These various polynomials of $y$ and $r$ variables which vanish on
the boundary $\Pi$ have an important property: they have direct
analogues in general situation, beyond the rhombus example. They all
are elements of the boundary ring, to be further considered in
s.\ref{bori} below.

\subsection{$n=4$: Generic skew quadrilateral, $w\neq z$}

As already mentioned in the Introduction, the case of $n=4$ is
distinguished, because one can always rotate $\Pi$ to make
$y_3=0$.\footnote{It is also distinguished in other ways, for
example, by unambiguously fixed its form with a peculiar "dual
space" conformal symmetry, see \cite{CS1,CS2}. This subject, though
potentially important for our considerations, is, however, left
beyond the scope of this paper.} Further, shifts of $y_1$ and $y_2$
move coordinate system to the center of the circle at the
intersection of two bisectrices (between three edges). The fourth
edge is tangent to the same circle due to the condition
$l_1+l_3=l_2+l_4$. Common rescaling of all $y$'s makes the radius
unit.

However, for generic quadrilateral, different from rhombus,
$q=rw\neq 1$. Still this is not fatal for our simplified
consideration, based on the use of (\ref{anzaNG}) because actually
$q = \vec\alpha \vec y + \beta$, moreover, $q = \alpha y_0 + \beta$.
This means that an additional shift of $y_0$ (and an appropriate
rescaling) restores the ansatz (\ref{anzaNG}) for generic
quadrilateral $\Pi$.

Detailed formulas for the $\sigma$-model solutions are listed in
\cite{mmt1} and we do not repeat them here. Some of these solutions
-- satisfying the Virasoro constraints -- are also NG solutions, see
\cite{mmt2}. As in the rhombus case, they can be converted into
power series for $y_0(y_1,y_2)$, which are somewhat sophisticated
and we also do not present them here. Note that moduli of the
$\sigma$-model solutions completely disappear after such conversion
and there is a single series for $y_0(y_1,y_2)$ for any given skew
quadrilateral $\bar\Pi$.

\subsection{$n=\infty$: A circle \label{ninfty}}

\subsubsection{NG equations}

In this case the coordinate $y_0$ is fast fluctuating along the
boundary between $\pm l/2$, where $l$ is the polygon side which
tends to zero as $n\rightarrow \infty$. Therefore, $y_0$ gets
infinitely small in this limit and \be
r^2 = 1-y_1^2-y_2^2, \nn \\
y_0 = 0 \label{ninftyNG} \ee which is the corollary of
(\ref{anzaNG}).
Eq.(\ref{ninftyNG}) is indeed a solution to the NG equations, \be
L_{NG} = \frac{1}{r^3} \ee

Near the boundary \be r \sim \sqrt{y_\bot} \ee

\subsubsection{$\sigma$-model equations}

As to the $\sigma$-model equations, \be y_i = \frac{2u_i}{1+u^2},\ \
\ r=\frac{1-u^2}{1+u^2} \ee Indeed, for $u_1=u\cos\phi$,
$u_2=u\sin\phi$, $\frac{\partial}{\partial u_1} =
\cos\phi\,\partial_u - \frac{\sin\phi}{u}\,\partial_\phi$,
$\frac{\partial}{\partial u_2} = \sin\phi\,\partial_u +
\frac{\cos\phi}{u}\,\partial_\phi$ and for $\ y_1=Y\!\cos\phi$, $\
y_2=Y\!\sin\phi\ $ the equations $\ \partial_i
\frac{1}{r^2}\partial_i y_j = 0$ turn into \be Y'' +
\left(\frac{1}{u}-\frac{2r'}{r}\right)Y' - \frac{1}{u^2}Y = 0 \ee
or, taking (\ref{ninftyNG}) into account, \be \ddot Y - Y +
\frac{2Y\dot Y^2}{1-Y^2} = 0 \label{ddotY} \ee with $t = \log u $.
The relevant solution\footnote{ It is easy to write down a general
solution to (\ref{ddotY}), given by the elliptic integral
$$
\frac{dY}{\sqrt{1-Y^2 + c(1-Y^2)^2}} = \frac{du}{u}
$$
with arbitrary constant $c$, however this is irrelevant. For
example, $c=0$, i.e. $\dot Y^2=1-Y^2$ would also give a solution,
$Y=\sin (\log u)$, but it is obviously irrelevant to our problem. }
is the one with $\dot Y^2 = Y^2(1-Y^2)$ and $Y = \frac{2u}{1+u^2}$,
while
\be
L_\sigma = {8\over (1-u^2)^2}
\ee

\section{
The boundary ring  \label{bori}}

\subsection{Strategy of solving NG-equations in more detail}

Eqs.(\ref{barin4}) implies that the following object is very
important in construction of NG solutions:

The boundary ring ${\cal R}_\Pi$
is defined as a ring of polynomials of
$y$-variables, i.e. at the boundary of $AdS_5$,
which vanish at $\Pi$.
Clearly, the ansatz for $r$ should be looked for inside
this ring, and a relation between $y$-variables, which
defines the remaining ansatz for $y_0$, should also belong
to this ring. In practice one can need a closure of the
ring (power series made out of its elements),
if the answer is not polynomial.

To find a solution in the simplified setting,
described in the introduction, we need {\it three} ansatze.

$\bullet$ First ansatz: $y_3=0$.

$\bullet$ Restrict consideration to special classes of polygons
and make the second ansatz, see (\ref{anzaNG}).

$\bullet$ Explicitly construct the boundary ring of $\Pi$ and
look for the third ansatz in it.

In fact, one can lift the first two restrictions:
if the boundary ring is known, all the three ansatze should
be looked for inside it.
However, in this paper we oversimplify our problem:
in this setting
$y_3=0$ and $P_2=y_0^2+1-y_1^2-y_2^2$ are obvious elements of
${\cal R}_\Pi$, it remains only to find the third ansatz --
and this is not fully trivial.

In the remaining part of this section we construct boundary rings
for some simple types of polygons.

\subsection{Polygons of the special type \label{assum}}

The boundary consists of generic polygon consists of the straight
segments \be \alpha_{ia}  y_i = 1; \ \ \ \  i = 0,1,2,3; \ \
a=1,\dots,\,n \ee (only $n-1$ of the $n$ vectors $\alpha_a$ are
linearly independent).

If we impose the simplifying constraints,
described in the Introduction,
i.e. that

$\bullet$ $y_0$ switches from increase to decrease at each
vertex (this is possible only for $n$ even),

$\bullet$ $y_3=0$, and

$\bullet$ the projection $\bar\Pi$ of $\Pi$
on the $(y_1,y_2)$ plane
is a polygon with all edges tangent to unit circle
(for $n=4$ this follows from the condition that
$l_1-l_2+l_3-l_4=0$),
then $\alpha_{ia}$ are expressed through $n$ angles and
\be
c_a y_1 + s_a y_2 = 1, \nn\\
y_1 = (-)^as_ay_0 + c_a, \nn\\
y_2= (-)^{a-1}c_ay_0 + s_a
\label{boundeqs}
\ee
with $c_a^2+s_a^2=1$.
In this case one can impose the first constraint/ansatz
in the form:
\be
y_0^2 = r^2 + y_1^2 + y_2^2 - 1
\label{rquadra}
\ee
Without the third constraint we still could write
\be
c_a y_1 + s_a y_2 = h_a
\ee
instead of (\ref{boundeqs}), but only when all $h_a$ are equal
(and can rescaled to unity) the ansatz (\ref{rquadra}) can
be true, and we shall impose it in what follows.

\begin{figure}
\unitlength 1mm 
\linethickness{0.4pt}
\ifx\plotpoint\undefined\newsavebox{\plotpoint}\fi 
\begin{picture}(127.863,123.612)(0,0)
\put(10.25,67){\vector(1,0){100.75}}
\put(26.311,46.825){\circle{5.115}}
\put(60.055,27.129){\circle{5.115}}
\put(93.908,47.635){\circle{5.115}}
\put(93.908,87.135){\circle{5.115}}
\put(60.313,106.905){\circle{5.115}}
\put(26.272,86.094){\circle{5.115}}
\multiput(28.392,87.704)(.0553970037,.0336835206){534}{\line(1,0){.0553970037}}
\multiput(62.433,105.691)(.0574674556,-.0337179487){507}{\line(1,0){.0574674556}}
\put(26.311,83.393){\line(0,-1){33.744}}
\multiput(28.392,45.487)(.0571972656,-.0336777344){512}{\line(1,0){.0571972656}}
\multiput(62.285,28.392)(.0553951311,.0336835206){534}{\line(1,0){.0553951311}}
\put(60.204,24.379){\line(0,-1){6.987}}
\put(60.055,104.353){\line(0,-1){74.474}}
\put(59.927,109.602){\vector(0,1){13.435}} \thicklines
\put(93.799,84.731){\line(0,-1){34.338}}
\put(63.698,123.612){\makebox(0,0)[cc]{$y_2$}}
\put(110.157,62.226){\makebox(0,0)[cc]{$y_1$}}
\put(118.428,82.078){\makebox(0,0)[cc]{$y_0=+y_2$}}
\put(93.76,87.243){\makebox(0,0)[cc]{2}}
\put(60.104,106.95){\makebox(0,0)[cc]{3}}
\put(26.163,86.09){\makebox(0,0)[cc]{4}}
\put(26.163,46.846){\makebox(0,0)[cc]{5}}
\put(60.104,26.87){\makebox(0,0)[cc]{6}}
\put(93.868,47.553){\makebox(0,0)[cc]{1}}
\put(91.747,62.933){\makebox(0,0)[cc]{1}}
\put(71.418,97.227){\makebox(0,0)[cc]{2}}
\put(48.26,96.344){\makebox(0,0)[cc]{3}}
\put(28.638,60.458){\makebox(0,0)[cc]{4}}
\put(37.124,42.78){\makebox(0,0)[cc]{5}}
\put(82.555,43.84){\makebox(0,0)[cc]{6}} \thinlines
\put(118.211,82.201){\oval(19.304,6.435)[]}
\qbezier(108.541,82.731)(103.856,82.29)(101.647,79.373)
\put(93.868,77.251){\vector(-1,0){.07}}\qbezier(101.647,79.373)(99.525,77.605)(93.868,77.251)
\multiput(42.886,96.449)(.0326619,-.0562314){15}{\line(0,-1){.0562314}}
\multiput(43.866,94.762)(.0326619,-.0562314){15}{\line(0,-1){.0562314}}
\multiput(44.845,93.075)(.0326619,-.0562314){15}{\line(0,-1){.0562314}}
\multiput(45.825,91.388)(.0326619,-.0562314){15}{\line(0,-1){.0562314}}
\multiput(46.805,89.701)(.0326619,-.0562314){15}{\line(0,-1){.0562314}}
\multiput(47.785,88.014)(.0326619,-.0562314){15}{\line(0,-1){.0562314}}
\multiput(48.765,86.327)(.0326619,-.0562314){15}{\line(0,-1){.0562314}}
\multiput(49.745,84.64)(.0326619,-.0562314){15}{\line(0,-1){.0562314}}
\multiput(50.725,82.953)(.0326619,-.0562314){15}{\line(0,-1){.0562314}}
\multiput(51.704,81.266)(.0326619,-.0562314){15}{\line(0,-1){.0562314}}
\multiput(52.684,79.579)(.0326619,-.0562314){15}{\line(0,-1){.0562314}}
\multiput(53.664,77.892)(.0326619,-.0562314){15}{\line(0,-1){.0562314}}
\multiput(54.644,76.205)(.0326619,-.0562314){15}{\line(0,-1){.0562314}}
\multiput(55.624,74.518)(.0326619,-.0562314){15}{\line(0,-1){.0562314}}
\multiput(56.604,72.832)(.0326619,-.0562314){15}{\line(0,-1){.0562314}}
\multiput(57.584,71.145)(.0326619,-.0562314){15}{\line(0,-1){.0562314}}
\multiput(58.563,69.458)(.0326619,-.0562314){15}{\line(0,-1){.0562314}}
\multiput(59.543,67.771)(.0326619,-.0562314){15}{\line(0,-1){.0562314}}
\multiput(60.523,66.084)(.0326619,-.0562314){15}{\line(0,-1){.0562314}}
\multiput(61.503,64.397)(.0326619,-.0562314){15}{\line(0,-1){.0562314}}
\multiput(62.483,62.71)(.0326619,-.0562314){15}{\line(0,-1){.0562314}}
\multiput(63.463,61.023)(.0326619,-.0562314){15}{\line(0,-1){.0562314}}
\multiput(64.443,59.336)(.0326619,-.0562314){15}{\line(0,-1){.0562314}}
\multiput(65.422,57.649)(.0326619,-.0562314){15}{\line(0,-1){.0562314}}
\multiput(66.402,55.962)(.0326619,-.0562314){15}{\line(0,-1){.0562314}}
\multiput(67.382,54.275)(.0326619,-.0562314){15}{\line(0,-1){.0562314}}
\multiput(68.362,52.588)(.0326619,-.0562314){15}{\line(0,-1){.0562314}}
\multiput(69.342,50.901)(.0326619,-.0562314){15}{\line(0,-1){.0562314}}
\multiput(70.322,49.214)(.0326619,-.0562314){15}{\line(0,-1){.0562314}}
\multiput(71.302,47.527)(.0326619,-.0562314){15}{\line(0,-1){.0562314}}
\multiput(72.281,45.84)(.0326619,-.0562314){15}{\line(0,-1){.0562314}}
\multiput(73.261,44.153)(.0326619,-.0562314){15}{\line(0,-1){.0562314}}
\multiput(74.241,42.467)(.0326619,-.0562314){15}{\line(0,-1){.0562314}}
\multiput(75.221,40.78)(.0326619,-.0562314){15}{\line(0,-1){.0562314}}
\multiput(76.201,39.093)(.0326619,-.0562314){15}{\line(0,-1){.0562314}}
\multiput(77.004,96.803)(-.032661,-.0570733){15}{\line(0,-1){.0570733}}
\multiput(76.024,95.091)(-.032661,-.0570733){15}{\line(0,-1){.0570733}}
\multiput(75.044,93.378)(-.032661,-.0570733){15}{\line(0,-1){.0570733}}
\multiput(74.064,91.666)(-.032661,-.0570733){15}{\line(0,-1){.0570733}}
\multiput(73.084,89.954)(-.032661,-.0570733){15}{\line(0,-1){.0570733}}
\multiput(72.105,88.242)(-.032661,-.0570733){15}{\line(0,-1){.0570733}}
\multiput(71.125,86.53)(-.032661,-.0570733){15}{\line(0,-1){.0570733}}
\multiput(70.145,84.817)(-.032661,-.0570733){15}{\line(0,-1){.0570733}}
\multiput(69.165,83.105)(-.032661,-.0570733){15}{\line(0,-1){.0570733}}
\multiput(68.185,81.393)(-.032661,-.0570733){15}{\line(0,-1){.0570733}}
\multiput(67.205,79.681)(-.032661,-.0570733){15}{\line(0,-1){.0570733}}
\multiput(66.226,77.969)(-.032661,-.0570733){15}{\line(0,-1){.0570733}}
\multiput(65.246,76.256)(-.032661,-.0570733){15}{\line(0,-1){.0570733}}
\multiput(64.266,74.544)(-.032661,-.0570733){15}{\line(0,-1){.0570733}}
\multiput(63.286,72.832)(-.032661,-.0570733){15}{\line(0,-1){.0570733}}
\multiput(62.306,71.12)(-.032661,-.0570733){15}{\line(0,-1){.0570733}}
\multiput(61.326,69.408)(-.032661,-.0570733){15}{\line(0,-1){.0570733}}
\multiput(60.347,67.695)(-.032661,-.0570733){15}{\line(0,-1){.0570733}}
\multiput(59.367,65.983)(-.032661,-.0570733){15}{\line(0,-1){.0570733}}
\multiput(58.387,64.271)(-.032661,-.0570733){15}{\line(0,-1){.0570733}}
\multiput(57.407,62.559)(-.032661,-.0570733){15}{\line(0,-1){.0570733}}
\multiput(56.427,60.847)(-.032661,-.0570733){15}{\line(0,-1){.0570733}}
\multiput(55.447,59.134)(-.032661,-.0570733){15}{\line(0,-1){.0570733}}
\multiput(54.468,57.422)(-.032661,-.0570733){15}{\line(0,-1){.0570733}}
\multiput(53.488,55.71)(-.032661,-.0570733){15}{\line(0,-1){.0570733}}
\multiput(52.508,53.998)(-.032661,-.0570733){15}{\line(0,-1){.0570733}}
\multiput(51.528,52.286)(-.032661,-.0570733){15}{\line(0,-1){.0570733}}
\multiput(50.548,50.573)(-.032661,-.0570733){15}{\line(0,-1){.0570733}}
\multiput(49.569,48.861)(-.032661,-.0570733){15}{\line(0,-1){.0570733}}
\multiput(48.589,47.149)(-.032661,-.0570733){15}{\line(0,-1){.0570733}}
\multiput(47.609,45.437)(-.032661,-.0570733){15}{\line(0,-1){.0570733}}
\multiput(46.629,43.725)(-.032661,-.0570733){15}{\line(0,-1){.0570733}}
\multiput(45.649,42.012)(-.032661,-.0570733){15}{\line(0,-1){.0570733}}
\multiput(44.669,40.3)(-.032661,-.0570733){15}{\line(0,-1){.0570733}}
\multiput(43.69,38.588)(-.032661,-.0570733){15}{\line(0,-1){.0570733}}
\qbezier(57.452,71.594)(63.993,73.98)(64.877,66.821)
\qbezier(64.7,74.776)(69.031,72.92)(68.766,67.175)
\qbezier(66.998,66.821)(67.175,63.197)(63.816,60.634)
\put(73.892,76.543){\makebox(0,0)[cc]{$\phi_2={\pi \over 3}$}}
\put(81.317,69.296){\makebox(0,0)[cc]{$\phi_1=0$}}
\put(74.599,58.159){\makebox(0,0)[cc]{$\phi_6=-{\pi \over 3}$}}
\put(46.845,74.423){\makebox(0,0)[cc]{$\phi_3={2\pi \over 3}$}}
\put(39.244,64.7){\makebox(0,0)[cc]{$\phi_4=\pi$}}
\put(42.426,55.331){\makebox(0,0)[cc]{$\phi_5=-{2\pi \over 3}$}}
\end{picture}
\caption{{\footnotesize Convention for labeling sides, angles and
vertices of the $Z_n$-symmetric polygon $\bar\Pi$. Its counterpart
$\Pi$ is shown in Fig.\ref{regpoly0} (in contrast with that figure,
we draw here the perfect polygon). The dashed lines are normals to
sides.}} \label{Znpol}
\end{figure}
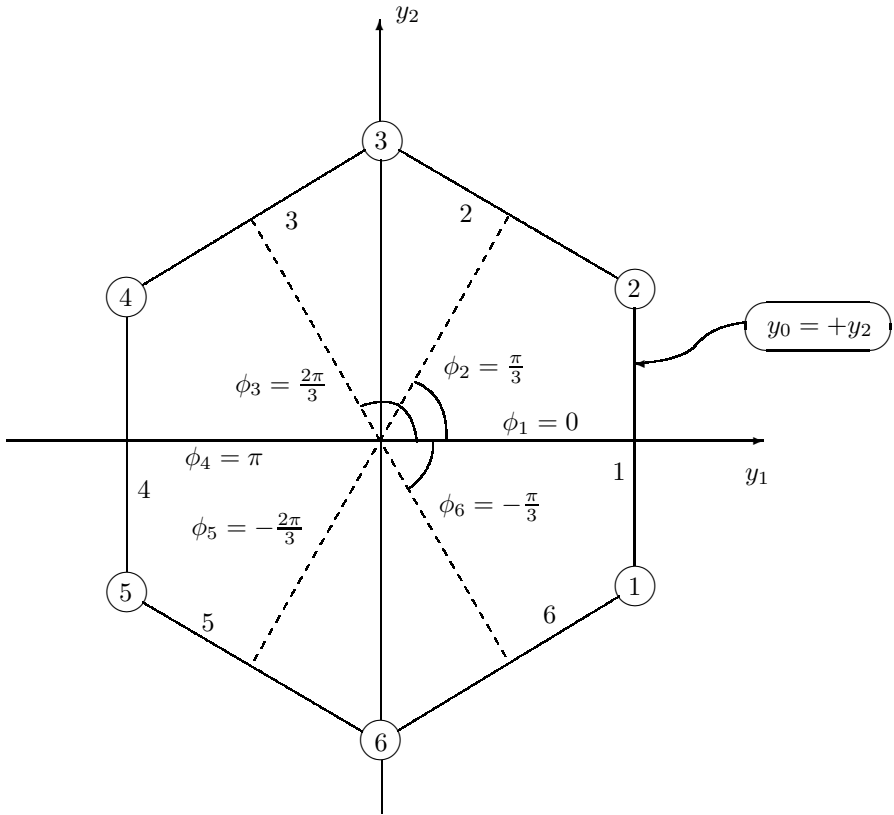

It is often convenient to represent (\ref{boundeqs}) in terms of
complex variable $z=y_1+iy_2$ and angles $\phi_a$ (see
Fig.\ref{Znpol}), $c_a=\cos\phi_a$, $s_a=\sin\phi_a$: \be z =
e^{i\phi_a}\left(1+i(-)^{a-1}y_0\right) \label{compbc} \ee

For $Z_n$-symmetric polygon $\bar\Pi$ all $h_a=1$, furthermore \be
c_a = \cos\frac{2\pi (a-1)}{n}, \ \ s_a = \sin\frac{2\pi (a-1)}{n}
\label{Znsines} \ee and the values of $(y_1,y_2;y_0)$ at the
vertices\footnote{ We assume that $a$-th vertex is at intersection
of $a$-th and $(a+1)$-st segments, see Fig.\ref{Znpol}. } are: \be
y_1^a = \frac{\cos\frac{\pi(2a-3)}{n}}{\cos\frac{\pi}{n}},\ \ \ \
y_2^a = \frac{\sin\frac{\pi(2a-3)}{n}}{\cos\frac{\pi}{n}},\nn\\
y_0^a = (-)^{a}\frac{1-\cos\frac{2\pi}{n}}{\sin\frac{2\pi}{n}}=
(-)^{a}\tan \frac{\pi}{n} \ee

Non-vanishing $y_0$ breaks the $Z_n$-symmetry when $\bar\Pi$
is lifted to $\Pi$. However, if we additionally assume that

\bigskip

$\bullet$ $y_0$ switches between increase and decrease at every
vertex, like in Fig.\ref{regpoly0}, then the symmetry is actually
preserved: $\Pi$ and thus the solution of interest possess the
$Z_{n/2}$-symmetry under rotation of $(y_1,y_2)$ plane by the angle
$\frac{4\pi}{n}$, while rotation by $\frac{2\pi}{n}$ is accompanied
by a flip $y_0\rightarrow -y_0$. The boundary ring also inherits
this symmetry.

\subsection{Polynomials that vanish at the boundary
(the boundary ring of $\Pi$)}

Three such polynomials are immediately read from (\ref{boundeqs})
\be
P_\Pi(y_1,y_2) = \prod_{a=1}^n \Big( h_a- c_a y_1 - s_a y_2\Big),\nn\\
\tilde P_\Pi(y_0,y_1) =
\prod_{a=1}^n \Big( y_1 + (-)^{a+1}s_ay_0 - c_ah_a\Big),\nn\\
\widetilde{\tilde P}_\Pi(y_0,y_2) =
\prod_{a=1}^n \Big( y_2 + (-)^{a}c_ay_0 - s_ah_a\Big)
\label{Pnpoly}
\ee
In what follows all $h_a=1$, and
\be
P_2=y_0^2+1-y_1^2-y_2^2
\ee
is also vanishing at the boundary.
Then one can consider division of polynomials (\ref{Pnpoly})
by $P_2$:
\be
P(\vec y) = P_2(\vec y) Q(\vec y) + S(\vec y)
\ee
Then $S$ is also vanishing at the boundary.
In this way one
can produce more polynomials from the {\it boundary ring},
but in general they have the same power as original $P$'s.
Of real interest are situations when $S(\vec y)$ factorizes,
$S(\vec y) = S_+(\vec y)S_-(\vec y)$ and one of the two
factors happens to belong to ${\cal R}_\Pi$ (this does
not follow immediately from factorization, since it can
happen instead that $S_+$ vanishes at some segments of $\Pi$,
while $S_-$ -- at the other).

\subsubsection{$n=4$, square ($b=0$)}

\be
P_{\Box} = (1-y_1^2)(1-y_2^2) = P_2 + (y_1y_2-y_0)(y_1y_2+y_0),
\nn \\
\tilde P_\Box = (1-y_1^2)(y_0^2-y_1^2) = -y_1^2P_2 -
(y_1y_2-y_0)(y_1y_2+y_0), \nn \\
\widetilde{\tilde P}_\Box = (1-y_2^2)(y_0^2-y_2^2) = - y_2^2P_2 -
(y_1y_2-y_0)(y_1y_2+y_0)
\ee
The relevant new element of the boundary ring
(selected by the choice of overall sign for $y_0$) is
\be
\tilde P_2 = y_0 - y_1y_2,
\label{tildeP2}
\ee
and it  indeed can be used as the third ansatz, giving
rise to solution of the NG equations (this is exactly
the main Alday-Maldacena solution of \cite{am1}).

\subsubsection{$n=4$, rhombus (any $|b|<1$)}

\be B^4P_\diamondsuit =
\Big(B^2-(y_1+by_2)^2\Big)\Big(B^2-(by_1+y_2)^2\Big)
=(1-b^2)^2P_2-S_+S_-
\nn \\
B^4\tilde P_\diamondsuit =
\Big(B^2y_1^2-(1-by_0)^2\Big)\Big(B^2y_1^2-(y_0+b)^2\Big)=\\
=S_+S_--2b(y_1^2+y_2^2-2)S_-+(b^2y_0^2-2b(1-b^2)y_0+b^2y_2^2-
(1+b^2+b^4)y_1^2-3b^2)P_2
\nn \\
B^4\widetilde{\tilde P}_\diamondsuit = \Big(B^2y_2^2-(y_0+b)^2\Big)
\Big(B^2y_2^2-(1-by_0)^2\Big)=\\
=S_+S_--2b(y_1^2+y_2^2-2)S_-+(b^2y_0^2-2b(1-b^2)y_0+b^2y_1^2-
(1+b^2+b^4)y_2^2-3b^2)P_2
\ee where \be S_\pm =(1-b^2)y_0\pm (b(y_1^2+y_2^2-2)+By_1y_2)\ee
Here $S_-\in {\cal R}_\Pi$, while this is not true for $S_+$.
Moreover, (\ref{y0y1y2b}) can be rewritten as \be S_-=bP_2=br^2\ee

\subsubsection{$n=6$, $Z_6$-symmetric polygon}

\be \tilde P_6 = y_1^6 - \frac{3}{2}y_1^4(y_0^2+1) +
\frac{9}{16}y_1^2(y_0^2+1)^2-
\frac{1}{16}(1-3y_0^2)^2 = \nn \\
= -\Big(y_1^4-\frac{1}{2}y_1^2(y_0^1+1) +
\frac{1}{16}(y_0^2+1)^2\Big) P_2 -
y_2^2\Big(y_1^2-\frac{1}{4}(y_0^2+1)\Big)^2+
\frac{1}{16}y_0^2(3-y_0^2)^2 = \nn \\ = \frac{1}{16}S_+S_-
-\Big(y_1^4-\frac{1}{2}y_1^2(y_0^1+1) + \frac{1}{16}(y_0^2+1)^2\Big)
P_2 \ee \be \widetilde{\tilde P}_6 = y_2^6 -
\frac{3}{2}y_2^4(y_0^2+1) + \frac{1}{2}y_2^3y_0(3-y_0^2) +
\frac{9}{16}y_2^2(y_0^2+1)^2 -\frac{3}{8}y_2y_0(y_0^2+1)(3-y_0^2) +
\frac{1}{16}y_0^2(3-y_0^2)^2 =\\= {1\over
16}\left(S_--4y_2P_2\right)^2 \ee  where \be S_\pm = y_0(3-y_0^2)\pm
y_2(4y_1^2 - y_0^2-1) \ee One can explicitly check that $S_-\in
{\cal R}_\Pi$ (while this is not true for $S_+$). $Z_3$-symmetric
version of this polynomial $S_+$ is obtained by subtracting
$y_2P_2$: \be P_3=y_0(3-y_0^2)-y_2(3y_1^2 - y_2^2) \label{P3} \ee
Further addition of $y_0P_2$ converts this $P_3$ into (for further
convenience, we also rescale this polynomial by ${1\over 4}$)\be
{\cal P}_3 = {1\over 4}y_0(4-y_1^2-y_2^2)-{1\over 4}y_2(3y_1^2 -
y_2^2) \label{calP3} \ee

\subsubsection{$n=8$, $Z_8$-symmetric polygon}

\be
\tilde P_8 = y_1^8 - 2y_1^6(y_0^2+1) +\frac{5}{4}y_1^4(y_0^2+1)^2-
\frac{1}{4}y_1^2(y_0^2+1)^3 + \frac{1}{4}y_0^2(1-y_0^2) = \nn \\
= -y_1^2
\Big(y_1^4-y_1^2y_2^2-y_1^2(y_0^2+1) + \frac{1}{4}(y_0^2+1)^2\Big)
P_2 + \nn \\
+ \Big(y_1y_2-\frac{1}{2}(y_0^2-1)\Big)
\Big(y_1y_2+\frac{1}{2}(y_0^2-1)\Big) (y_1y_2-y_0)(y_1y_2+y_0) \ee
The residual polynomial $S$ factorizes, but too strongly: particular
factors do not belong to the boundary ring (do not vanish at all the
boundaries), as it happened for $n=4$.

Instead the boundary ring contains a $Z_4$-symmetric polynomial of
degree $4$: \be P_4=y_0(1-y_0^2) - y_1y_2(y_1^2-y_2^2) \label{P4}
\ee Adding $y_0P_2$ and rescaling, we obtain \be {\cal P}_4 =
{1\over 2}y_0(2-y_1^2-y_2^2) - {1\over 2}y_1y_2(y_1^2-y_2^2)
\label{calP4} \ee

\subsubsection{Arbitrary even $n$, $Z_n$-symmetric polygon
\label{brZnsympol}}

The low-degree elements (\ref{tildeP2}), (\ref{P3}) and (\ref{P4})
of the boundary rings have an obvious generalization to
arbitrary $Z_n$-symmetric situation with even $n$:
the corresponding boundary rings always contain a polynomial
(generator) of degree $n/2$:
\be
P_{n/2} = \prod_{a=1}^{n/2}(s_a + c_ay_0) -
\prod_{a=1}^{n/2} (s_a y_1+c_ay_2) =
K_{n/2}(1,y_0) - K_{n/2}(y_1,y_2),
\ee
where $c_a$ and $s_a$ are given in (\ref{Znsines})
and the product
\be
K_{n/2}(y_1,y_2) = \prod_{a=1}^{n/2} (s_a y_1+c_ay_2)
= (-)^{n/2-1}\prod_{a=1}^{n/2} (-s_a y_1+c_ay_2)
= \prod_{a=1}^{n/2} {\rm Im} \left( e^{i\phi_a} z \right)
\ \stackrel{(\ref{Znsines})}{=} \
\frac{1}{2^{n/2-1}} {\rm Im}\left(z^{n/2}\right)
\label{Kn2}
\ee
is over the $n/2$ symmetry axes of $\bar\Pi$,
orthogonal to the $n/2$ pairs of polygon edges.
It is easy to see that
\be
K_{n/2}(1,y_0) = \frac{1}{2^{n/2-1}}
{\rm Im}\left((1+iy_0)^{n/2}\right) =
y_0 \tilde K(y_0^2)
\ee
where $\tilde K$ is a polynomial of degree
$entier(\frac{n-2}{4})$ of its variable.
By subtraction of appropriate powers of $P_2$ multiplied by
$y_0$ we can finally convert $P_{n/2}$ into
\be
{\cal P}_{n/2} = y_0Q_{(n)}(y^2) - K_{n/2}(y_1,y_2)
\label{calPn2}
\ee
with $y^2=y_1^2+y_2^2$ and

$$
\begin{array}{|c|c|c|}
\hline
&&\\
n & Q_{(n)} & K_{n/2} \\
&&\\
\hline
&&\\
2 & 1 &  y_2  \\
&&\\
4 & 1 & y_1y_2  \\
&&\\
6 &  \frac{4-y^2}{4} &
\frac{y_2(3y_1^2-y_2^2)}{4}\\
&&\\
8 & \frac{2-y^2}{2} &
\frac{y_1y_2(y_1^2-y_2^2)}{2}  \\
&&\\
10 & \frac{(y^2-2y-4)(y^2+2y-4)}{16}
= \frac{16-12y^2+y^4}{16}
&\frac{y_2(5y_1^4-10y_1^2y_2^2+y_2^4)}{16}\\
&&\\
12 & \frac{(4-y^2)(4-3y^2)}{16}
= \frac{16-16y^2+3y^4}{16} &
\frac{y_1y_2(3y_1^2-y_2^2)(y_1^2-3y_2^2)}{16}\\
&&\\
14 & -\frac{(y^3+4y^2-4y-8)(y^3-4y^2-4y+8)}{64}
= \frac{64-80y^2+24y^4-y^6}{64}
&\frac{y_2(7y_1^6-35y_1^4y_2^2+21y_1^2y_2^4-y_2^6)}{64}\\
&&\\
16 & -\frac{(y^2-2)(8-8y^2+y^4)}{16}
= \frac{16-24y^2+10y^4-y^6}{16}
&\frac{y_1y_2(y_1^2-y_2^2)(y_1^4-6y_1^2y_2^2+y_2^4)}{16}\\
&&\\
18 & -\frac{(4-y^2)(y^3-6y^2+8)(y^3+6y^2-8)}{256}
= \frac{256-448y^2+240y^4-40y^6 +y^8}{256}
&\frac{y_2(3y_1^2-y_2^2)(3y_1^6-27y_1^4y_2^2+33y_1^2y_2^4
-y_2^6)}{256}\\
&&\\
20 & \frac{(y^2-2y-4)(y^2+2y-4)(5y^4-20y^2+16)}{256}
=\frac{256-512y^2+336y^4-80y^6+5y^8}{256}
&\frac{y_1y_2(5y_1^8-60y_1^6y_2^2+126y_1^4y_2^4
-60y_1^2y_2^6+5y_2^8)}{256}\\
&&\\
& \ldots &\\
\hline
\end{array}
$$
In general
\be
Q_{(n)} = 
\frac{(1+\sqrt{1-y^2})^{n/2}-(1-\sqrt{1-y^2})^{n/2}}
{2^{n/2} \sqrt{1-y^2}} = \nn \\
= \left(1- \frac{n-4}{8}\,y^2 +
\frac{(n-6)(n-8)}{128}\,y^4
- \frac{(n-8)(n-10)(n-12)}{3072}\,y^6 + \ldots\right)
\ +\ O(y^{n})
\label{Qexpan}
\ee
The role of the last term at the r.h.s. is to eliminate $y^2$ for
$n\leq 4$, including $n=2$;
$y^4$ for $n\leq 8$,
including $n=2,4$ and so on.

It follows from (\ref{Qexpan}) that near the point $y^2=1$
\be
Q_{(n)}(y^2) =
\frac{n}{2^{n/2}}\sum_{k=0}^{\frac{n-2}{4}}
\frac{(\frac{n}{2}-1)!}{(2k+1)!(\frac{n}{2}-1-2k)!}(1-y^2)^k =
\frac{n}{2^{n/2}} + O(y^2-1)
\label{Q1}
\ee

\section{Power series solutions in $Z_n$-symmetric case
\label{exa}}
\setcounter{equation}{0}

\subsection{Recurrent relations
\label{exaid}}

With our four assumptions, listed in s.\ref{assum}, in the case of
the $Z_n$-symmetric $\bar\Pi$ the boundary conditions -- and thus
the solution of interest -- lies entirely at $Y_3=Y_4=0$ (i.e.
essentially in $AdS_3$) and has a number of discrete symmetries. We
list the symmetries in detail in the section \ref{hexa}, devoted to
the first non-trivial case of $n=6$. Here we just use the result of
symmetry analysis: it allows to look for the remaining unknown
function $y_0(y_1,y_2)$ in the form:\footnote{ Of course, one can
look at the power series solution to NG equations without imposition
of any symmetries:
$$
y_0 = \sum_{i,j\geq 0} a_{ij}y_1^iy_2^j
$$
The recurrence relations for coefficients $a_{ij}$
are somewhat complicated: already the at level two
$$
a_{02} = -\frac{a_{20}(1+a_{00}^2-a_{01}^2) + a_{11}a_{01}a_{10}}
{1+a_{00}^2-a_{10}^2}
$$
with $a_{00}$, $\ a_{01}, a_{10}$ and
$a_{11}, a_{20}$
remaining as free parameters, while at level three we have
$$
a_{21} = -\frac{1}{2a_{10}a_{01}( 1+a_{00}^2-a_{10}^2)}
\Big\{
3a_{30}(1+a_{00}^2-a_{01}^2)(1+a_{00}^2-a_{10}^2)
a_{12}(1+a_{00}^2-a_{10}^2)^2
+ $$ $$
+ 4a_{20}^2a_{10}(1+a_{00}^2-a_{01}^2)
+ 2a_{20}a_{10}\Big(a_{00}(1+a_{00}^2-a_{10}^2)
+2a_{01}a_{11}a_{10}\Big)
+a_{11}(1+a_{00}^2-a_{10}^2)(a_{00}a_{01}+a_{10}a_{11})
\Big\}
$$
and
$$
a_{03}=-\frac{1}{3}\Big\{
a_{21}(1+a_{00}^2-a_{01}^2)(1+a_{00}^2-a_{10}^2)
+(a_{00}a_{10}a_{11} + 2a_{01}a_{10}a_{12}+ a_{01}a_{11}^2)
(1+a_{00}^2-a_{10}^2)
$$ $$
+2a_{01}(2a_{20}-a_{00})\big[
a_{20}(1+a_{00}^2-a_{01}^2) + a_{01}a_{10}a_{11}\big]
\Big\}
\frac{1}{(a_{00}^2-a_{10}^2)(2+a_{00}^2-a_{10}^2)}
$$
with additional free parameters $a_{12},a_{30}$.

One can also lift the restriction $r^2=y_0^2+1-y_1^2-y_2^2$
and also substitute it by a power series expansion:
$$
r = 1 + \sum_{i,j\geq 0} \rho_{ij} y_1^iy_2^j
$$
Further analysis of these options is beyond the scope of the present
paper. } \be y_0 = K_{n/2} \sum_{i,j\geq 0} c_{ij}^{(n)}
K_{n/2}^{2i}y^{2j} = \sum_{i,j\geq 0} c_{ij}^{(n)}
K_{n/2}^{2i+1}y^{2j} \label{posey0} \ee The coefficients $c_{ij}$
are defined by NG equations, with $r^2 = P_2 = y_0^2+1-y^2$
substituted as another part of our ansatz. The NG equations produce
$c_{ij}$ in recursive form: all coefficients at the given level
$k=i\frac{n}{2}+j$ are determined by solving a linear system of
equations through the coefficients of the previous levels, for
example,\footnote{For $n=2$ and $n=4$ already the first of these
relations are slightly more involved:
$${\rm for}\ \ n=2\ \ \ c_{01} = \frac{3c_{10}}{c_{00}^2-4},$$
$${\rm for}\ \ n=4\ \ \ c_{01} = \frac{c_{00}(1-c_{00}^2)}{6}$$
This illustrates the general phenomenon: generic relations at level
$k$ arise in their most simple form for large enough $n$, while for
the lowest values of $n$ formulas include additional contributions.
If not this kind of complication, the series could be partly summed,
for example,
$$
\sum_{j\geq 0} c_{0j}y^{2j} \approx 2^{n/2}
\frac{n(n-2)c_{00}}{16\Gamma\left(\frac{1}{2}\right)}
\sum_{j=0}^\infty \frac{\Gamma\left(j+\frac{n-2}{4}\right)
\Gamma\left(j+\frac{n}{4}\right)}
{j!\,\Gamma\left(j+\frac{n+2}{2}\right) }\ y^{2j} +
O(c_{00}^2,c_{10},c_{20},\ldots)
$$
Explicitly written series is a hypergeometric function, however such
an expression has a limited value exactly because for given $n$ the
omitted terms at the r.h.s. are significant. \label{hyperg}} \be
c_{01} = \frac{(n-2)n}{8(n+2)}\,c_{00}, \ \ \ n\geq 6, \nn\\
c_{02} = \frac{(n-2)n}{128}\,c_{00},
\ \ \ n\geq 8, \nn \\
c_{03} = \frac{(n-2)n(n+8)}{3072}\,c_{00},
\ \ \ n\geq 10, \nn \\
c_{04} = \frac{(n-2)n(n+10)(n+12)}{8^4\cdot 4!}\,c_{00},
\ \ \ n\geq 12,
\nn\\
c_{05} = \frac{(n-2)n(n+12)(n+14)(n+16)}{8^5\cdot 5!}\,c_{00},
\ \ \ n\geq 14,
\nn\\
\ldots \nn \\
c_{0j} =  \frac{(n-2)n\cdot(n+4j-4)!!}{8^j j!\,(n+2j)!!}\,
c_{00}, \ \ \ n\geq 4+2j \nn \\
\ldots
\label{coj}
\ee
As illustrated by this example,
recursion relations depend on $n$ and we list the first few
relations below in subsections, devoted to consideration
of particular lowest even values $n$.

The lowest values of coefficients $c_{ij}$ are listed in the
table:

\bigskip

$$
\begin{array}{|c|c|c|c|c|c|c|}
\hline
&&&&&&\\
n & \frac{c_{01}}{c_{00}} & \frac{c_{02}}{c_{00}}
& \frac{c_{03}}{c_{00}} &\frac{c_{04}}{c_{00}}
&\frac{c_{05}}{c_{00}}& \ldots \\
&&&&&&\\
\hline
&&&&&&\\
&\frac{n(n-2)}{8(n+2)}&\frac{n(n-2)}{8^2\cdot 2!}
&\frac{(n-2)n(n+8)}{8^3 \cdot 3!}
&\frac{(n-2)n(n+10)(n+12)}{8^4\cdot 4!}
&\frac{(n-2)n(n+12)(n+14)(n+16)}{8^5\cdot 5!}&\\
&+\ {\rm corrections}&+\ {\rm corrections}&+\ {\rm corrections}
&+\ {\rm corrections}&+\ {\rm corrections}&\\
&{\rm at}\ n<6 &{\rm at}\ n<8 &{\rm at}\ n<10
&{\rm at}\ n<12 &{\rm at}\ n<14 & \\
&&&&&&\\
\hline
\hline
&&&&&&\\
4 & \frac{1}{6} - \frac{1}{6}c_{00}^2
& \frac{1}{16} - \frac{5}{48}c_{00}^2 +  &
{\rm see} \ (\ref{cceofn4})
&  {\rm see} \ (\ref{cceofn4})  & {\rm see} \ (\ref{cceofn4})   & \\
&&&&&&\\
&&+\frac{1}{24}c_{00}^4 -
\frac{3}{16}\frac{c_{10}}{c_{00}}&&&&\\
&&&&&&\\
\hline
&&&&&&\\
6 & \frac{3}{8} & \frac{3}{16} -\frac{27}{320}c_{00}^2
&\frac{7}{64}-\frac{133}{1280}c_{00}^2
-\frac{3}{64}\frac{c_{10}}{c_{00}}
& {\rm see} \ (\ref{cceofn6})     &{\rm see} \ (\ref{cceofn6})  & \\
&&&&&&\\
\hline
&&&&&&\\
8 &\frac{3}{5} &\frac{3}{8}&\frac{1}{4} -  \frac{1}{28} c_{00}^2
&  \frac{45}{256} -  \frac{603}{8960}c_{00}^2
- \frac{3}{256}\frac{c_{10}}{c_{00}}
&   \frac{33}{256} - \frac{50247}{582420} c_{00}^2
- \frac{99}{3328}\frac{c_{10}}{c_{00}} &\\
&&&&&&\\
\hline
&&&&&&\\
10 &\frac{5}{6}&\frac{5}{8}&\frac{15}{32}&
\frac{275}{768} - \frac{125}{9216}c_{00}^2
&\frac{143}{512}-\frac{107}{3072}c_{00}^2-\frac{3}{1024}
\frac{c_{10}}{c_{00}}&\\
&&&&&&\\
\hline
&&&&&&\\
12 &\frac{15}{14}&\frac{15}{16}&\frac{25}{32}
&  \frac{165}{256} = \frac{3\cdot5\cdot11}{2^8} &
 \frac{273}{512} - \frac{27}{5632} c_{00}^2 &\\
&&&&&&\\
\hline
&&&&&&\\
14 &\frac{21}{16}&\frac{21}{16}&\frac{77}{64}
&   \frac{273}{256}=\frac{3\cdot 7\cdot 13}{2^8}
&  \frac{1911}{2048}=\frac{3\cdot7^2\cdot13}{2^{11}}  &\\
&&&&&&\\
\hline
&&&&&&\\
16 &\frac{14}{9}&\frac{7}{4}&\frac{7}{4}
&   \frac{637}{384}= \frac{7^2\cdot 13}{3\cdot 2^7}
&  \frac{49}{32}= \frac{7^2}{2^5}  & \\
&&&&&&\\
\hline
&&&&&&\\
\ldots &&&&&&\\
&&&&&&\\
\hline
\end{array}
$$

\bigskip

NG equations do not fix all the coefficients $c_{ij}$
unambiguously: solution to the equations should depend
on arbitrary function of a single variable and indeed
recurrence relations do not determine some free parameters,
namely, $c_{i0}$ for all $i$. This freedom needs to be fixed
by boundary conditions.

\subsection{Boundary conditions and sum rules}

The problem is that the boundary conditions are imposed at
$\Pi$, i.e. at finite (rather than infinitesimally small)
values of $K$ and $y^2$, and one
needs to sum the whole series (\ref{posey0}) in order to
take them into account. At $\Pi$ we have
\be
y_0 = \frac{K}{Q_{(n)}(y^2)}, \nn \\
y_0 = \sqrt{y^2-1}
\ee
i.e.
\be
K^2 = (y^2-1)Q^2_{(n)}(y^2)
\ee
and
\be
\sum_{i,j\geq 0} c_{ij}^{(n)} (y^2-1)^iQ^{2i+1}_{(n)}(y^2)y^{2j}
= 1
\label{presumru}
\ee
For example,
expanding the l.h.s. of this relation in powers of $y^2$,
we obtain an infinite set of "sum rules" for the coefficients
$c_{ij}$:
\be
\sum_{i\geq 0}^{N} c_{i0}q_{i0} = 1, \nn \\
\sum_{i\geq 0}^{N} c_{i0}q_{i1}
+ \sum_{i\geq 0}^{N-1} c_{i1}q_{i0} = 0, \nn \\
\ldots \nn \\
\sum_{k=0}^m
\left( \sum_{i\geq 0}^{N-m} c_{ik}q_{i,m-k}\right) =
\delta_{m0}
\label{sr}
\ee
where $q_{ij}^{(n)}$ are expansion coefficients of the known
quantities
\be
\sum_{j\geq 0} q_{ij}^{(n)}y^{2j}
= (1-y^2)^iQ^{2i+1}_{(n)}(y^2)
= \left(1 -
\Big(i\,\frac{n}{4}+\frac{n-4}{8}\Big)y^2
+ \ldots\right)
\ee
and $N=\infty$.
The choice of the upper limits in
sums over $i$ in (\ref{sr}) can be made in different ways,
we present an example which treats $c_{i0}$ as small corrections
of $i$-th order, while exactly known coefficients are {\it not}
considered small -- as we shall see in examples below, this
is not a bad approximation to reality.
Expansion can also be made in other parameters, for example
in powers of $y^2-1$, see (\ref{sry21}) in s.\ref{y21} below.
However, in order to convert such formulas
to the form (\ref{sr}) one needs resummation
of series $\sum_j c_{ij}y^{2j}$, which can, probably, be
performed in the future as outlined in footnote \ref{hyperg}.

This illustrates the general problem:
it is not immediately clear how the two
ingredients of the problem --
the recurrence relations for $c_{ij}$, implied by NG equations
(which, additionally, we do not know in full yet),
and the sum rules (\ref{presumru}) and (\ref{sr}), implied
by boundary conditions,
-- can be combined to produce an answer
in a self-consistent analytical form.

\subsection{Approximate treatment of the $Z_n$-symmetric case}

What we can do, however, is to consider approximations.
This can of course be done in various ways, preserving
or optimizing one or another property of the problem.
Not surprisingly, they give different
-- even parametrically different --
estimates for the free parameters $c_{i0}$,
still for $Z_n$-symmetric polygons $\bar\Pi$ an
impressively good match can be found.

\subsubsection{Truncating sum rules (\ref{sr})
\label{tru}}

From power series point of view the most straightforward
approximation would be to cut the sums in (\ref{sr}) at some level
$N$, then only a limited number of coefficients $c_{ij}^{(n)}$ will
contribute, thus the recurrence relations for them are explicitly
available. Take the first $N$ of these truncated equations and solve
them to determine approximate values of the free parameters
$c_{i0}$, with $0\leq i<N$. Then the series (\ref{posey0}),
truncated to the level $N$ in sums over $i$, will produce an
approximate solution to our problem: a minimal surface in $AdS_3
\subset AdS_5$, bounded by the $Z_{n/2}$-symmetric polygon $\Pi$
(and $Z_n$-symmetric $\bar\Pi$). For example, for truncation at the
level $N=0$ we have simply \be {\bf level-zero\ truncation:} \ \ \
c_{00}=1, \ \ {\rm all\ other}\ \  c_{ij}=0 \label{lev0trun} \ee In
order to specify the next {\it free} parameters one can increase $N$
in (\ref{sr}). In the next approximation, for truncation at level
$N=1$, we have: \be
c_{00} + c_{10} = 1, \nn \\
c_{00}q_{01} + c_{10}q_{11} + c_{01}q_{00} =
-\frac{n-4}{8}\,c_{00} - \frac{3n-4}{8}\,c_{10} + c_{01} = 0,
\nn \\
c_{01} \ \stackrel{(\ref{coj})}{=}\
\frac{(n-2)n}{8(n+2)}\, c_{00} - \frac{1}{6}
\,\delta_{n,4} c_{00}^3\ \ \ {\rm for} \ \
n\geq 4,
\ee
implying that for $n\geq 6$
\be
{\bf level-one\ truncation:} \ \ \
c_{00}=\frac{(n+2)(3n-4)}{n(3n+2)}= 1-\frac{8}{n(3n+2)},\nn \\
c_{10}= \frac{8}{n(3n+2)}, \nn \\
c_{01}= \frac{(n-2)(3n-4)}{8(3n+2)},\nn \\
{\rm all\ other}\ \  c_{ij}=0
\label{lev1trun}
\ee
and so on.

As we shall see, this approach, at least with the low-level
truncations, does not produce a good enough match: even for $n=6$
the deviation from boundary conditions will be well seen by bare
eye.

\subsubsection{Expansion in the vicinity of $y^2=1$
\label{y21}}

The reason for this failure is obvious: as we already mentioned,
boundary conditions are imposed at finite values of $y^2$, and
$Q_{(n)}(y^2)$ changes rather fast with the change when $y^2$ goes
away from zero. For polygons $\Pi$ of arbitrary shape the variable
$y^2$ can change in broad range, however if $Z_n$-symmetry is
imposed, we are more lucky: at $\Pi$ the  variable $y^2$ takes
values between $1$ at the tangent points between sides of $\bar\Pi$
and the inscribed circle and $\left(\cos\frac{\pi}{n}\right)^{-2}$
at the vertices. For large enough $n$ the upper limit is practically
indistinguishable from the lower, i.e. $y^2\approx 1$ at $\Pi$. At
the same time $y_0^2$ on $\Pi$ varies between plus and minus
$\tan\frac{\pi}{n}$, i.e. is rather small at least at large enough
$n$. Actually, deviations of $y^2$ from $1$ and $y_0^2$ from $0$ are
below $\left(\frac{\pi}{n}\right)^2$, i.e. within 25\% at most
already at $n=6$.

All this implies that a much better approximation
can be based on expansion near $y^2=1$, instead of
$y^2=0$ considered in s.\ref{tru}.
At the same time expansion in $y_0^2$ can still be
taken around $y_0^2=0$.
The leading estimate in this approach
-- a substitute of (\ref{lev0trun}) -- is easily derived
from (\ref{presumru}):
\be
Q_{(n)}(1)\sum_{j=0}^{N=\infty} c^{(n)}_{0j} = 1
\label{sry21}
\ee
Unfortunately, we do not yet know how to calculate the
sum at the r.h.s. (see comments in footnote \ref{hyperg}).
What we can do, we can -- unjustly -- truncate the sum.
To distinguish the result from (\ref{sr}) we call it
by "approximation" rather than "truncation" and label
the free parameters, obtained at a given approximation
level by appropriate number of primes.
Putting $N=0$, we get
\be
{\bf the\ zeroth\ approximation:} \ \ \
c_{00}^{(n)} = \left(Q_{(n)}(1)\right)^{-1}
\ \stackrel{(\ref{Q1})}{=} \
\frac{2^{n/2}}{n} \equiv C_{00}^{(n)}
\label{brutru}
\ee
This clearly differs parametrically from (\ref{lev0trun}),
though for low values of $n$, the difference is not so
dramatic: from (\ref{brutru}) we have
$$
\begin{array}{|c|c|c|c|c|c|c|}
\hline
&&&&&&\\
n&4&6&8&10&12&\ldots\\
&&&&&&\\
\hline
&&&&&&\\
C_{00}^{(n)} &1 &\frac{4}{3}&2&\frac{16}{5}&\frac{16}{3}&\\
&&&&&&\\
\hline
\end{array}
$$
One can numerically improve this approximation by
increasing $N$, i.e. by taking into account other
coefficients $c_{0j}$. Remaining free parameters
$c_{i0}$ are defined from other sum rules from the
chain, which begins with (\ref{sry21}).
Remarkably, these parameters do not affect
(\ref{sry21}) itself and thus do not affect our prediction
for $c_{00}$ -- this is different from the situation
in s.\ref{tru} and can be important for further investigation,
because general formulas for $c_{0j}$ are much simpler
than those for $c_{ij}$ with $i\geq 1$.
In particular, for $N=1$ we get instead of
(\ref{brutru}):
\be
c_{00}^{(n)}+c_{01}^{(n)} = \left(Q_{(n)}(1)\right)^{-1}
\ee
and making use of the first line in (\ref{sry21}) we obtain
that in this approximation (for $n\geq 6$)
\be
{\bf the\ first\ approximation:} \ \ \
c_{00}^{(n)} = \frac{\left(Q_{(n)}(1)\right)^{-1}}{
1 + \frac{n(n-2)}{8(n+2)}}
= \frac{C_{00}^{(n)}}{1 + \frac{n(n-2)}{8(n+2)}} =
\frac{2^{n/2}}{n\left(1 + \frac{n(n-2)}{8(n+2)}\right)}
 \equiv {C'}^{(n)}_{00}
\label{brutru'}
\ee
and
$$
\begin{array}{|c|c|c|c|c|c|c|}
\hline
&&&&&&\\
n&4&6&8&10&12&\ldots\\
&&&&&&\\
\hline
&&&&&&\\
\frac{{C'}_{00}^{(n)}}{C_{00}^{(n)}}
&1 &\frac{8}{11}=0.(72)&\frac{5}{8}=0.625&\frac{6}{11}=0.(54)
&\frac{14}{29}=0.4827586207\ldots &\\
&&&&&&\\
\hline
&&&&&&\\
{C'}_{00}^{(n)} &1
&\frac{32}{33}=0.9(69)
&\frac{5}{4}=1.25 &\frac{96}{55} = 1.7(45)
&\frac{224}{87}=2.574712644\ldots &\\
&&&&&&\\
\hline
\end{array}
$$
Similarly one can evaluate $C''_{00}$ for second-level
truncation and so on.

\subsubsection{Straightening of edges
\label{straight}}

One can try to further improve estimate (\ref{brutru}) by a somewhat
different method. Since it is based on expansion near $y^2=1$, it is
clear that (\ref{sry21}) and thus (\ref{brutru}) optimize the
matching of boundary conditions in the vicinity of this point -- a
tangent point with inscribed circle.

However, one can think of other optimization criteria.
For example, one can rather minimize the deviation of
$y_0(y_1=1,y_2)$
from the boundary condition -- a straight line $y_0=y_2$ --
in average, i.e. "globally" rather than locally, in vicinity
of a middle point.
This can be easily achieved by the mean square
method, adjusting $c_{00}$ to minimize the integral
\be
\int_{{\rm segment}} \Big(c_{00}K_{n/2}-y_2\Big)^2 dy_2
\ee
One can also take, say, $c_{01}$ into account, by minimizing
\be
\int_{{\rm segment}} \Big((c_{00}+c_{01}y^2)K_{n/2}-y_2\Big)^2 dy_2
\ee
and substituting $c_{01}$ from (\ref{coj}).
These mean-square values of $c_{00}$ are
\be
c_{00} = 2^{n/2-1} \frac{
\int_{-t_n}^{t_n} t\, {\rm Im}\, (1+it)^{n/2}dt }
{\int_{-t_n}^{t_n} \left\{ {\rm Im}\, (1+it)^{n/2}\right\}^2dt }
\label{meas}
\ee
and
\be
c_{00}' = 2^{n/2-1}
\frac{
\int_{-t_n}^{t_n} t\, {\rm Im}\, (1+it)^{n/2}
\left(1+\frac{n(n-2)}{8(n+2)}(1+t^2)\right)dt }
{\int_{-t_n}^{t_n} \left\{ {\rm Im}\, (1+it)^{n/2}
\left(1+\frac{n(n-2)}{8(n+2)}(1+t^2)\right)
\right\}^2dt }, \ \ \ \ n>4
\label{meas'}
\ee
and they are slightly different from $C_{00}$ in (\ref{brutru})
and $C'_{00}$ in (\ref{brutru'}) respectively:
$$
\begin{array}{|c|c|c|c|c|c|c|}
\hline
&&&&&&\\
n&4&6&8&10&12&\ldots\\
&&&&&&\\
\hline
&&&&&&\\
\xi_n = \frac{c_{00}}{C_{00}}&\ 1\ &1.070&1.112&1.140&1.159&\\
&&&&&&\\
\hline
&&&&&&\\
\xi'_n = \frac{c'_{00}}{{C'_{00}}^{\phantom{5^5}}\!}
&1&1.017&1.073&1.110&1.136&\\
&&&&&&\\
\hline
\end{array}
$$

Looking at the plots confirms our expectation that
the choice $c_{00}=C_{00}$
minimizes the deviation at $y_2=0$, while the mean square
method allows to diminish the "global" deviation.
It is also clear that taking corrections into account
makes the difference between local and global smaller,
i.e. indeed improves the approximation.

\subsubsection{Sharpening angles
\label{angles}}

Of course, optimization of boundary conditions "in average"
is not the only alternative to that of behavior at a tangent
point. One more interesting option is to optimize the
behavior of solutions at the angles of $\Pi$,
responsible for quadratic divergencies of regularized area.
This is straightforward application of discriminantal
technique \cite{NOLAL}, but lies beyond the scope of the
present paper.
We list only a few typical values of $c_{00}^{(n)}$,
produced by this optimization criterium
in the leading approximation (i.e. in neglect
of  corrections due to $c_{ij}^{(n)}$ with $i,i\neq 0$):
$$
\begin{array}{|c|c|c|c|c|c|c|}
\hline
&&&&&&\\
n&4&6&8&10&12&\ldots\\
&&&&&&\\
\hline
&&&&&&\\
c_{00}^{(n)}& 1
&\frac{8}{3\sqrt{3}}=1.5396\ldots 
&\frac{3\sqrt{3}}{2} = 2.5980\ldots 
&
\frac{256\sqrt{5}}{125}
= 4.5794\ldots 
&\frac{100\sqrt{5}}{27} = 8.2817\ldots 
&\\
&&&&&&\\
\hline
&&&&&&\\
\eta_n = \frac{c_{00}^{(n)}}{C_{00}^{(n)}}& 1
&\frac{2}{\sqrt{3}} = 1.1547\ldots 
&\frac{3\sqrt{3}}{4} = 1.2990\ldots 
&\frac{4.579\ldots}{3.2} = 1.431\ldots 
&\frac{8.2817\ldots}{16/3} = 1.5528\ldots 
&\\
&&&&&&\\
\hline
\end{array}
$$
Corrections -- though somewhat ugly --
are also relatively easy to include.
For example, for the coefficient in
$c_{00}^{(n)}K_{n/2}\left(1+
\left[\frac{n(n-2)}{8(n+2)}-\frac{1}{6}\delta_{n,4}
c_{00}^2\right]
z\bar z\right)$
the angle-existence criterium gives:
$$
\begin{array}{|c|c|c|c|c|c|c|}
\hline
&&&&&&\\
n&4&6&8&10&12&\ldots\\
&&&&&&\\
\hline
&&&&&&\\
&&\frac{  \sqrt{ 2386309-59\cdot 1153^{3/2}}}{48\sqrt{33} }
&\frac{\sqrt{94\cdot(31\cdot 39)^{3/2}-18\cdot 29\cdot 7529}}{100}
&\frac{25\sqrt{6}}{27}&&\\
c_{00}^{(n)}
& 1
&
= 1.0023\ldots
&
= 1.4632\ldots 
&=2.2680\ldots
&3.6566\ldots 
&\\
&&&&&&\\
\hline
&&&&&&\\
\eta'_n = \frac{c_{00}^{(n)}}{{C'}^{(n)}_{00}}
& 1
&1.0337\ldots
&1.1705\ldots
&1.2994\ldots
&1.4202\ldots
&\\
&&&&&&\\
\hline
\end{array}
$$
Comparing with s.\ref{straight}, we see that both straightening
sides of the polygon and sharpening its angles requires slight
increase of $c_{00}$, naturally, sharpening requires a stronger
increase because it involves vertices of $\Pi$ which are mostly
remote from the tangent points.

\subsubsection{Comparison table\label{compta}}

It is instructive to summarize our discussion of approximation
approach in the form of the following table.
The table lists optimal values of the most important free parameter
$c_{00}^{(n)}$.
Different lines in it correspond to different optimization
criteria, considered in the previous subsections.
Different columns correspond to truncations at different level,
to be concrete, in the $N$-th of this table contributions
from $c_{0j}$ with $j\leq N$ are taken into account,
all $c_{ij}$ with $i>0$ are neglected. they can also be
incorporated, but this will unnecessarily overload the formulas.

The difference between the first two lines can be shortly
illustrated as follows. They both use (\ref{presumru}) in the
schematic form of
\be
\sum_{j=0}^N c_{0j}y^{2j} = \Big(Q(y^2)\Big)^{-1}
\ee
In the first line we take $y^2\approx 0$ and obtain
\be
c_{00} \approx \frac{1}{Q(0)} = 1
\ee
with negligible corrections dues to $c_{0j}$, since they
are multiplied by small $y^2$.
In the second line we take instead $y^2\approx 1$ and obtain
\be
c_{00} + c_{01} + \ldots + c_{0N} \approx \frac{1}{Q(1)} =
\frac{2^{n/2}}{n} \equiv C_{00}
\label{c2line}
\ee
Thus the resulting $c_{00}$ differs from unity for two reasons:
$Q(1)\neq Q(0)$ and the sum at the l.h.s. multiplies $c_{00}$
by a factor $1 + \frac{c_{01}}{c_{00}} + \ldots +
\frac{c_{0N}}{c_{00}}$, which can be easily evaluated with the
help of (\ref{coj}).

$$
\begin{array}{|c|c|c|c|c|}
\hline
&&&&\\
\ \ \ \ \ \ \ \ \ \ \ \ \ N&0&1&2&\ldots \\
{\rm optimization}&&&&\\
{\rm criterium}&&&&\\
\hline
&&&&\\
s.\ref{tru}&1  &\ \ \ {\rm corrections}
&{\rm are\ small}\ \ \ \ \ \ \ \ \ \ & \\
&&&&\\
\hline
&&&&\\
s.\ref{y21}&C_{00}^{(n)} = \frac{2^{n/2}}{n}&
{C'}^{(n)}_{00}
&{C''}^{(n)}_{00}
&\\
&&&&\\
\hline
&&&&\\
s.\ref{straight}& \xi_n C_{00}^{(n)}& \xi_n'{C'}_{00}^{(n)} &
\xi''_n{C''}_{00}^{(n)} &\\
&&&&\\
\hline
&&&&\\
s.\ref{angles}&\eta_n C_{00}^{(n)}&\eta'_n {C'}_{00}^{(n)}&
\eta''_n {C''}_{00}^{(n)} &\\
&&&&\\
\hline
\end{array}
$$
Note that $\xi''_n$ and $\eta''_n$ are not
presented in ss.\ref{straight} and \ref{angles}, but they
can be easily evaluated by the same methods.

\bigskip

As demonstrated in the following sections this approach
works surprisingly well. Even without promoting it further
to exact analytical solution, one can try to use these
approximations for the study of regularized $NG$ and
$\sigma$-model actions and approximate comparison with the
BDS/BHT formulas. For this purpose one needs to extend our
consideration from $Z_{n/2}$-symmetric to generic polygons $\Pi$
(at the first stage the boosting procedure of \cite{am1} can
be enough to produce some non-trivial results), what requires
construction of the corresponding boundary rings and finding
the adequate counterparts of the ansatz (\ref{anzaNG}) in
${\cal R}_\Pi$. Regularization issues would be the next
(note that one should be also careful with the difference
between $NG$ and $\sigma$-model actions which can arise
after $\epsilon$-regularization \cite{popo}, despite this did
not happen at $n=4$, one can not
a priori exclude the possibility that this difference
depends on the shape of $\Pi$). All these issues are left to
the future work. In what follows we present only some examples
of approximate solutions.

\subsection{Examples \label{cexa}}

\subsubsection{$n=4$, a $Z_4$-symmetric $\bar\Pi$, i.e. a square}

We already considered this example among the known ones
in the previous sections. Here we use it to illustrate
the power series consideration.

Taking the symmetry-dictated representation (\ref{posey0}),
\be
y_0 = \sum_{i,j\geq 0}^{N=\infty} c_{ij}(y_1y_2)^{2i+1}y^{2j}
\label{squan}
\ee
and substituting it (together with $r^2=P_2=1+y_0^2-y^2$)
into the NG equations, we obtain:
\be
\begin{array}{rcl}
c_{01} &=& \frac{c_{00}(1-c_{00}^2)}{6},  \nn \\
c_{02} &=&  \frac{1}{16} c_{00} - \frac{5}{48}c_{00}^3 +
 \frac{1}{24}c_{00}^5 - \frac{3}{16}c_{10},  \nn \\
c_{03} &=&  \frac{1}{32} c_{00}
 - \frac{113}{1680} c_{00}^3 + \frac{151}{3360} c_{00}^5
 - \frac{1}{112} c_{00}^7 - \left(\frac{45}{224}
 - \frac{11}{224} c_{00}^2\right)c_{10},   \nn \\
c_{04} &=&  \frac{7}{384} c_{00}
- \frac{1591}{34560} c_{00}^3
 + \frac{1921}{48384} c_{00}^5 - \frac{1709}{120906} c_{00}^7
 + \frac{1}{448} c_{00}^9 -\nn \\
  &-&  \left(\frac{45}{256} -  \frac{113}{1344} c_{00}^2
 + \frac{115}{5376} c_{00}^4\right)c_{10}
 + \frac{5}{256} c_{20}, \nn \\
c_{05} &=& \frac{3}{256} c_{00} - \frac{127}{3840} c_{00}^3
 + \frac{445111}{13305600} c_{00}^5- \frac{321599}{19958400} c_{00}^7
 + \frac{35671}{7983360} c_{00}^9 - \frac{181}{399168} c_{00}^{11}-
 \nn \\
  &-&  \left(\frac{75}{512} - \frac{36163}{354816} c_{00}^2
 + \frac{75223}{1774080} c_{00}^4
 - \frac{1583}{354816} c_{00}^6\right)c_{10}
 - \frac{513}{39424} c_{10}^2c_{00}
 + \left(\frac{225}{5632}
 +\frac{5}{5632} c_{00}^2\right)c_{20}, \nn \\
c_{06} &=&  \frac{33}{4096} c_{00} - \frac{1517}{61440} c_{00}^3
 + \frac{810469}{29030400} c_{00}^5
  - \frac{1156597}{70963200} c_{00}^7
  + \frac{58061}{9580032} c_{00}^9
  - \frac{449623}{383201280} c_{00}^{11}
  + \frac{4883}{38320128} c_{00}^{13}
 -\nn \\
  &-&  \left(\frac{495}{4096}
  -\frac{2327}{21504} c_{00}^2 + \frac{2438803}{42577920} c_{00}^4
  - \frac{75953}{6082560} c_{00}^6
  +\frac{8423}{4257792} c_{00}^8 \right)c_{10} \ -\
   \left(\frac{227}{7168} - \frac{15}{4928} c_{00}^2\right)
   c_{00}c_{10}^2 \ + \nn \\
  &+& \ \left(\frac{225}{4096}  + \frac{1}{45056}c_{00}^2
  + \frac{355}{135168} c_{00}^4  \right)c_{20}
  \ -\  \frac{7}{4096} c_{30}, \nn \\
   & &  \ldots
   \end{array}
   \ee
   \be
\begin{array}{rcl}
c_{11} &=&  - \frac{1}{126} c_{00}^3
  +\frac{1}{63} c_{00}^5 - \frac{1}{126} c_{00}^7
  + \left(\frac{15}{14} - \frac{3}{7} c_{00}^2\right) c_{10}, \nn \\
c_{12} &=&   - \frac{1}{72} c_{00}^3
+ \frac{19}{560} c_{00}^5
 - \frac{37}{1680} c_{00}^7
  + \frac{1}{504} c_{00}^9 +
 +\left(\frac{15}{16} - \frac{79}{112} c_{00}^2
 +\frac{13}{56} c_{00}^4\right)c_{10}
 - \frac{5}{16} c_{20}, \nn \\
c_{13} &=&
 - \frac{5}{288} c_{00}^3 + \frac{173}{3564} c_{00}^5
 - \frac{118817}{2993760} c_{00}^7 +\frac{3659}{374200} c_{00}^9
 - \frac{95}{74844} c_{00}^{11}
   \nn \\
   &+& \left(\frac{25}{32}
   - \frac{1939}{2376} c_{00}^2 + \frac{14989}{33264} c_{00}^4
  - \frac{4595}{66528} c_{00}^6   \right)c_{10}
  + \frac{45}{352} c_{00}c_{10}^2
  -\left( \frac{225}{352}-\frac{95}{1056} c_{00}^2 \right)c_{20},\nn \\
c_{14} &=& -\frac{11}{576}c_{00}^3 +\frac{611}{10368}c_{00}^5
-\frac{382643}{6652800}c_{00}^7 +\frac{1350247}{59875200}c_{00}^9
-\frac{59797}{11975040}c_{00}^{11}+\frac{149}{1197504}c_{00}^{13}
+\nn \\
&+&\left( \frac{165}{256}  -\frac{1895}{2304}c_{00}^2
+\frac{1552423}{2661120}c_{00}^4
-\frac{469631}{2661120}c_{00}^6
+\frac{4327}{133056}c_{00}^8
\right)c_{10}
+\left(\frac{555}{1792}-\frac{93}{1232}c_{00}^2
\right)c_{00}c_{10}^2
-\nn\\
&-&\left(\frac{225}{256} - \frac{2219}{8448}c_{00}^2
+\frac{105}{1408}c_{00}^4\right)c_{20}
+\frac{7}{128}c_{30}, \nn\\
& & \ldots
\end{array}
\ee
\be
\begin{array}{rcl}
c_{21} &=&
   \frac{1}{462} c_{00}^5 - \frac{31}{6930} c_{00}^7
  - \frac{2}{3465} c_{00}^9  + \frac{2}{693} c_{00}^{11} \nn \\
  &-& \left( \frac{37}{462} c_{00}^2
  - \frac{3}{22} c_{00}^4 + \frac{47}{462} c_{00}^6 \right)c_{10}
  - \frac{9}{22} c_{00}c_{10}^2
  + \left(+ \frac{45}{22} - \frac{5}{11} c_{00}^2
  \right)c_{20},  \nn \\
c_{22} &=&
+\frac{61}{9072}c_{00}^5 -\frac{1181}{71280}c_{00}^7
+\frac{7141}{1496880}c_{00}^9
+\frac{1889}{299376}c_{00}^{11}
-\frac{185}{149688}c_{00}^{13}
+\nn \\
&+&\left(  -\frac{37}{168}c_{00}^2+\frac{71497}{166320}c_{00}^4
-\frac{7009}{20790}c_{00}^6 + \frac{1429}{33264}c_{00}^8
\right)c_{10}
-\left(\frac{111}{112}-\frac{261}{616}c_{00}^2
\right)c_{00}c_{10}^2
+\nn\\
&+&\left( \frac{45}{16}  -\frac{111}{88}c_{00}^2
+\frac{85}{264}c_{00}^4
\right)c_{20}
-\frac{7}{16}c_{30}, \nn\\
& & \ldots
\end{array}
\label{cceofn4} \ee Note that sums over powers of $c_{00}$ are often
alternated, what could be a signal about the nice convergence
properties of the $c$-series, -- but not always(!), see, for
example, the $c_{20}$-terms in $c_{05}$ or the first line in
$c_{22}$ (this can be our error, but not a misprint).

Remaining $c_{i0}$ are the free parameters (moduli)
of NG solutions, which should be fixed by boundary conditions.

Remarkably, these recurrence relations possess a solution
$c_{ij}=0$, which corresponds to the $n=\infty$ solution
from s.\ref{ninfty}, approached from the side of $Z_4$-symmetric
configurations in the $(y_1,y_2)$ plane.
The corresponding choice of the free parameters is $c_{i0}=0$.
What is much less trivial, they possess another exact solution
when moduli are chosen to be $c_{i0}=\delta_{i0}$:
\be
c_{00}=1, \ \ \
{\rm all\ other}\ c_{ij}=0
\label{n4solexa}
\ee
what is the standard square solution $y_0=y_1y_2$,
see s.\ref{square}.
The first ($n=\infty$) limiting solution $c_{ij}^{(n)}=0$
will exist for all even values of $n$, while exact solutions
with some $c_{ij}$ non-vanishing still remain to be found
(unfortunately, not in this paper).

Now, one can construct plots of $y_0(y_1,y_2)$ and
the corresponding $r(y_1,y_2)$ for various choices
of free parameters with the help of truncated series,
i.e. for finite $N$ in (\ref{squan}).
It is clear that the change of free parameters change the
boundary conditions, and a special choice needs to be made
to match the right ones. Of course, in this case we know
the answer: it is (\ref{n4solexa}). What is important
for our approach, is that (\ref{n4solexa}) is also
reproduced by the truncated sum rules (\ref{sr}):
see (\ref{lev0trun}).

\subsubsection{$n=6$, a $Z_6$-symmetric $\bar\Pi$
\label{hexa} }

{\bf Symmetries}

\bigskip

The problem possesses
the following discrete symmetries,
see Figs.\ref{regpoly0} and \ref{Znpol}):

$Z_3$ ($120^\circ$ rotation):
\be
\begin{array}{lcl}
y_1 \rightarrow -\frac{1}{2}y_1 + \frac{\sqrt{3}}{2}y_2,
&\ \ \ \ \ \ \ & P_2 \rightarrow P_2, \\
&& P_3 \rightarrow P_3, \\
y_2 \rightarrow -\frac{\sqrt{3}}{2}y_1 - \frac{1}{2}y_2,
&&\\
&& K \rightarrow K, \\
y_0 \rightarrow y_0, && L \rightarrow L
\end{array}
\ee

$\tilde Z_3$ ($60^\circ$ rotation):
\be
\begin{array}{lcl}
y_1 \rightarrow \frac{1}{2}y_1 + \frac{\sqrt{3}}{2}y_2,
&\ \ \ \ \ \ \ & P_2 \rightarrow P_2, \\
&& P_3 \rightarrow -P_3, \\
y_2 \rightarrow -\frac{\sqrt{3}}{2}y_1 + \frac{1}{2}y_2,
&&\\
&& K \rightarrow -K, \\
y_0 \rightarrow y_0, && L \rightarrow -L
\end{array}
\ee

$Z_2$ (reflection w.r.t. the horizontal axis):
\be
\begin{array}{lcl}
y_1 \rightarrow y_1, & \ \ \ \ \ \ \
&P_2 \rightarrow P_2,\\
&& P_3 \rightarrow -P_3, \\
y_2 \rightarrow -y_2, && \\
&& K \rightarrow -K,  \\
y_0 \rightarrow -y_0, && L \rightarrow L
\end{array}
\ee

$\tilde Z_2$ (reflection w.r.t. the vertical axis):
\be
\begin{array}{lcl}
y_1 \rightarrow -y_1, & \ \ \ \ \ \ \
&P_2 \rightarrow P_2,\\
&& P_3 \rightarrow P_3, \\
y_2 \rightarrow y_2, && \\
&& K \rightarrow K,  \\
y_0 \rightarrow y_0, && L \rightarrow -L
\end{array}
\ee
Here $P_2$ and $P_3$ are the generators of the $Z_3$-invariant
boundary ring (i.e. they vanish at $\Pi$), given by
\be
P_2 = y_0^2+1-y_1^2-y_2^2 = y_0^2+1-y^2, \nn \\
{\cal P}_3 = \frac{1}{4}\Big(
y_0(4-y_1^2-y_2^2) - y_2(3y_1^2-y_2^2)\Big) =
y_0\left(1-\frac{1}{4}y^2\right)-K_3
\ee
and
\be
K_3 = \frac{1}{4}y_2(3y_1^2-y_2^2) , \nn \\
L_3 = \frac{1}{4}y_1(3y_2^2-y_1^2), \nn \\
y^2 = y_1^2+y_2^2 \ee Note that $L_3^2+K_3^2=\frac{1}{16}y^6$ and
$L$ by itself does not appear in the boundary ring.

It is now clear that
\be
y_0 = K_3 \sum_{i,j\geq 0}^{N=\infty} c_{ij}\, K_3^{2i}y^{2j}
\label{hexan}
\ee
is the most general power series consistent with the symmetries.

\bigskip

\noindent
{\bf Recurrence relations}

\bigskip

Recurrence relations, implied by NG equations, this time are
\be
\begin{array}{rcl}
c_{01} &=& \frac{3}{8}\,c_{00},\nn \\
c_{02} &=&  \frac{3}{16}\,c_{00} - \frac{27}{320}\,c_{00}^3, \nn \\
c_{03} &=&  \frac{7}{64}\,c_{00} - \frac{133}{1280}\,c_{00}^3
- \frac{3}{64}\,c_{10}, \nn \\
c_{04} &=& \frac{9}{128}\,c_{00} - \frac{14193}{143360}\,c_{00}^3
  + \frac{243}{51200}\,c_{00}^5 - \frac{27}{320}\,c_{10}, \nn \\
c_{05} &=&  \frac{99}{2048}\,c_{00}
- \frac{100269}{1146880}\,c_{00}^3
+ \frac{16659}{1638400}\,c_{00}^5
  - \left(\frac{27}{256}- \frac{1539}{225280}\,c_{00}^2
   \right)c_{10},  \nn \\
c_{06} &=&  \frac{143}{4096}\,c_{00}
 - \frac{172623}{2293760}\,c_{00}^3
 + \frac{660307}{45875200}\,c_{00}^5
 - \frac{2187}{6553600}\,c_{00}^7
 - \left(\frac{117}{1024}- \frac{16789}{901120}\,c_{00}^2
 \right)c_{10}
 + \frac{5}{4096}\,c_{20}, \nn \\
  & &   \ldots
  \end{array}
  \ee
  \be
\begin{array}{rcl}
c_{11} &=&  - \frac{3}{320}\,c_{00}^3
+ \frac{243}{1600}\,c_{00}^5+ \frac{9}{5}\,c_{10}, \nn \\
c_{12} &=&  - \frac{3}{128}\,c_{00}^3
+ \frac{81}{256}\,c_{00}^5 + \left(\frac{9}{4}
   - \frac{81}{352}\,c_{00}^2\right)c_{10}, \nn \\
c_{13} &=&  - \frac{39}{1024}\,c_{00}^3
+ \frac{12741}{28672}\,c_{00}^5
- \frac{2187}{51200}\,c_{00}^7
    + \left(\frac{39}{16}- \frac{4373}{7040}\,c_{00}^2
  \right)c_{10}
  - \frac{5}{64}\,c_{20}, \nn \\
   & &  \ldots
\end{array}
\label{cceofn6}
\ee
These formulas look a little simpler than (\ref{cceofn4}).
The reason is that the same level of complexity will be
now achieved in higher-order corrections: complicated
non-linear term lie over diagonal in the table in s.\ref{exaid},
and $c_{ij}$ with low $i+j$ get contributions only from the
first columns of the table -- thus they do not contain too
many non-linearities.

The recurrence relations possess a solution $c_{ij}=0$,
associated with the $n=\infty$ solution, but they do not
have any obvious non-trivial solution, like
(\ref{n4solexa}) at $n=4$.

\bigskip

\noindent
{\bf Approximations and plots}

\bigskip

Therefore we need to turn to our various approximate methods,
which we analyze both theoretically and experimentally --
with the help of computer simulations.
The results are summarized in the table
from s.\ref{compta} which is now filled for
$n=6$ and has one more -- experimental -- line  added.
We remind that it lists the values of a single free
parameter $c_{00}$,
adjusted under different assumptions with different
accuracy.

$$
\begin{array}{|c|c|c|c|}
\hline
&&&\\
n=6 \ \ \ \ \ \ \ \ \ \ \ N&0&1&\ldots \\
{\rm optimization}&&&\\
{\rm criterium}&&&\\
\hline
&&&\\
s.\ref{tru}&1  & \frac{14}{15}=0.9(3)
&  \\
&&&\\
\hline
&&&\\
s.\ref{y21}&C_{00}^{(6)} = \frac{4}{3}=1.(3)&
{C'}^{(n)}_{00} = \frac{32}{33}=0.(96)
&\\
&&&\\
\hline
&&&\\
s.\ref{straight}& \xi_6 C_{00}^{(6)}=1.4271\ldots
& \xi_6'{C'}_{00}^{(6)} = 0.9858\ldots
&\\
&&&\\
\hline
&&&\\
s.\ref{angles}
&\eta_6 C_{00}^{(6)}=\frac{8}{3\sqrt{3}}=1.5396\ldots 
&\eta'_6 {C'}_{00}^{(6)} = 1.0023\ldots
&\\
&&&\\
\hline
\end{array}
$$

The rest of this section is a set of comments to this table.

\bigskip

{\bf The first line} results from comparison of
reliable expansion of NG solutions at small values of $y^2$
with similar expansion of the boundary ring generators.

In the first column contains the value $c_{00}=1$
from (\ref{lev0trun}): the most naive approximation to
both NG equations and boundary conditions,
which basically takes nothing but $Z_6$ symmetry
into account.

At truncation level $N=1$, represented in the second
column, we have from (\ref{lev1trun}):
\be
\sum_{i=0}^{N=1} c_{i0} = c_{00} + c_{10} = 1, \nn \\
\sum_{i=0}^{N-1=0} c_{i1} +
\sum_{i=0}^{N=1} \left(-\frac{1+6i}{4}\right)c_{i0} =
c_{01} - \frac{1}{4}c_{00} - \frac{7}{4}c_{10}
\ \stackrel{(\ref{cceofn6})}{=}\
\frac{1}{8}\Big(c_{00} - 14c_{10}\Big) = 0
\label{n6level1tru}
\ee
what means that in this approximation
\be
c_{00} = \frac{14}{15} =
\left.\frac{(n+2)(3n-4)}{n(3n+2)}\right|_{n=6}, \ \ \
c_{10} = \frac{1}{15} =
\left.\frac{8}{n(3n+2)}\right|_{n=6}
\label{lev1trun6}
\ee
We see that already at this low level
$c_{00}$ is indeed very close to $1$, while
$c_{10}$ is negligibly small. This last fact can be used
for a posteriori justification of truncation procedure:
the second terms in (\ref{lev1trun6}) are much smaller than
the first terms.
Thus inclusion of additional free parameter ($c_{10}$)
appears inessential, while $c_{01}$, though large enough,
$c_{01}=\frac{3}{8}c_{00}$ does not actually affect the
value of $c_{00}$, because it does not show up in the
first equation in (\ref{n6level1tru}).

{\bf Second line} results from comparison of expansions
with typical $y^2\sim 1$. This is expected to considerably
improve the matching with boundary conditions, at expense
of a worse control over NG equation. Exact criterium, adopted
in this line, is optimized behavior at the tangent points
between $\bar\Pi$ and its inscribed circle (i.e. at
$z=e^{i\pi k\over 3}$). First and second column differ by the
choice of $y_0(y_1,y_2)$ for this adjustment:
it is
\be
y_0 = c_{00}K_3
\label{n6cK}
\ee
in the first column and
\be
y_0 = c_{00}K_3\left(1+\frac{3}{8}y^2\right)
\label{n6cKc00}
\ee
in the second one.

{\bf Third line} differs from the second one by a slight
change of optimization criterium: now we adjust $c_{00}$
in (\ref{n6cK}) and (\ref{n6cKc00}) in the first and second
columns in order to make $y_0(y_1,y_2)$ closer to the
segments of $\bar\Pi$ "in average", at expense of
weakening the condition at the middle (tangent) points.
As seen from the table this implies a slight increase in
optimal $c_{00}$.

{\bf Forth line} results from shifting the emphasize
in optimization criterium further from the tangent points
-- this time to the vertices of $\bar\Pi$. It is now
requested that angles -- the origins of the main (quadratic)
divergencies of the regularized action -- are really
angles and not some smoothened curves of with large curvature.
This implies an even stronger increase of optimal $c_{00}$.



\subsubsection{$n = 8$}


In this and the two next subsubsections we show the Tables for
$n=8$, $n=10$ and $n=12$.

$$
\begin{array}{|c|c|c|c|}
\hline
&&&\\
n=8 \ \ \ \ \ \ \ \ \ \ \ N&0&1&\ldots \\
{\rm optimization}&&&\\
{\rm criterium}&&&\\
\hline
&&&\\
s.\ref{tru}&1
& \frac{25}{26} = 0.9615\ldots
& \\
&&&\\
\hline
&&&\\
s.\ref{y21}&C_{00}^{(8)} = 2&
{C'}^{(8)}_{00} = \frac{5}{4}=1.25
&\\
&&&\\
\hline
&&&\\
s.\ref{straight}& \xi_8 C_{00}^{(8)}=2.2239\ldots
&\xi_8'{C'}_{00}^{(8)} = 1.3412\ldots
&\\
&&&\\
\hline
&&&\\
s.\ref{angles}&
\eta_8 C_{00}^{(8)}=\frac{3\sqrt{3}}{2} = 2.5980\ldots 
&\eta'_8 {C'}_{00}^{(8)}= 1.4632\ldots 
&\\
&&&\\
\hline
\end{array}
$$

\subsubsection{$n = 10$}

$$
\begin{array}{|c|c|c|c|}
\hline
&&&\\
n=10 \ \ \ \ \ \ \ \ \ \ N&0&1&\ldots \\
{\rm optimization}&&&\\
{\rm criterium}&&&\\
\hline
&&&\\
s.\ref{tru}&1  &\frac{39}{40}=0.975
& \\
&&&\\
\hline
&&&\\
s.\ref{y21}&C_{00}^{(10)} = \frac{16}{5}=3.2&
{C'}^{(10)}_{00} = \frac{96}{55}=1.7(45)
&\\
&&&\\
\hline
&&&\\
s.\ref{straight}& \xi_{10} C_{00}^{(10)}=3.6459\ldots
&\xi_{10}'{C'}_{00}^{(10)} = 1.9372\ldots
&\\
&&&\\
\hline
&&&\\
s.\ref{angles}&\eta_{10} C_{00}^{(10)}=\frac{256\sqrt{5}}{125}
= 4.5794\ldots
&\eta'_{10} {C'}_{00}^{(10)}=2.2680\ldots
&\\
&&&\\
\hline
\end{array}
$$

\subsubsection{$n = 12$}

$$
\begin{array}{|c|c|c|c|}
\hline
&&&\\
n=12 \ \ \ \ \ \ \ \ \ \ N&0&1&\ldots \\
{\rm optimization}&&&\\
{\rm criterium}&&&\\
\hline
&&&\\
s.\ref{tru}&1  &\frac{56}{57}=0.9824\ldots
& \\
&&&\\
\hline
&&&\\
s.\ref{y21}&C_{00}^{(12)} = \frac{16}{3}=5.(3)&
{C'}^{(12)}_{00} = \frac{224}{87} = 2.5747\ldots
&\\
&&&\\
\hline
&&&\\
s.\ref{straight}& \xi_{12} C_{00}^{(12)}=6.1801\ldots
&\xi_{12}'{C'}_{00}^{(12)} = 2.9242\ldots
&\\
&&&\\
\hline
&&&\\
s.\ref{angles}&\eta_{12} C_{00}^{(12)}=\frac{100\sqrt{5}}{27}
= 8.2817\ldots 
&\eta'_{12} {C'}_{00}^{(12)}=3.6566\ldots 
&\\
&&&\\
\hline
\end{array}
$$

\section{A better use of the boundary ring:
the idea and the problem
\label{puzz}}
\setcounter{equation}{0}

\subsection{Boundary ring as a source of ansatze}

A serious drawback of above considerations was that the
power series ansatz (\ref{posey0}),
\be
y_0 = K_{n/2} \sum_{i,j\geq 0} c_{ij}^{(n)} K_{n/2}^{2i}y^{2j}
= \sum_{i,j\geq 0} c_{ij}^{(n)} K_{n/2}^{2i+1}y^{2j},
\label{posey0a}\label{Csol}
\ee
while explicitly taking into account all the symmetries
of the problem, is not {\it a priori} adjusted to satisfy
boundary conditions:
we first solve NG equations to define $c_{ij}$ and
impose boundary conditions at the very end,
considering them an {\it a posteriori} restriction on the
free parameters of NG solutions.
This is of course a usual procedure in differential
equations theory, however, one can attempt to improve it
and impose boundary conditions {\it a priori}, building
them into the ansatz for NG solution.

Such possibility seems to be immediately provided by
the knowledge of boundary ring.
Indeed, all our ansatze should actually belong to
(a completion of) ${\cal R}_\Pi$, and we can require this
at the very beginning, but not at the very end of the
calculation.
This means that instead of (\ref{posey0a}) we can rather
write, in addition to $r^2 = P_2$,
\be
{\cal P}_{n/2} = y_0P_2 B^{(n)}
\label{bcanz}
\ee
where ${\cal P}_{n/2}$ and $P_2$ are elements of our
${\cal R}_\Pi$,
\be
{\cal P}_{n/2} \ \stackrel{(\ref{calPn2})}{=}\
y_0Q_{(n)}(y^2) - K_{n/2}(y_1,y_2)
\ee
and
\be
P_2 = y_0^2+1 -y_1^2-y_2^2 = y_0^2+1-z\bar z,
\ee
while $B^{(n)}$ is some power series,
restricted only by discrete symmetries and by NG equations.
Whatever $B^{(n)}$, eq.(\ref{bcanz}) guarantees that the
resulting $y_0(y_1,y_2)$
automatically satisfies boundary conditions.

Symmetry implies that we can put
\be
B^{(n)} = \sum_{i,j\geq 0} b^{(n)}_{ij}
y_0^{2i}y^{2j}
\label{Bsern}
\ee
and it remains to adjust coefficients $b_{ij}$
to satisfy the NG equations.
Remarkably, this can be done, but, what is worse,
not in a single way
-- and this puts this kind of approach into question.

\subsection{NG equations as recurrence relations for $b_{ij}$
\label{exaB}}

Making use of (\ref{calPn2}),
\be
{\cal P}_{n/2} = y_0Q_{(n)}(y^2) - K_{n/2}(y_1,y_2)
\label{calPn2a}
\ee
we can rewrite (\ref{bcanz}) as
\be
y_0 = \frac{K}{Q -(1-y^2)B - y_0^2B}
\label{Bsol}
\ee
and solve it iteratively for $y_0$, converting
power series $B$ into a new power series for
$y_0(y_1,y_2)$, or, in other words, expressing
coefficients $c_{ij}$ in (\ref{posey0a}) through
$b_{ij}$ in (\ref{Bsern}).
One can develop a diagram technique in the spirit of
\cite{MS} to describe these interrelations.
A very important thing about (\ref{bcanz}),
which allows us to make this trick, is that it
has a structure
\be
y_0(1+ \ldots) = F(y_1,y_2) + O(y_0^2)
\label{linearity}
\ee
with a term which is {\it linear} in $y_0$.
It is this structure that guarantees that $y_0$ is
a single-valued function of $y_1$ and $y_2$, at least in
the vicinity of $y_1=y_2=0$.  There can be even more
interestingly-looking ansatze, like a Riemann-\-surface-\-style
$P_3^2 = P_2F(P_2)$, which are also consistent with boundary
conditions and symmetries, but not with (\ref{linearity}),
and thus they can not be used to provide the simplest
minimal surfaces (though they can describe some less trivial
extremal configurations, at least in principle).
$P_{n/2}$ in (\ref{bcanz}) satisfy the criterium
(\ref{linearity}) for all $n$,
because $s_{n/2}=0$, while all other $s_a\neq 0$ in the products
(\ref{Kn2}).

One can now find $b_{ij}$ either directly, by substituting
our $y_0(y_1,y_2)$ into NG equations or by expressing them
through $c_{ij}$ which we already know, see s.\ref{bexa}
below for some examples.
In this way we discover, first, that ansatz (\ref{bcanz})
is nicely consistent with NG equations: $b_{ij}$ can indeed
be adjusted to satisfy them, and, like in the case of
$c_{ij}$, NG equations become recurrence relations for
$b_{ij}$. Moreover, there are free parameters, and, furthermore,
the set of free parameters is as large as it was in the
case of $c_{ij}$.
In fact, the mapping $\{b\}\rightarrow\{c\}$ appears triangle
and invertible: it looks like (\ref{bcanz}) does not
restrict formal series NG solutions at all!

\subsection{The problem}

This looks like an apparent contradiction.
Boundary conditions, explicitly imposed on NG solutions by
(\ref{bcanz}) should restrict the set of solutions to
a small variety, presumably, consisting of a single function
$y_0(y_1,y_2)$.
However, this does not happen at the level of formal series.
This means that convergence problems can be far more
severe when we switch from the $c$-expansions to $b$-expansions.
Making this promising approach into a working one
remains a puzzling open problem.

\subsection{Toy example and resolution of the puzzle
\label{puzzreso}}

The following toy example sheds light on both the resolution
of the "paradox" and possible ways out.

Consider the simplest possible equation
\be
\dot x = 0
\label{xd0}
\ee
with the boundary condition $x=1$ at $t=1$,
\be
x(t=1)=1
\label{bcxt}
\ee
The variables $x$ and $t$ can be thought of as modeling
$y_0$ and $y_2$ respectively, and since there is no
analogue of $y_1$ the freedom in the choice of solutions
is just single-parametric: $x(t) = x_0$ is arbitrary constant.
Generalizations to higher derivatives and to non-linear
equations are straightforward, but unnecessary:
all important aspects of the problem
are well seen already at the level of (\ref{xd0}).

As an analogue of (\ref{bcanz}) we can write, for example
\be
x = t + (1-x)B(t)
\ee
Indeed, whatever is $B(t)$, solution of this algebraic
equation,
\be
x = \frac{t+B(t)}{1+B(t)}
\label{xvsB}
\ee
always satisfies our boundary condition (\ref{bcxt}).
For example, very different choices of $B$, even
$x$-dependent, like
\be
B=0 \Rightarrow x=t, \nn \\
B=1 \Rightarrow x=\frac{t+1}{2}, \nn \\
B=x \Rightarrow x = \sqrt{t}, \nn \\
B=t^2 \Rightarrow x=\frac{t+t^2}{1+t^2}, \nn\\
\ldots
\ee
all provide $x(t)$, which satisfy (\ref{bcxt}).

Of course, these choices do not provide solutions to
the equation of motion (\ref{xd0}).
However, we can apply all the same methods that we
used in our consideration of Plateau problem.
Eq.(\ref{xd0}) implies an equivalent equation for $B$:
\be
\dot B = \frac{B+1}{t-1}
\ee
which can be either solved explicitly:
\be
B(t) = b_0 - (b_0+1)t
\label{Bxsol}
\ee
or rewritten as recurrence relations
\be
b_1 = -1-b_0, \nn \\
b_2 = 0, \nn \\
b_3 =0, \nn \\
\ldots
\ee
for the coefficients $b_k$ of power series
\be
B(t) = \sum_{k=0} b_k t^k
\label{Bd0}
\ee
Moreover, the coefficients $b_k$ can be easily mapped
to $x_k$ in
\be
x(t) = \sum_{k=0} x_kt^k
\ee
by
\be
x_0 = \frac{b_0}{1+b_0}, \nn\\
x_1 = \frac{1+b_0+b_1}{(1+b_0)^2}, \nn \\
\ldots
\ee
and equations of motion leave exactly one free parameter
in both series: (\ref{xd0}) does not fix $x_0$, while
(\ref{Bd0}) -- $b_0$.
The map $x(t) \leftrightarrow B(t)$ looks one-to-one.

Advantage of this toy example is that here we can resolve
our "paradox".
The answer is that {\it exact} solution to equation of
motion for $B$ belongs to the {\it rare} class of functions
$B$ which violate the relation
\be
(\ref{xvsB}) \Rightarrow (\ref{bcxt})
\label{xNfolbc}
\ee
Namely they all possess the property $B(t=1) = -1$,
which makes (\ref{xNfolbc}) unjust:
\be
x(t)   \ \stackrel{(\ref{xvsB})}{=}\ \frac{t+B(t)}{1+B(t)}
\ \stackrel{(\ref{Bxsol})}{=}\
\left.\frac{t+b_0+b_1t+b_2t^2}{1+b_0+b_1t+b_2t^2}\right|_{
\stackrel{b_1 = - b_0-1}{b_2=0}}
= \frac{b_0(1-t)}{(1+b_0)(1-t)} = \frac{1+b_1}{b_1}\neq 1
\label{xtsol}
\ee
and in particular $x(t=1)\neq 1$.
However, as soon as we substitute {\it exact} solution
for $B$ by any approximation, for example, keep $b_2\neq 0$
in above calculation, (\ref{bcxt}) is immediately recovered:
$x(t=1)=1$ for any $b_2\neq 0$, whatever small!

\begin{figure}\begin{center}
{\includegraphics[width=150pt,height=150pt] {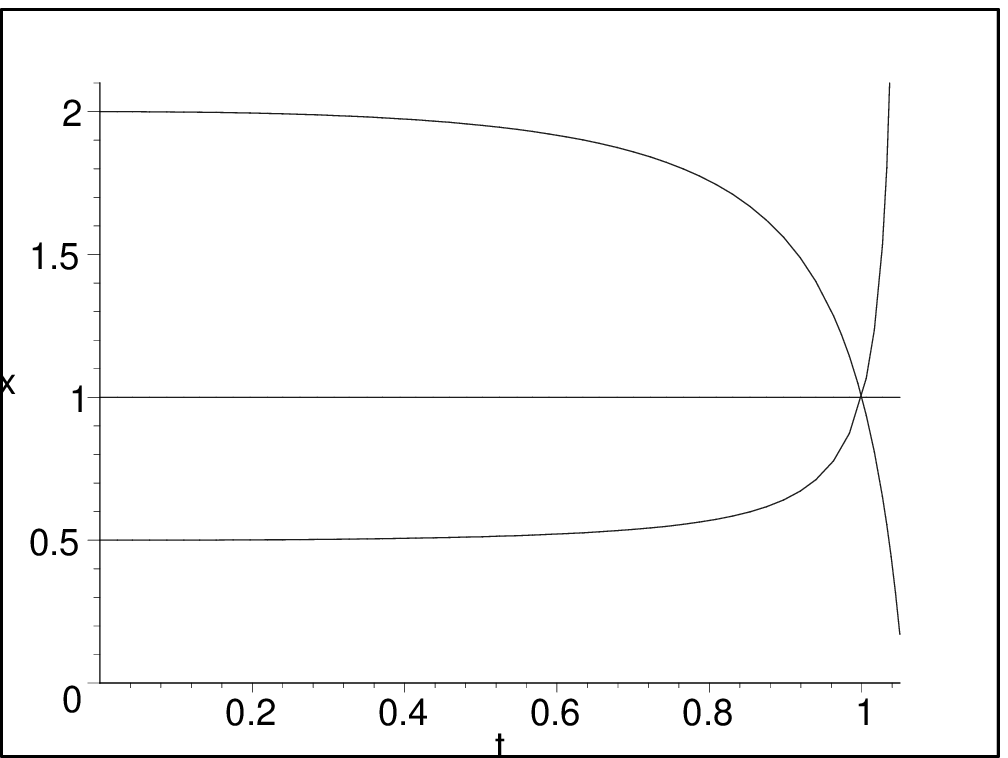}}
{\includegraphics[width=150pt,height=150pt] {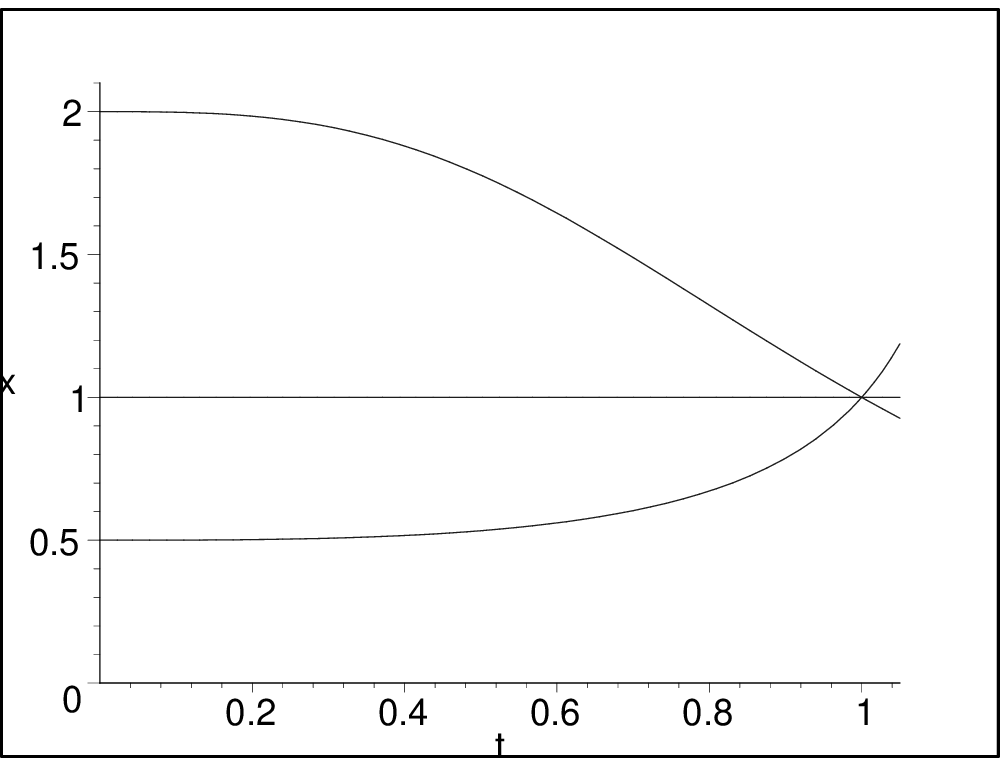}}
{\includegraphics[width=150pt,height=150pt] {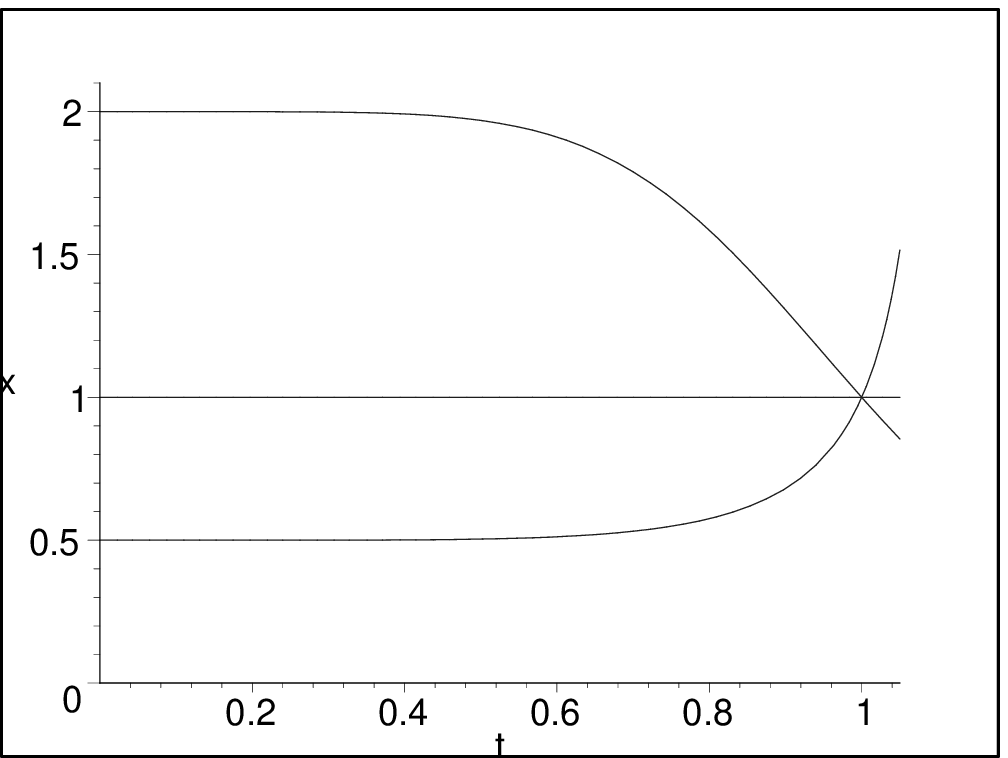}}
\caption{\footnotesize{ Left picture shows $x(t)$ as given by
eq.(\ref{xtsol}) at different values of parameters: $b_1 = 1$,
$b_2=-0.1$ (upper line), $b_1 = \infty $, $b_2=0$ (middle line) and
$b_1=-2$, $b_2=0.1$ (lower line). At $t<0.7-0.8$ this $x(t)$ behaves
as $x(t) = const = \frac{b_1+1}{b_1}$, i.e. $2$, $1$ and
$\frac{1}{2}$ respectively and satisfies equation of motion $\dot x
= 0$. In the vicinity of $t=1$, however, equation is violated so
that $x(t)$ satisfies the boundary condition $x(1)=1$ instead. The
size of the deviation domain is regulated by $b_2$: the smaller
$b_2$ the smaller the region of deviation -- and the more drastic is
jump of $x(t)$ from exact solution to the boundary condition. The
true solution $x(t)=1$ -- which satisfies both equation of motion
and boundary condition -- is the only one which behaves smoothly
under the variation of $b_2$. The two other pictures show the same
phenomenon for eq.(\ref{xtsolalt}), for different number of
iterations, $N=3$ (middle picture) and $N=6$ (right picture), and
for different values of the free parameter, $x_0=2$ (upper line),
$x_0=1$ (middle line) and $x_0=1\over 2$ (lower line). Again, the
true solution with $x_0=1$ is distinguished by smooth dependence on
the iteration number $N$. The difference with example in the left
picture is that $x(t)$ is non-singular even beyond the segment
$t=[0,1]$, instead $B(t)$ develops singularity at $t=1$. }}
\label{Figxt}
\end{center}\end{figure}

What happens is that for small $b_2$ this $x(t)$ changes
abruptly from $\frac{b_1+1}{b_1}$ to $1$ in a small
(of the size $b_2/b_1$) vicinity of $t=1$, see Fig.\ref{Figxt}.
Thus violation of equations of motion is large,
but it takes place in a small domain: the series for $B$
are not {\it uniformly} convergent.

This explains our observations in s.\ref{exaB}:
consideration of approximate solutions with ansatz (\ref{bcanz})
should and does provide a perfect description of boundary
conditions -- but at expense of NG equations (what is not
so easy to observe in pictures). Equations will not be violated
only for appropriately fixed free parameters, and now we
understand the criterium: the free parameters should be
adjusted so that there is no abrupt change of our
would-be solutions in close vicinity of the boundary.

\bigskip

One can also analyze other toy examples, which can be
closer to realistic boundary rings.
It can make sense to substitute (\ref{bcxt}) say, by
\be
x = t + x(1-t)B
\label{bcxt1}
\ee
keeping in mind that $x$ models the ratio $y_0/y_2$
and $t$ models $y_1^2$, so that (\ref{bcxt1}) resembles
(\ref{bcanz}) at small $y_2$.
Looking at this example one can see that {\it exact} solutions
for $B(t)$ blow up at $t=1$, what is a slight additional
complication, though it does not change our conclusions
implied by analysis of (\ref{bcxt}).
In particular, if exact $B(t)$ is substituted by its
truncation at $N$-th level (i.e. if the first $N$ terms of
$t$-expansion of $B(t)$ are kept), then the corresponding
\be
x_N(t) = \frac{x_0}{1+t^N(x_0-1)}
\label{xtsolalt}
\ee
and we see that the domain of deviation from exact
solution $x=x_0$ is getting closer and closer to $t=1$ with
increase of $N$. A typical behavior for $x_0\neq 1$
in the vicinity of the boundary $t=1$ has strong $N$-dependence,
and the true solution $x=x_0=1$, which satisfies {\it both} the
differential equation and boundary condition is
distinguished by the {\it lack} of such $N$-dependence.

\subsection{On continuation of solutions beyond $\Pi$
\label{solcontin}}

To further emphasize the difference between
approximate solutions from s.\ref{exa} and s.\ref{puzz},
it deserves mentioning also that they have principally
different behavior {\it outside} the polygon-bounded domain.
Implication of  (\ref{bcanz})  is that $r^2=P_2$ vanishes not only
on  $\Pi$ but also on all their continuations into the outside:
on entire $n$ straight lines in $y$-space, which contain the $n$
segments of $\Pi$. This can easily be an artefact of polynomial-based
consideration behind (\ref{bcanz}), which is not necessarily
preserved in transition to functional analysis, like in s.\ref{exa}.
On the other hand, solutions to Plateau problem in the flat Euclidean
space are known to have this property \cite{Oss}.
Still, one should be cautious about this analogy, because in
Euclidean case the basic equation is ordinary Laplace and
minimal surfaces are too closely associated with complex analytic
functions and the Schwarz reflection principle.

\subsection{Recurrence relations for
$b_{ij}^{(n)}$ from NG equations at $n=4$ and $n=6$
\label{bexa}}

\subsubsection{Relation between $b_{ij}$ and $c_{ij}$
at the level of generating functions}

After providing some arguments in favor of (\ref{bcanz})
-- and before showing impressive pictorial confirmation
in s.\ref{bcexa} below -- we need to address the
most difficult issue in this approach: solution of NG
equations.
In this paper we restrict ourselves to description of
recurrence relations for coefficients of $B$ in (\ref{bcanz})
and to demonstration of one-to-one correspondence between
$b$- and $c$-series from s.\ref{exaid}.
Problems of (uniform) convergence and
theoretically-reliable approximation methods will be
addressed elsewhere.

There are different possibilities to {\it define} a
formal series for $B$, for example, one can take
\be
B = \sum_{i,j\geq 0} b_{ij}^{(n)} y_0^{2i}y^{2j}
\label{Bexpan}
\ee
or, given the symmetries of the problem
(in our $Z_n$-symmetric case),
\be
B = \sum_{i,j\geq 0} \tilde b_{ij}^{(n)} K_{n/2}^{2i} y^{2j}
\label{Bexpantilde}
\ee
The advantage of the first representation is that it
does not have explicit $n$-dependence and symmetry restrictions,
however, these advantages can easily become {\it disadvantages}
in particular numerical considerations (making calculations
in symmetric situations more tedious than really needed),
then the second representation can be used.
In particular, with the second representation for $B$
eq.(\ref{bcanz}) is always (for all $n$)
a cubic equation in $y_0$,
what allows to formally rewrite it as analytical expression
for $y_0(y_1,y_2)$
(despite based on low-efficient Cardano's formulas,
it drastically simplifies MAPLE calculations).

Substituting (\ref{Bexpan}) into (\ref{bcanz}), we can
solve for $y_0$ iteratively:
\be
y_0 = \frac{K}{Q(y^2) - (1-y_1^2-y_2^2+y_0^2)B} =
\frac{K(y_1,y_2)}{Q(y^2) - (1-y^2)B_0(y^2)} + O(K^3),
\ee
where
\be
B_0 = \sum_{j\geq 0} b_{0j}y^{2j} =
\sum_{j\geq 0} \tilde b_{0j}y^{2j}
\ee
(obviously, at $i=0$ coefficients $\tilde b_{0j} = b_{0j}$).
Comparing this to
\be
y_0 = \sum_{ij} c_{ij} K^{2i+1}y^{2j} =
K(y_1,y_2) Y_0(y^2) + O(K^3), \ \ \ \
Y_0(y^2) = \sum_{j\geq 0} c_{0j}y^{2j}
\ee
we immediately obtain:
\be
Y_0(y^2) = \frac{1}{Q(y^2) - (1-y^2)B_0(y^2)}
\label{YthrB}
\ee
what provides a general expression for the coefficients
$c_{0j}$ through $b_{0j}$ or -- vice versa,
of $b_{0j}$ through $c_{0j}$, if (\ref{YthrB})
is rewritten as
\be
B_0(y^2) = \frac{Q(y^2)Y_0(y^2)-1}{(1-y^2)Y_0(y^2)}
\label{BthrY}
\ee
For example,
\be
c_{00} = \frac{1}{1-b_{00}}, \nn\\
c_{01} = \frac{\frac{n-4}{8} + b_{01}-b_{00}}{(1-b_{00})^2},\nn\\
\ldots
\label{cvbn}
\ee
and
\be
b_{00} = \frac{c_{00}-1}{c_{00}},\nn\\
b_{01} = \frac{12-n}{8} +\frac{c_{01}-c_{00}}{c_{00}^2}, \nn \\
\ldots
\label{bcvn}
\ee
This demonstrates that we indeed get a one-to-one relation,
but when expressed in terms of generating functions,
it is an equation with singularities at finite points,
what signals about existence of potential convergence problems.

It is easy to find in a similar way the relations between generating
functions $Y_i(y^2)$ and $B_i(y^2)$ (note that for $i\geq 1$ $\tilde
B_i(y^2)$ differs from $B_i(y^2)$ -- by an easily derived relation).
However, we can also proceed in a more primitive way: substitute
(\ref{Bexpan}) into NG equation and obtain recurrence relations for
$b_{ij}$ -- just in the same way as we did for $c_{ij}$ in
s.\ref{exaid}. As in that case these relations depend on $n$, and we
list the first few for $n=4$ and $n=6$. We give also explicit
examples of triangular invertible  relations between individual
$b_{ij}^{(n)}$ and $c_{ij}^{(n)}$. Like above, index $n$ is often
omitted to avoid further overloading of formulas.

\subsubsection{$n=4$. Recurrence relations
\label{breren4}}

\be
b_{01} = \frac{\big(4-5b_{00}\big)b_{00}}{6(1-b_{00})},\nn\\
b_{02} = \frac{\big(63-262b_{00}+316b_{00}^2-115b_{00}^3\big)b_{00}}
{144(1-b_{00})^3} - \frac{3b_{10}}{18(1-b_{00})^2},\nn\\
b_{03} = \frac{\big(1188 -9138b_{00} +23969b_{00}^2
-28626b_{00}^3+15990b_{00}^4 - 3385b_{00}^5\big)b_{00}}
{4320(1-b_{00})^5} - \nn \\
-\frac{\big(84 - 178b_{00} +89b_{00}^2\big)b_{10}}{480(1-b_{00})^4}
-\frac{3b_{11}}{10(1-b_{00})^2}, \nn\\
\ldots \nn \\
b_{11}=-\frac{\big(3 -86b_{00} +137b_{00}^2 -53b_{00}^3 \big)b_{00}
}{126(1-b_{00})^3} +
\frac{\big(69-118b_{00}+59b_{00}^2\big)b_{10}}{42(1-b_{00})^2}
\nn \\
 \ldots \ee These relations express all $b_{ij}$ in terms of the
free parameters $b_{i0}$, which are not fixed by NG equations and
boundary conditions. At $n=4$ they admit a solution $b_{ij}=0$,
corresponding to $c_{ij} = \delta_{i1}\delta_{j1}$, i.e. to
Alday-Maldacena square solution $y_0=y_1y_2$, see s.\ref{square}
above. However, $b_{ij}^{(n)} = 0$ will not be a solution at higher
$n\geq 6$. The limit $c^{(n)}_{ij}=0$, leading to the $n=\infty$
solution (\ref{ninftyNG}) with a unit-circle boundary from the
$Z_n$-symmetric ansatz, looks complicated in $b$-variables, even at
$n=4$.



\subsubsection{$n=6$. Recurrence relations for $b_{ij}$
\label{breren6}}

Similarly, for $n=6$ we get:
\be
b_{01}=\frac{1+5b_{00}}{8},\nn\\
b_{02}=\frac{28+130b_{00}-185b_{00}^2}{80(1-b_{00})},\nn\\
b_{03} = \frac{35+127b_{00}-471b_{00}^2+285b_{00}^3 - 24b_{10}}
{512(1-b_{00})^2},\nn \\
b_{04}=\frac{38284 + 93700b_{00} -766310b_{00}^2 + 1022700b_{00}^3
-390075b_{00}^4}{716800(1-b_{00})^3}
+\frac{123 b_{10}}{1280(1-b_{00})^2},\nn\\
b_{05}=\frac{5204441 +5826459b_{00}-136463910b_{00}^2
+301180330b_{00}^3-243508650b_{00}^4+67625250b_{00}^5
}{126156800(1-b_{00}^4)} - \nn \\
-\frac{3\big(10630 -21098b_{00} +10549b_{00}^2\big) b_{10}}
{225280(1-b_{00})^4}, \nn\\
\ldots \nn \\
b_{11} = \frac{28-290b_{00}+505b_{00}^2}{1600(1-b_{00})} +
  \frac{13b_{10}}{10}, \nn\\
b_{12} = \frac{249 -18609b_{00} +73260b_{00}^2 -47740b_{00}^3}
{70400(1-b_{00})^2}
+\frac{\big( 2840-5302b_{00}+2651b_{00}^2\big)b_{10}}
{1760(1-b_{00})^2},\nn\\
\ldots
\label{bcoef}
\ee
These formulas express all $b_{ij}$ in terms of the free
parameters (moduli) $b_{i0}$,
which are not immediately fixed by NG equations and the
boundary conditions.
As explained in \cite{mmt1,mmt2} these moduli (whenever they exist)
are not necessarily
inessential in consideration of $\epsilon$-regularized NG actions
(areas) in the study of Alday-Maldacena program.
As also explained in these papers, there are two ways to
deal with such moduli: either understand their {\it raison d'etre}
and eliminate in a rigorous way (say, using Virasoro constraints
in the case of \cite{mmt1,mmt2}, or analysis from
s.\ref{puzzreso} in our present situation)
-- what can be quite a tedious thing to do,--
or simply minimize the {\it answer}, i.e. regularized area,
w.r.t. the variation of moduli
-- this can be a simpler thing to do in practice and,
even more important, this
can also reveal some additional hidden structures
behind our problem (like the height function in \cite{mmt1,mmt2}).

\subsubsection{$n=6$: Relation between $b_{ij}$ and $c_{ij}$}


As a simple example of this relation we present a few first
formulas for $n=6$.
The first two lines coincide with (\ref{cvbn}).
\be
c_{00}= \frac{1}{1-b_{00}}, \nn\\
c_{01} = \frac{(1-4b_{00}+4b_{01})}{4(1-b_{00})^2}
= \frac{(1-4b_{00})}{4(1-b_{00})^2}
+\frac{b_{01}}{(1-b_{00})^2},\nn\\
c_{02} = \frac{(1-4b_{00})^2}{16(1-b_{00})^3}
-\frac{(1+2b_{00}-4b_{01})b_{01}}{16(1-b_{00})^3}
+\frac{b_{02}}{(1-b_{00})^2}, \nn\\
c_{03} = \frac{ (1-4 b_{00})^3}{64(1-b_{00})^4}
-\frac{(5+4b_{00})(1-4b_{00}+4b_{01})b_{01}}{16(1-b_{00})^4}
+\frac{b_{01}^3}{(1-b_{00})^4}
-\frac{ (1+2b_{00}-4b_{01})b_{02}}{2(1-b_{00})^3}
+\frac{b_{03}}{(1-b_{00})^2},\nn \\
\ldots \nn \\ \nn \\
c_{10} = \frac{b_{00}+b_{10}}{(1-b_{00})^4},\nn \\
\ldots
\ee


\subsection{Approximate NG solutions with exact
boundary conditions
\label{bcexa}}

Thus, one is finally prepared for the final set of examples.
Similarly to s.\ref{cexa}, one can build a set of plots in order
to see how the approximation works..
The difference is that now one has to use
\be
y_0 = \frac{K_3}{1-\frac{y^2}{4}}
\label{y0Kbc}
\ee
instead of (\ref{n6cK}) and
\be
y_0 = \frac{K_3}{1-\frac{y^2}{4} - (1+y_0^2-y^2)b_{00}}
\label{y0Ky2bc}
\ee
instead of (\ref{n6cKc00}).
Similar modifications has to be made for other values of $n$.
Note that the r.h.s. of (\ref{y0Kbc}) can not
be multiplied by any constant without breaking (\ref{bcanz}):
coefficient at the r.h.s. is strictly unity.
As to (\ref{y0Ky2bc}), it contains a free parameter $b_0$,
but equation still needs to be resolved w.r.t. $y_0$
(actually, this is a cubic equation).

In this case, any plot confirms that the boundary conditions are
exactly satisfied -- what looks impressive after comparison
with the results of s.\ref{exa}.
Moreover, in accordance with expectations of s.\ref{solcontin},
matching extends to entire straight lines, beyond $\Pi$ itself.
Unfortunately, we did not yet invent an equally nice visualization
of deviations from the NG equations -- which, as we discussed,
can be strong in the vicinity of the boundary $\Pi$,
unless the remaining free parameters (like $b_{00}$) are
adjusted to some unique true value.
Therefore, it remains unclear whether this type of criterium
can be effective for the practical search of these true values.
Still even the very rough approximation like (\ref{y0Kbc})
can already be applied to the study of string/gauge duality.
The next step to be made is evaluation of regularized areas
for configurations like (\ref{y0Kbc}).

\section{Conclusion
\label{conc}}
\setcounter{equation}{0}

In this paper we discussed a systematic approach to
construction of NG solutions in AdS backgrounds with
polygons, consisting of null vectors,
in the role of bounding contour at infinity.

It is suggested to look for NG solutions in the form of
formal series, restricted by symmetries (if any) and
boundary conditions.
Boundary conditions can be explicitly taken
into account by expanding formal-series in elements of
the boundary ring, which consists of all polynomials
vanishing at the boundary polygon.
NG equations provide recurrence relations for the
coefficients of formal series.

Actually, boundary conditions can be imposed on
formal series both before and after their substitution
into NG equations.

While the first options (it is considered in s.\ref{puzz})
seems to be conceptually and
aesthetically better, it does not provide immediate practical
way to fix the remaining free parameters from the first
principles.
For application purposes this is not obligatory a problem,
because approximately evaluated regularized area can be simply
minimized w.r.t. to such parameters -- resembling the way
the $z$-variables have been handled in \cite{mmt1}.

The second option
(considered in s.\ref{exa})
is less attractive, instead it
produces spectacularly accurate approximations to
the would-be exact solutions, and even the boundary
conditions seem to be matched pretty well.
Inaccuracies seem to increase in the vicinities of the
polygon angles, which give dominant contributions to
the IR divergencies of regularized areas.
This is one of the problems which should be addressed
when one tries to make use of these methods in the
further studies of string/gauge dualities.
Note that this problem (at least at the level of quadratic
divergencies) is {\it a priori} avoided if
the first option is chosen, because boundary conditions
are imposed {\it exactly}.

To conclude, our concrete suggestion for further
development of Alday-Maldacena program
is to take the $B=0$ version of (\ref{bcanz}),
i.e.
\be
y_0 = \frac{K_{n/2}}{Q_{(n)}(y^2)}\, , \nn \\
r = \sqrt{y_0^2+1-y_1^2-y_2^2}
\label{finalanza}
\ee
as the first approximation to the minimal surface,
and concentrate on developing technique for evaluating
regularized areas for such surfaces
(see table in s.\ref{brZnsympol} for a list
of $K_{n/2}$ and $Q_{(n)}$).
After this is done,
one can begin including corrections to (\ref{finalanza}),
implied by NG equations, which can be systematically
found by the methods of the present paper.
NG equations fix functional form ($y_1$ and $y_2$ dependence)
of corrections in any given order, and remaining free
parameters can be fixed by the general method of \cite{mmt1}:
by extremizing the resulting {\it integral}
(see also comments at the end of s.\ref{breren6}).
Generalization of (\ref{finalanza}) beyond the $Z_n$-symmetric
polygons $\Pi$ will be considered elsewhere.

\section*{Acknowledgements}

We are grateful to T.Mironova for help with the pictures. H.Itoyama
acknowledges the hospitality of ITEP during his visit to Moscow when
this paper started. A.Morozov is indebted for hospitality to Osaka
City University and for support of JSPS during the work on this
paper. The work of H.I. is partly supported by Grant-in-Aid for
Scientific Research 18540285 from the Ministry of Education, Science
and Culture, Japan and the XXI Century COE program "Constitution of
wide-angle mathematical basis focused on knots" (H.I.), the work of
A.M.'s is partly supported by Russian Federal Nuclear Energy Agency,
by the joint grant 06-01-92059-CE,  by NWO project 047.011.2004.026,
by INTAS grant 05-1000008-7865, by ANR-05-BLAN-0029-01 project and
by the Russian President's Grant of Support for the Scientific
Schools NSh-8004.2006.2, by RFBR grants 07-02-00878 (A.Mir.) and
07-02-00645 (A.Mor.).


\begin{thebibliography}{12}

\bibitem{BDS} Z.Bern, L.Dixon and V.Smirnov,
{\it Iteration of Planar Amplitudes in Maximally Supersymmetric
Yang-Mills Theory at Three Loops and Beyond}, Phys.Rev. {\bf D72}
(2005) 085001, hep-th/0505205

\bibitem{BA} L.Lipatov, {\it Evolution Equations in CQD}, ICTP
Conference, May, 1997\\
J.Minahan and K.Zarembo, {\it The Bethe-Ansatz for N=4 Super
Yang-Mills},
JHEP {\bf 0303} (2003) 013, hep-th/0212208\\
N.Beisert, C.Kristjansen and M.Staudacher, {\it The Dilatation
Operator of Conformal N=4 Super Yang-Mills Theory},
Nucl.Phys. {\bf B664} (2003) 131-184, hep-th/0303060\\
N.Beisert, B.Eden and M.Staudacher, {\it Transcendentality and
Crossing}, J.Stat.Mech. {\bf 0701} (2007) P021, hep-th/0610251\\
M.Staudacher, {\it Dressing, Nesting and Wrapping in AdS/CFT},
Lecture at RMP Workshop, Copenhagen, 2007

\bibitem{BHT} A.Brandhuber, P.Heslop and G.Travaglini,
{\it MHV Aplitudes in $N=4$
Super Yang-Mills and Wilson Loops}, arXiv:0707.1153

\bibitem{mmt1} A.Mironov, A.Morozov and T.N.Tomaras, {\it On n-point Amplitudes
in N=4 SYM}, JHEP {\bf 11} (2007) 021, arXiv:0708.1625

\bibitem{am1} L.Alday and J.Maldacena,
{\it Gluon Scattering Amplitudes at Strong Coupling},
arXiv:0705.0303

\bibitem{amfirst} S.Abel, S.Forste and V.Khose, {\it Scattering Amplitudes in
Strongly Coupled $N=4$ SYM from Semiclassical Strings in AdS},
arXiv:0705.2113

\bibitem{AMoth2} E.Buchbinder, {\it Infrared Limit of Gluon Amplitudes at Strong
Coupling}, arXiv:0706.2015

\bibitem{AMoth3} J.Drummond, G.Korchemsky and E.Sokatchev,
{\it Conformal properties of four-gluon planar amplitudes and Wilson
loops}, arXiv:0707.0243

\bibitem{AMoth5} F.Cachazo,
M.Spradlin and A.Volovich, {\it  Four-Loop Collinear Anomalous
Dimension in N = 4 Yang-MillsTheory}, arXiv:0707.1903

\bibitem{AMoth6} M.Kruczenski,
R.Roiban, A.Tirziu and A.Tseytlin, {\it Strong-Coupling Expansion of
Cusp Anomaly and Gluon Amplitudes from Quantum Open Strings in
$AdS_5\times S^5$}, arXiv:0707.4254

\bibitem{AMoth7} Z.Komargodsky and S.Razamat,
{\it  Planar Quark Scattering at Strong Coupling and Universality},
arXiv:0707.4367

\bibitem{AMoth8} L.Alday and J.Maldacena, {\it Comments on Operators
with Large Spin}, arXiv:0708.0672; {\it  Comments on gluon
scattering amplitudes via AdS/CFT}, arXiv:0710.1060

\bibitem{AMoth9} A.Jevicki,
C.Kalousios, M.Spradlin and A.Volovich, {\it Dressing the Giant
Gluon}, arXiv:0708.0818

\bibitem{MMT} A.Mironov, A.Morozov and T.N.Tomaras, {\it On n-point Amplitudes
in N=4 SYM}, arXiv:0708.1625

\bibitem{AMoth10} H.Kawai and T.Suyama, {\it Some Implications
of Perturbative Approach to AdS/CFT Correspondence}, arXiv:0708.2463

\bibitem{AMoth11}
S.G.Naculich and H.J.Schnitzer, {\it  Regge behavior of gluon
scattering amplitudes in N=4 SYM theory}, arXiv:0708.3069

\bibitem{AMoth12} R.Roiban
and A.A.Tseytlin, {\it  Strong-coupling expansion of cusp anomaly
from quantum superstring}, arXiv:0709.0681

\bibitem{CS1} J.M.Drummond, J.Henn, G.P.Korchemsky and E.Sokatchev,
{\it On planar gluon amplitudes/Wilson loops duality},
arXiv:0709.2368

\bibitem{CS2} D.Nguyen, M.Spradlin and A.Volovich, {\it  New Dual Conformally
Invariant Off-Shell Integrals}, arXiv:0709.4665

\bibitem{AMoth14} J.McGreevy and
A.Sever, {\it  Quark scattering amplitudes at strong coupling},
arXiv:0710.0393

\bibitem{AMoth15} S.Ryang, {\it  Conformal SO(2,4) Transformations of
the One-Cusp Wilson Loop Surface}, arXiv:0710.1673

\bibitem{AMoth16} D.Astefanesei,
S.Dobashi, K.Ito and H.S.Nastase, {\it  Comments on gluon 6-point
scattering amplitudes in N=4 SYM at strong coupling},
arXiv:0710.1684

\bibitem{mmt2} A.Mironov, A.Morozov and T.Tomaras,
{\it Some properties of the Alday-Maldacena minimum},
arXiV:0711.0192 (hep-th), to appear in Physics Letters {\bf B}

\bibitem{popo} A.Popolitov, {\it On coincidence of
Alday-Maldacena-regularized $\sigma$-model and Nambu-Goto areas of
minimal surfaces}, arXiv:0710.2073

\bibitem{GY} Gang Yang, {\it  Comment on the Alday-Maldacena solution in
calculating scattering amplitude via AdS/CFT}, arXiv:0711.2828

\bibitem{amlast} K.Ito, H.S.Nastase and K.Iwasaki, {\it Gluon scattering
in ${\cal N}=4$ Super Yang-Mills at finite temperature},
arXiv:0711.3532

\bibitem{KT} R.Kallosh and A.Tseytlin, {\it Simplifying
Superstring Action on $AdS_5\times S^5$}, JHEP {\bf 9810} (1998)
016, hep-th/9808088

\bibitem{pream} N.Drukker, D.Gross and H.Ooguri,
{\it Wilson Loops and Minimal Surfaces}, Phys.Rev. {\bf D60} (1999)
125006, hep-th/9904191\\
Yu.Makeenko,{\it  Light-Cone Wilson Loops
and the String/Gauge Correspondence}, JHEP {\bf 0301} (2003) 007,
hep-th/0210256

\bibitem{ka} M.Kruczenski,
{\it A Note on Twist Two Operators in $N=4$ SYM and Wilson Loops in
Minkowski Signature}, JHEP {\bf 0212} (2002) 024, hep-th/0212115

\bibitem{KLOV} A.Kotikov, L.Lipatov and V.Velizhanin,
{\it Anomalous Dimensions of Wilson Operators in $N=4$
SYM Theory}, Phys.Lett. {\bf B557} (2003) 114-120, hep-ph/0301021 \\
A.Kotikov, L.Lipatov, A.Onishchenko and V.Velizhanin, {\it
Three-Loop Universal Anomalous Dimension of the Wilson Operators in
$N=4$ SUSY Yang-Mills Model}, Phys.Lett. {\bf B595} (2004) 521-529;
Erratum-ibid. {\bf B632} (2006) 754-756, hep-th/0404092

\bibitem{NOLAL}
{\footnotesize For modernized presentation of the subject see}\\
V.Dolotin and A.Morozov,
{\it The Universal Mandelbrot Set, Beginning of the Story},
World Scientific, 2006; hep-th/0501235; hep-th/0701234; \\
Andrey Morozov, {\it JETP Letters}, {\bf 86}, N11 (2007); arXiv:0710.2315 \\
V.Dolotin and A.Morozov,
{\it Introduction to non-linear Algebra}, World Scientific, 2007;
hep-th/0609022; \\
Sh.Shakirov, {\it Theor.Math.Phys.}, {\bf 153(2)} (2007) 1477-1486;
math/0609524\\
{\footnotesize For traditional textbooks see}\\
S.Lang, {\sl Algebra}, Addison-Wesley Seires in Mathematics, 1965\\
B.L.Van der Varden, {\sl Algebra}, I, II, Springer-Verlag, 1967,
1971

\bibitem{MS} A.Morozov and M.Serbyn,
{\it Theor.Math.Phys., to appear},
hep-th/0703258

\bibitem{Oss}
{\footnotesize See discussion of Schwarz reflection
principle at the bottom of page 16 of English edition or at
page 28 of Russian edition in} \\
P.Hoffman and H.Karcher, {\it Complete Embedded Minimal
Surfaces of Finite Total Curvature}, in
{\it Encyclopaedia of Math.Science}, {\bf 90},
{\it Geometry V. Minimal Surfaces}, ed.R.Osserman,
Springer

\end{thebibliography}
\end{document}